\documentclass{article}

\pdfoutput=1

\usepackage{arxiv}
\usepackage[utf8]{inputenc} 
\usepackage[T1]{fontenc}    
\usepackage{hyperref}       
\usepackage{booktabs}       
\usepackage{amsfonts}       
\usepackage{nicefrac}       
\usepackage{microtype}      
\usepackage{graphicx}
\usepackage[flushleft]{threeparttable}
\usepackage{subfigure}
\usepackage{caption}
\usepackage[square,numbers]{natbib}
\usepackage{doi}

\title{Evaluation of Force Fields for Molecular Dynamics Simulations of Platinum in Bulk and Nanoparticle Forms}

\date{}	

\author{
 Ingrid M. Padilla Espinosa \\
  Department of Mechanical Engineering\\
  University of California, Merced\\
  Merced, CA 95340, USA \\
  \texttt{ipadillaespinosa@ucmerced.edu} \\
   \And
 Tevis D.B. Jacobs \\
  Department of Mechanical Engineering
and Materials Science\\
  University of Pittsburgh\\
  Pittsburgh, PA 15261, USA \\
  \texttt{tjacobs@pitt.edu} \\
  \And
 Ashlie Martini \\
  Department of Mechanical Engineering\\
  University of California, Merced\\
  Merced, CA 95340, USA \\
  \texttt{amartini@ucmerced.edu} \\
}

\renewcommand{\headeright}{}
\renewcommand{\undertitle}{Published in: \textit{J. Chem. Theory Comput.} 2021, 17, 7, 4486–4498}

\hypersetup{
    pdftitle={Evaluation of Force Fields for Molecular Dynamics Simulations of Platinum in Bulk and Nanoparticle Forms},
    pdfsubject={cond-mat.mes-hall, cond-mat.mtrl-sci},
    pdfauthor={I.M. Padilla Espinosa, T.D.B. Jacobs, A. Martini},
    pdfkeywords={Platinum, Nanomaterials, Molecular Dynamics, Force Fields}
}

\begin{document}
\maketitle
\begin{abstract}
Understanding the size- and shape-dependent properties of platinum nanoparticles is critical for enabling design of nanoparticle-based applications with optimal and potentially tunable functionality. Towards this goal, we evaluated nine different empirical potentials with the purpose of accurately modeling faceted platinum nanoparticles using molecular dynamics simulation. First, the potentials were evaluated by computing bulk and surface properties -- surface energy, lattice constant, stiffness constants, and the equation of state -- and comparing these to experimental measurements and quantum mechanics calculations. Then, the potentials were assessed in terms of the stability of cubic and icosahedral nanoparticles with faces in the \{100\} and \{111\} planes, respectively. Although none of the force fields predicts all the evaluated properties with perfect accuracy, one potential -- the embedded atom method formalism with a specific parameter set -- was identified as best able to model platinum in both bulk and nanoparticle forms.
\end{abstract}

\section{Introduction}
The catalytic, electronic, optical, and thermal properties of platinum nanoparticles make them valuable materials for the pharmaceutical and specialty chemistry industries,~\cite{Singh2001,Armor2011} for fuel cells,~\cite{Firouzjaie2020, Stephen2019, Jung2019} and solar energy conversion,~\cite{ Bruix2012, Calogero2011} as well as for biomedical applications,~\cite{Brondani2009} among others.  However, platinum is an expensive transition metal and its reserves are limited. Therefore, studies have focused on improving platinum nanoparticle properties in an effort to reduce the amount of platinum required and decrease the associated costs.~\cite{Wang2020,Garlyyev2019}

Studies on catalysis have demonstrated that the efficiency of platinum nanoparticles increases as the particle size decreases ~\cite{Rioux2008} as a result of the increased surface-to-volume ratio.~\cite{Zang2015}
When the size of the particle is reduced to just a few nanometers, the nanoparticles form facets and remain stable in shapes such as icosahedra,~\cite{Rodriguez1996, Wu2017} tetrahedra,~\cite{Narayanan2016,Chiu2011} cuboctahedra or “quasi spherical”,~\cite{Ahmadi2019, Narayanan2016,Song2004} cubes,~\cite{Song2004, Narayanan2016, Chiu2011, Fu2013} and truncated octahedra,~\cite{Li2000, Ahmadi2019, Song2004} 
These different nanoparticle shapes have different surface facets (mostly \{111\}, \{110\}, and \{100\} ~\cite{NilssonPingel2018, Song2004}) and different numbers of atoms at corners and edges that affect their overall properties.~\cite{Cao2016, Zang2015} Since nanoparticle properties are dependent on shape and size,~\cite{RoldanCuenya2013} it is desirable to understand and predict the evolution of their shape- and size-dependent properties throughout the service life of a given application.

Due to their small size, a comprehensive understanding of the shape- and size-dependence of nanoparticle properties is difficult to achieve experimentally.  
For example, the dependence of strength and deformation on particle size and shape has been investigated. Studies on the deformation of faceted gold nanoparticles using an anvil cell~\cite{Parakh2020} have shown the occurrence of Shockley dislocations along facet vertices. However, to prevent sintering, the maximum load achieved with this method is limited, and the stress cannot be calculated directly, instead it is inferred from lattice changes. Separate \textit{in situ} studies on the compression of silicon~\cite{DeneenNowak2007} and gold~\cite{Casillas2012} nanoparticles using transmission electronic microscopy (TEM)--atomic force microscopy (AFM) have characterized reversible and irreversible deformation, as well as hardening--deformation relationships. But, these experiments are limited by AFM tip tilting, slip off of the nanoparticle from the tip, and lack of controllable strain rate. Additionally, the strain distribution caused by the substrate support in platinum nanoparticles has been studied with a scanning transmission electron microscope (STEM)~\cite{Daio2015} and has shown that the nanoparticles experience larger strain if the substrate is SnO\textsubscript{2} compared to C, affecting the electrochemical activity.~\cite{Wang2009} However, image resolution was adversely affected by the fast scans required to prevent beam damage and due to contamination from outgassing. For platinum nanoparticles, structural changes and strain during catalytic reactions have been observed using Bragg coherent diffraction imaging.~\cite{Choi2020} 
However, these measurements are sensitive to fluctuations in the scans from temperature differences between the nanoparticles and the media.
All of these examples demonstrate the significant challenges to using experimental approaches to link the material properties and behavior of nanoparticles.

An alternative to experimental methods is simulation. The stability of clusters and very small nanoparticles, including platinum nanoparticles, has been studied previously~\cite{Haberlen1997,Xiao2004,Chepkasov2018, Nanba2017, Wei2016} using density functional theory (DFT). 
Such DFT calculations can predict interatomic distances, cohesive energy, and binding energy with high accuracy, as well as the shape-stability and surface reconstruction of nanoparticles.
However, DFT is severely limited in the number of atoms, and therefore nanoparticle size, that can be modeled.~\cite{Verga2018}
The size scale of DFT also precludes investigation of the effects of strain due to substrate interactions or stress due to external loading on nanoparticle properties. Therefore, such calculations are usually limited to bulk materials or small clusters of atoms, and are difficult to apply to nanoparticles larger than few nanometers.

Instead, molecular dynamics (MD) simulations offer a valuable tool to study the shape- and size-dependence of small faceted metallic nanoparticles properties at scales large enough to be relevant to applications. Although molecular dynamics simulations do not account explicitly for molecular orbitals and so cannot reproduce electronic effects,~\cite{Scheerschmidt2007} they provide reasonable accuracy for modeling fundamental microstructural mechanisms in equilibrium or for a deformed structure, the equation of state of a system, phase equilibrium, transport properties, etc.~\cite{Steinhauser2009}
In MD, the selection of potential, or force field, is fundamental for accurate characterization and prediction.~\cite{Hale2018,Harrison2018}
Force fields can be categorized as pair potentials or many-body potentials. While pair potentials are simpler to implement, they have drawbacks for platinum and other FCC metals, often requiring  the inclusion of volume dependent correction factors to properly match experimental data.~\cite{Baskes1979} Also, pair potentials incorrectly estimate the relative magnitudes of vacancy formation and cohesive energies, resulting in incorrect stacking fault energies, surface structure, and relaxation properties.~\cite{Cleri1993} In general, many-body potentials describe metallic systems more accurately than pair potentials and are able to capture the Cauchy discrepancy of elastic constants.~\cite{Cleri1993}

Multiple different formalisms for many-body potentials exist.
For metals, the embedded atom method (EAM),~\cite{Foiles1986, Daw1983} and other potentials based on it, such as Finnis-Sinclair,~\cite{Finnis1984} concentration-dependent EAM,~\cite{Caro2005} and modified EAM (MEAM),~\cite{Baskes1992} are widely used. Another many-body potential is the effective medium theory (EMT) for bonding in metallic systems,~\cite{Jacobsen1987} which uses a simple form of the effective medium theory of condensed matter to define the potential energy. Additionally, sets of parameters for metallic systems are based on bond-order potentials extended from the Tersoff formulation,~\cite{Tersoff1988} such as the Brenner potential ~\cite{Brenner1990} and the charge optimized many body (COMB) potential.~\cite{Shan2010}
Metallic systems have also been modeled using reactive force field (ReaxFF)~\cite{C.T.vanDuin2001} which is based on bond-orders. ReaxFF accounts for dynamic partial charge equilibrations in the system and captures chemical reactions. ReaxFF parameters have been developed for chemical systems of platinum interacting with carbon~\cite{Fantauzzi2014} and oxygen.~\cite{Sanz-Navarro2008} 
While parameters sets for all these potential formalisms have been developed for platinum, their relative accuracy, particularly for nanoparticles, has not been characterized. 

Here, we evaluated nine different readily available and easily implementable potentials based on their ability to predict bulk and surface material properties: lattice constant, stiffness constants, equations of state, and surface energies of \{100\}, \{110\}, \{111\} facets. Model predictions were compared to previously reported experimental data and quantum mechanics calculations. Then, the potentials were further evaluated in terms of their ability to model the stability of platinum nanoparticles. Analysis of the calculated bulk and surface properties, as well as nanoparticle stability, show that, although none of the force fields can accurately model all properties, one of the potentials was identified as best able to model platinum in both bulk and nanoparticle forms. 
Our results provide the basis for selection of a force field to model platinum and platinum nanoparticles properties and behavior in future studies.

\section{Methods}
The nine different potentials for platinum evaluated in this study are listed in Table~\ref{tab:forcefields}.
Each potential is designated by the force field type in all capital letters hyphenated with the year that it was reported, as shown in the right most column of Table~\ref{tab:forcefields}.

\begin{table} [ht!]
\centering
  \begin{threeparttable}
    \begin{tabular}{c c c c c} 
        \hline \hline
        Ref. & Year & Force Field Type & Designation \\ [0.5ex] 
        \hline \hline
        \citenum{Foiles1987} & 1987 & EAM & EAM-(1987) \textsuperscript{*} \\ 
        \hline
        \citenum{Zhou2004} & 2004 & EAM & EAM-(2004) \textsuperscript{*}\\
        \hline
        \citenum{Baskes1992} & 1992 & MEAM & MEAM-(1992)\\
        \hline
        \citenum{Lee2003} & 2003 &	MEAM & MEAM-(2003) \textsuperscript{*}\\
        \hline
        \citenum{Sanz-Navarro2008} & 2008	& ReaxFF & REAX-(2008) \\
        \hline
        \citenum{Fantauzzi2014} & 2014	& ReaxFF & REAX-(2014)\\ 
        \hline
        \citenum{Jacobsen1996} & 1996 &	EMT & EMT-(1996) $^\zeta$\\
        \hline
        \citenum{Albe2002}	& 2002	& Tersoff-Brenner & TERSOFF-(2002) $^\zeta$\\
        \hline
        \citenum{Antony2017} & 2017 &	COMB & COMB-(2017) \\ [1ex] 
        \hline
    \end{tabular}
    \caption{Force fields evaluated in this work, organized by force field type.}
    \label{tab:forcefields}

    \begin{tablenotes}
        \small
        \item
        EAM: Embedded atom method, MEAM: modified embedded atom method, COMB: Charge optimized many body, EMT: Effective medium theory.
        The superscript \textsuperscript{*} indicates force field parameters obtained from the NIST repository~\cite{Hale2018} and $^\zeta$ indicates parameters obtained from the openKIM project.~\cite{Tadmor2011}
    \end{tablenotes}
 \end{threeparttable}
\end{table}

Other force field formulations have been proposed for platinum~\cite{Yun2012, Cai1996, Ludwig2006, Panagiotides2010, Papanicolaou2009, Cleri1993, DeClercq2016, Panizon2015, Dai2006, Januszko2015} and machine learning has been used recently to generate sets of force field parameters.~\cite{Hernandez2019, Chapman2020} 
However, here we considered only readily available and easily implementable force fields and parameters.

These force fields differ both in functional form as well as how they were parameterized, i.e., with what data the potential parameters were fit. The EAM and EAM-based force fields are robust potentials with only modest demands for computational resources and their formulation is particularly well suited to model pure metals and alloys. EAM is empirically fit to the sublimation energy, equilibrium lattice constants, elastic constants, and vacancy and interstitial formation energies. EAM potentials have been shown to reproduce surface reconstructions observed experimentally.~\cite{Foiles1987} The EMT potential, like EAM, is well suited for metallic systems involving pure metals and metal alloys. By using the simplest medium description for the interatomic interactions, calculations with this force field are computationally inexpensive. The EMT potential was fit to cohesive energies, lattice constants, and bulk and shear moduli. The Tersoff-like potentials have been successful describing a wide range of materials, including covalently bonded and metallic systems. The set of parameters of TERSOFF-(2002) was fit for a metallic-covalent system Pt-C, and is capable predicting the structural and cohesive properties for both elements (Pt and C) individually in different crystal phases, as well as the interatomic interactions between the elements. The COMB-(2017) was fit to defect formation energies, surface energies and stacking fault energies, and its parameterization particularly focused on platinum nanoparticles. The REAX-(2014) parameters were fit to describe Pt-O interactions, oxygen adsorption, and oxide formation. These interactions are particularly valuable to study catalytic processes. The parameterization of bulk platinum was fit to the following bulk phases of platinum: face centered cubic; ideal hexagonal close-packed; body centered cubic; simple cubic; diamond cubic; and b-tungsten. The REAX-(2008) was focused on the parameters for platinum nanoclusters interacting with carbon platelets and hydrogen for electrocatalysis studies. One of the advantages of COMB and ReaxFF potentials is the possibility of coupling the single element parameters with different multi-component systems as well as the ability to capture the formation and breaking of chemical bonds. However, these potentials tend to be computationally more expensive due to the recalculation of partial charges, the determination of bond-orders, and because a very small time steps (0.1~fs - 0.25~fs) is required to properly capture the dynamics of the system.

As described in the prior paragraph, all of these potentials have been designed and fit to the properties of platinum; however, this fitting was primarily done for bulk systems. Therefore the open research question is how accurately do these various potentials describe the behavior of platinum nanoparticles, and which potential is most accurate. To answer this question, the potentials described in Table~\ref{tab:forcefields} were evaluated by calculating the lattice constant, surface energies, and stiffness constants, and modeling of the equation of state of bulk platinum; then these results were compared to previously reported experimental properties and density functional theory (DFT) calculations. Finally, the force fields were used to model a cubic nanoparticle with faces in \{100\} planes and an icosahedron with faces in \{111\} planes at room temperature to compare the accuracy of predictions of nanoparticle stability.

All simulations were performed using the large scale atomic/molecular massively parallel simulator (LAMMPS) package.~\cite{Plimpton1995} For the dynamics simulations, the time step for ReaxFF and COMB force fields was 0.2~fs, and a time step of 1~fs was used for all the other force fields. The temperature of the dynamics simulations was controlled using a Nosé-Hoover thermostat with a damping parameter of 0.1~ps. And the pressure was controlled using a Nosé-Hoover barostat with a damping parameter of 1~ps.

For simulations of bulk and surface properties, two system sizes with approximately 16,000 and 19,000 atoms were modeled to evaluate possible size effects. The models were FCC single-crystal structures, created with the LAMMPS software. For all force fields, the maximum difference between the surface energies, stiffness constants, and bulk modulus calculated using the two different-size models was less than 0.09\%. This minimal difference showed that the smaller system was large enough and was used subsequently for all bulk and surface property simulations. 

\subsection{Lattice Constant}
The lattice parameter of "bulk" platinum was calculated by generating supercells based on the experimental lattice parameter for FCC platinum 0.392~nm.~\cite{Kittel2005} Periodic boundary conditions were applied in all directions. The geometry was minimized using the conjugate gradient method while allowing the simulation box to relax in all directions to account for possible differences between the experimental lattice parameter and the lattice predicted by the force field. 
The temperature of the system was increased to 298~K and equilibrated for 100 ps using a canonical ensemble. This was followed by an isobaric-isothermal equilibration at 298~K and atmospheric pressure for an additional 100~ps, to allow changes in the dimensions of the systems.
The force field-predicted lattice constant was calculated from the last 20 ps of the equilibrated structure of the isobaric-isothermal equilibration step.

\subsection{Surface Energies}
Small nanoparticles of varying shapes can form facets in different orientations and therefore different surface energies determine their catalytic, electronic, optical, and thermal properties.~\cite{Pal2015} 
Therefore, the ability of the potentials to predict the surface energy in different (hkl) faces was evaluated. In particular we calculated the surface energies for the (111), (110), and (100) planes. Most of the common nanoparticle geometries form facets oriented in the families of these planes. Also, the availability of DFT data for these surfaces in the literature enabled direct comparison. 

A typical "slab" model as described below was used to calculate the surface energy of platinum. First, the atomic energy in "bulk" platinum was calculated by generating supercells as in the lattice parameter calculations. The geometry was minimized using the conjugate gradient method and the temperature of the system was increased to 298~K and equilibrated for 100~ps using a canonical ensemble. The increase in temperature introduces a small disturbance of the energy of the system to avoid energetic local minima. After the equilibration process, the energy of the system was minimized again until the difference in energy between iterations divided by average energy was less than 1x10\textsuperscript{-12}. 
To simulate free surfaces, the supercells were oriented such that the (hkl) planes of the relevant surface were perpendicular to the x direction. A schematic of the orientation of the faces in the slab model can be seen in the \href{https://pubs.acs.org/doi/10.1021/acs.jctc.1c00434}{Supporting Information}. Vacuum layers of 5~nm in the x direction were constructed to provide exposed surfaces in both sides of the slab. Periodic boundary conditions were imposed in all directions. Then, the same MD simulation steps and settings as described for the “bulk” systems were followed, with the difference that the length of the simulation box was kept fixed in the x direction to preserve the free surface. The surface energy $\gamma$ of a facet with Miller index (hkl) was calculated with the eq~ \ref{eq:SurfaceEnergy}~\cite{Tran2016}
\begin{equation}\label{eq:SurfaceEnergy}
  \gamma_{hkl}=\frac{E_{slab}-E_{bulk/atom}.N_{slab}}{2A_{slab}}   
\end{equation}
where \textit{E\textsubscript{slab}} is the total energy of the slab, \textit{E\textsubscript{bulk/atom}} is the energy per atom of the oriented bulk system, \textit{N\textsubscript{slab}} is the number of atoms in the slab, and \textit{A\textsubscript{slab}} is the exposed surface area of the slab in one direction, assuming that the area remains constant at both exposed surfaces. 

\subsection{Stiffness Constants}
A direct static method was used to statistically calculate elastic properties. In this analysis, the platinum supercell was relaxed using an energy minimization coupled with box adjustments towards zero pressure. Twelve infinitesimal deformations were introduced by changing the dimensions of the simulation box. These deformations correspond to six tensile and six pure shear strains of magnitude $\pm$5x10\textsuperscript{-6}. The system energy was minimized following each deformation. 
For infinitesimal strain, the stress-strain relationship can be assumed to be linear, so the stiffness constants were calculated according to Hooke’s law shown in eq~\ref{eq:Hooke'slaw}.
\begin{equation}\label{eq:Hooke'slaw}
    \sigma_{ij}={C_{ijkl}}{\varepsilon_{kl}}  			
\end{equation}

where $\sigma_{ij}$ is the stress tensor calculated from the Virial stress definition,~\cite{Zhou-2003} \textit{C\textsubscript{ijkl}} are the stiffness constants, and $\varepsilon_{kl}$ is the strain imposed to the system in each direction.
Two additional magnitudes of strain (±1x10\textsuperscript{-6} and ±5x10\textsuperscript{-5}) were tested and negligible difference (less than 0.01\%) between the stiffness constants calculated at the different strain magnitudes was observed. 

\subsection{Equation of State}
The platinum supercell with periodic boundary conditions was geometrically optimized using the conjugate gradient method and the temperature of the system was equilibrated at 298~K using a canonical ensemble for 3~ps.
This was followed by a pressure-temperature equilibration process at 0.1~MPa (1~atm) and 298~K, using an isothermal-isobaric (NPT) ensemble for 5~ps. A Nosé-Hoover barostat was used to control the pressure of the system with a damping parameter of 1~ps.

After the equilibration process, the system was hydrostatically compressed in increments of 2~GPa to up to 50~GPa, while keeping the temperature constant. The discrete increments were enforced in an isothermal-isobaric ensemble for 2~ps and the systems were further equilibrated for 2~ps after each increment. The pressure $P$ and volume $V$ were time averaged over the last 0.5~ps of the equilibration run.
Then, the bulk modulus was predicted by fitting the thermodynamic data to the Birch-Murnaghan equation of state~\cite{Birch1947, Murnaghan1944}    
\begin{equation}\label{eq:BM_EoS}
    P(V)=\frac{3}{2} B_{0}\Bigg(\bigg(\frac{V}{V_o}\bigg)^{-\frac{7}{3}}-\bigg(\frac{V}{V_o}\bigg)^{-\frac{5}{3}}\Bigg)\Bigg(1+\frac{3}{4}(B_{0}^\prime-4)\bigg(\bigg(\frac{V}{V_o}\bigg)^{-\frac{2}{3}}-1\bigg)\Bigg)
\end{equation}
where $V_0$ is the initial volume, $B_{0}$ is the bulk modulus and $B_{0}^\prime$ is the first derivative of the bulk modulus.

\subsection{Stability of Small Pt Nanoparticles}
The force fields described above were used to model 3.2~nm nanoparticles with cubic and icosahedral shapes. These shapes were selected because icosahedral and cubical platinum nanoparticles are stable, experimentally observed and theoretically predicted,~\cite{Ahmadi2019} and because their facets are oriented in the \{111\} and \{100\} planes. Shrink-wrapped boundary conditions (i.e. the boundaries of the simulation box extended to the limits of the model system) were used for the simulations. The cube nanoparticles were generated using lammps and the icosahedral nanoparticles were generated with OpenMD.~\cite{gezelter2010openmd} The nanoparticles were geometrically optimized using the conjugate gradient method to achieve an energy convergence between steps of 1x10\textsuperscript{-7}. Next, the temperature was equilibrated at 298~K using a canonical ensemble for 0.3~ns. The stability of the nanoparticles was evaluated in terms of the change of potential energy over time.
In addition to the Nosé-Hoover thermostat, the equilibration process was repeated with two different thermostats: Langevin and Berendsen. These additional simulations were performed to confirm that the performance of the force fields was independent of the thermostat.

\section{Results and Discussion}
Molecular dynamics simulations, using these nine different force fields, were used to calculate lattice constant, surface energies, stiffness constants, and bulk modulus derived from the equation of state of platinum. The results from each calculation are discussed below.

\subsection{Lattice Constant}
\begin{figure} [ht]
\centering
{\includegraphics[width=0.5\textwidth]{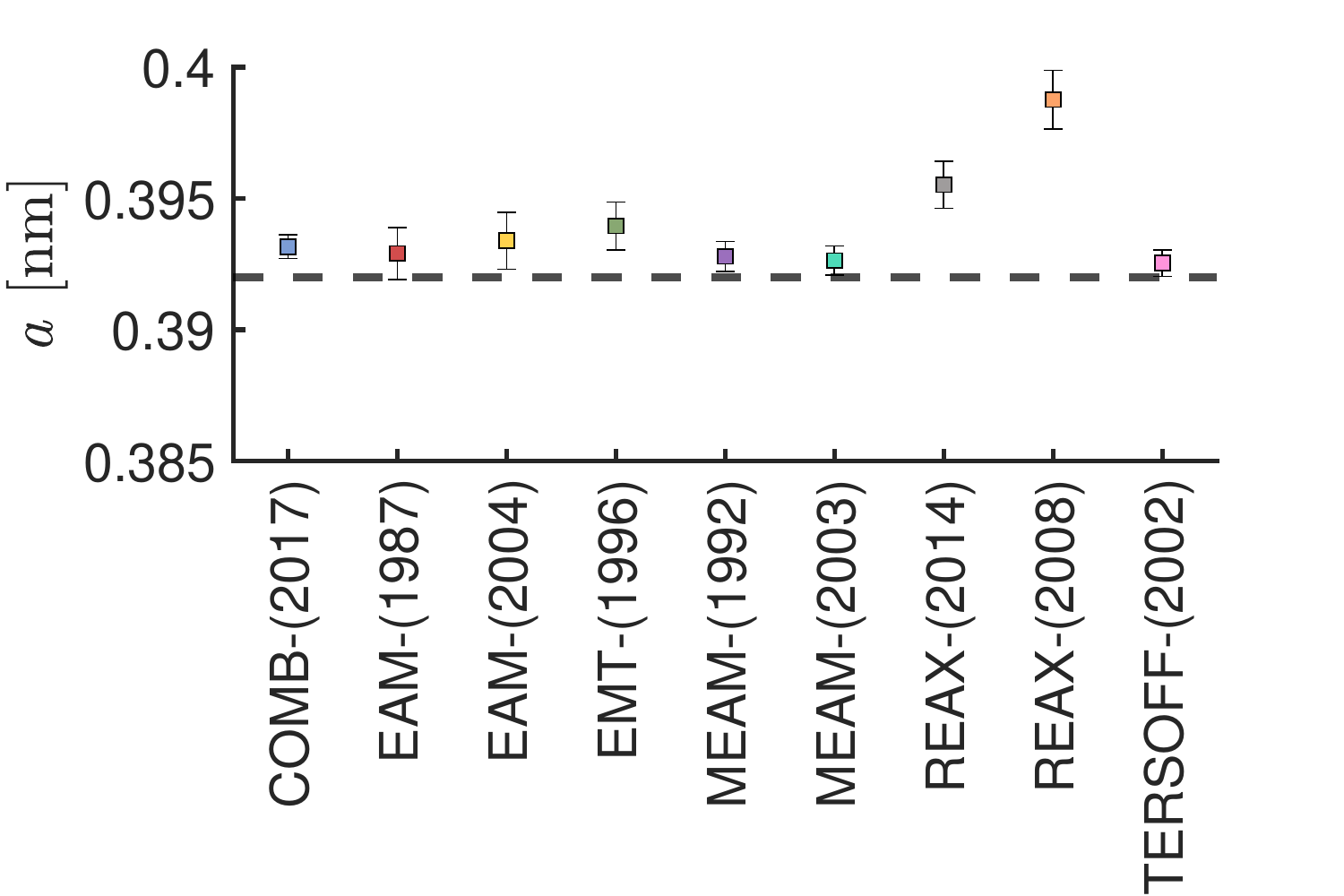}}
\caption{Lattice constant calculated by the force fields (symbols) compared to the experimental value (dashed line).}\label{fig:lattice-ff}
\end{figure}

First, after a geometry optimization process and relaxation at room temperature and atmospheric pressure, the lattice constant for FCC platinum was calculated for each force field.
The modeling results and standard deviation shown in Figure~\ref{fig:lattice-ff} are compared to the known experimental value of 0.392~nm.~\cite{Kittel2005}
It is observed that the predicted lattice constant of bulk platinum at room temperature, is close to the experimental value (error between 0.1\% to 1.7\% ) for most of the force fields used in this work. The least accurate are the two REAX-(2014) and REAX-(2008) potentials that overestimate the lattice constant, but still have a small error of 0.9 and 1.7\%, respectively.
These differences can be explained by the parameterization of each force field. For the EAM and MEAM force fields, the lattice parameter is used as an input for the parameterization and so matches the experimental value almost exactly. However, ReaxFF is parameterized to match the atomic energies of a system, so larger variation of the lattice parameter can be expected. Additionally, REAX-(2014) was parameterized to match DFT-calculated energies that predicted a lattice constant of 0.397~nm instead of the experimental value (0.392~nm).
On the other hand, the REAX-(2008) parameterization was focused on matching the adsorption energy of platinum clusters on carbon platelets and not on the prediction of bulk properties. In the parameterization,~\cite{Sanz-Navarro2008} the lattice mismatch between the carbon and platinum atoms at the interface caused desorption of some platinum atoms and consequent restructuring of the atoms in the Pt cluster. These processes lead to longer Pt-Pt bonds and, consequently, a larger lattice parameter when the force field is used to simulate bulk platinum.

\subsection{Surface Energies}
Differences in the energies of the different surfaces are responsible for the stability of nanoparticles and favor the formation of specific geometric configurations.
Experimental approaches to obtaining the surface energy of solids include calculations from contact angle measurements,~\cite{Calvimontes2017} extrapolation for the solid phase from measurements of surface tension in a solid-liquid-vapor system,~\cite{Tyson1977} and inverse gas chromatography,~\cite{P.Yla-Maihaniemi2008} among other methods.~\cite{Chen2020} Nevertheless, experimental calculation of the surface energy of specific facets is difficult and the results are often inconsistent among experiments.~\cite{Tran2016}

Quantum-mechanics calculations based on DFT have been widely used to calculate the surface energy of platinum. However, the calculated values are dependent on the exchange correlation function used and the selection of parameters.~\cite{Vega2018, Singh-Miller2009} The general gradient approximation (GGA) tends to underestimate the surface energies, while the local density approximation (LDA) may overestimate it. Vega et al.~\cite{Vega2018} evaluated different DFT exchange correlation functionals and found that, although there are limitations, the Vosko-Wilk-Nusair (VWN) functional~\cite{Vosko1980} within the local density approximation is well suited for predicting surface energies of platinum. In this study, we compare the energies predicted using different empirical potentials with the DFT values calculated by Vega et al. with the VWN functional.

\begin{figure} [ht]
\centering
\begin{tabular}{c c}
\subfigure[]{}{\includegraphics[width=0.5\textwidth]{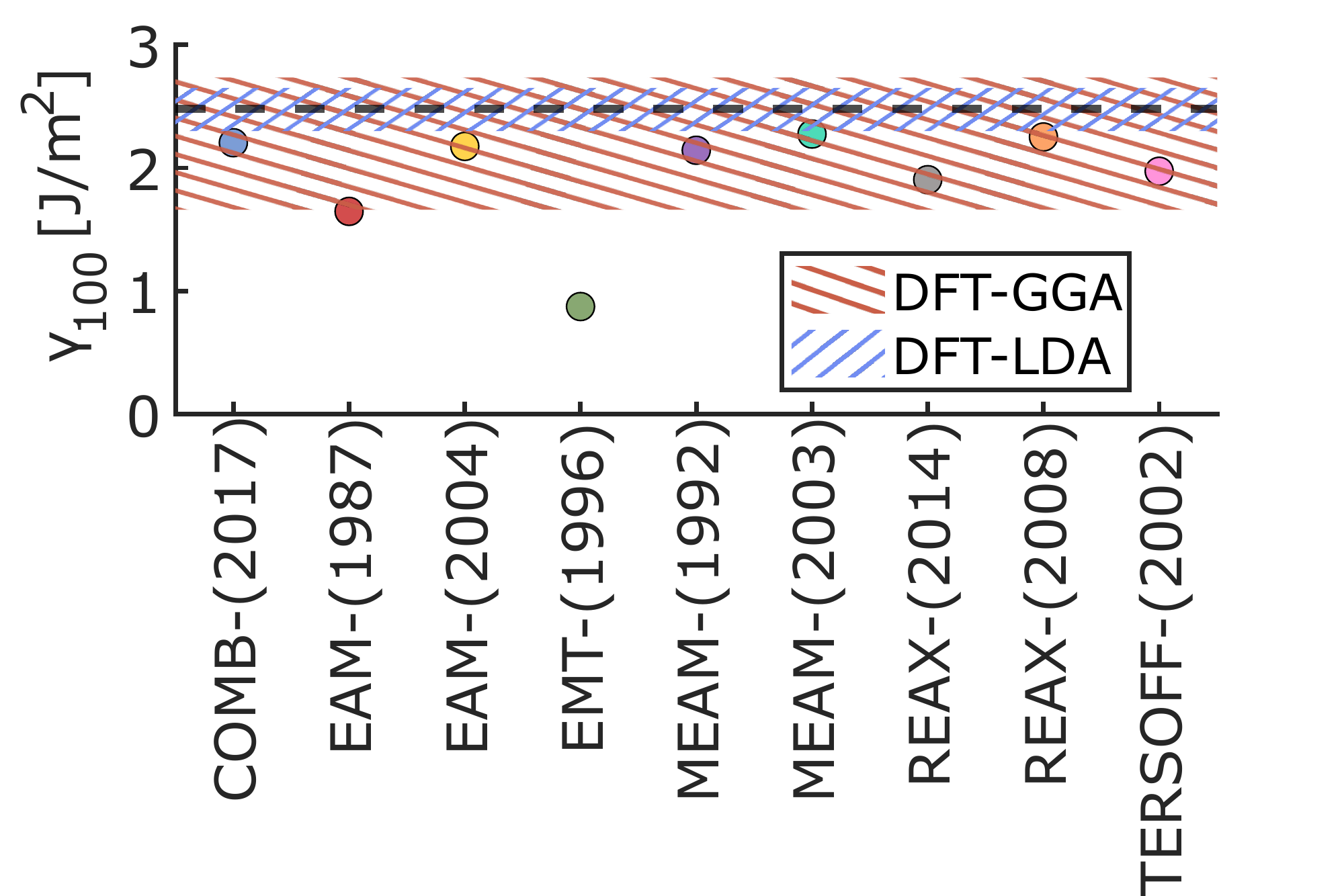}}
&
\subfigure[]{}{\includegraphics[width=0.5\textwidth]{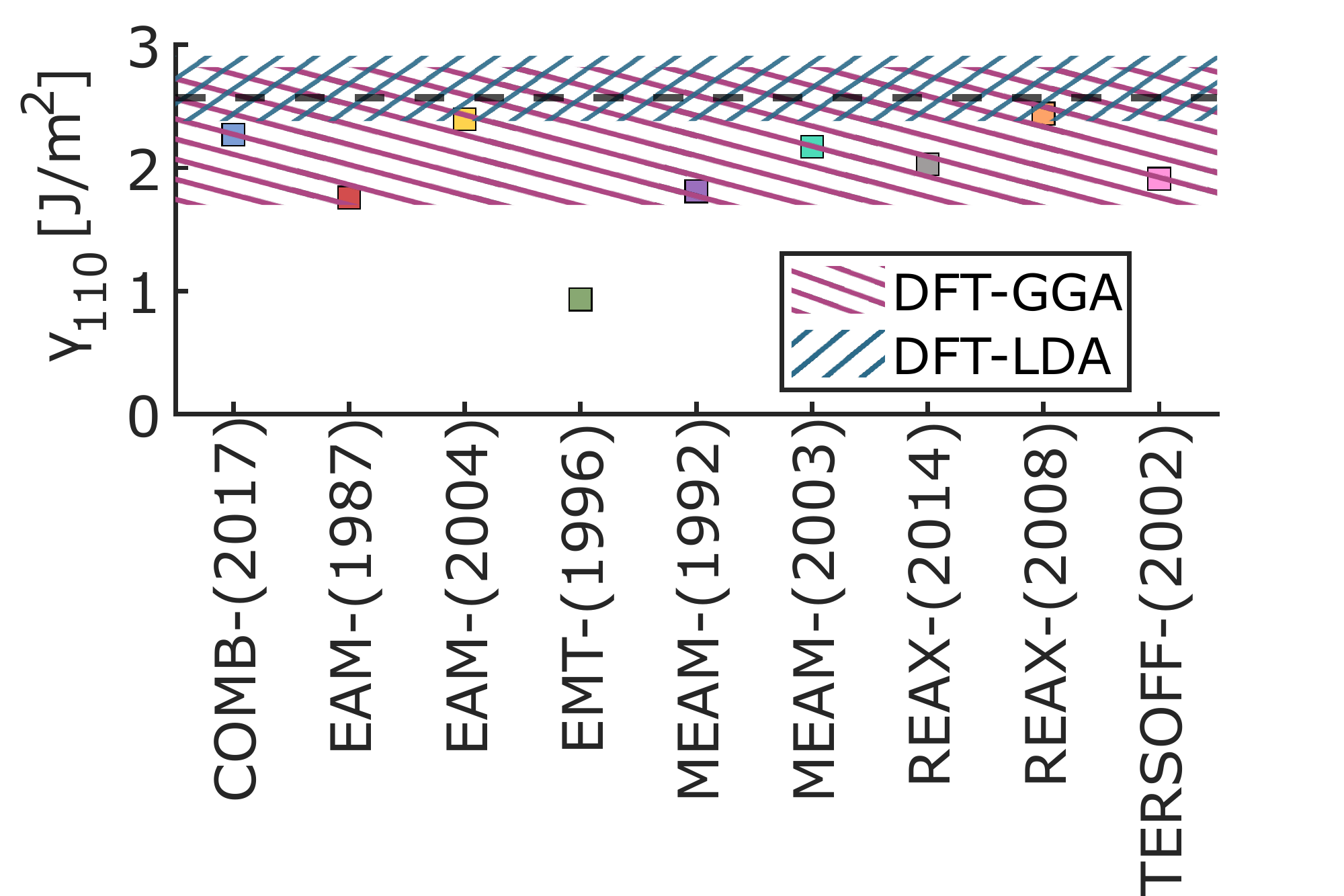}}
\\
\subfigure[]{}{\includegraphics[width=0.5\textwidth]{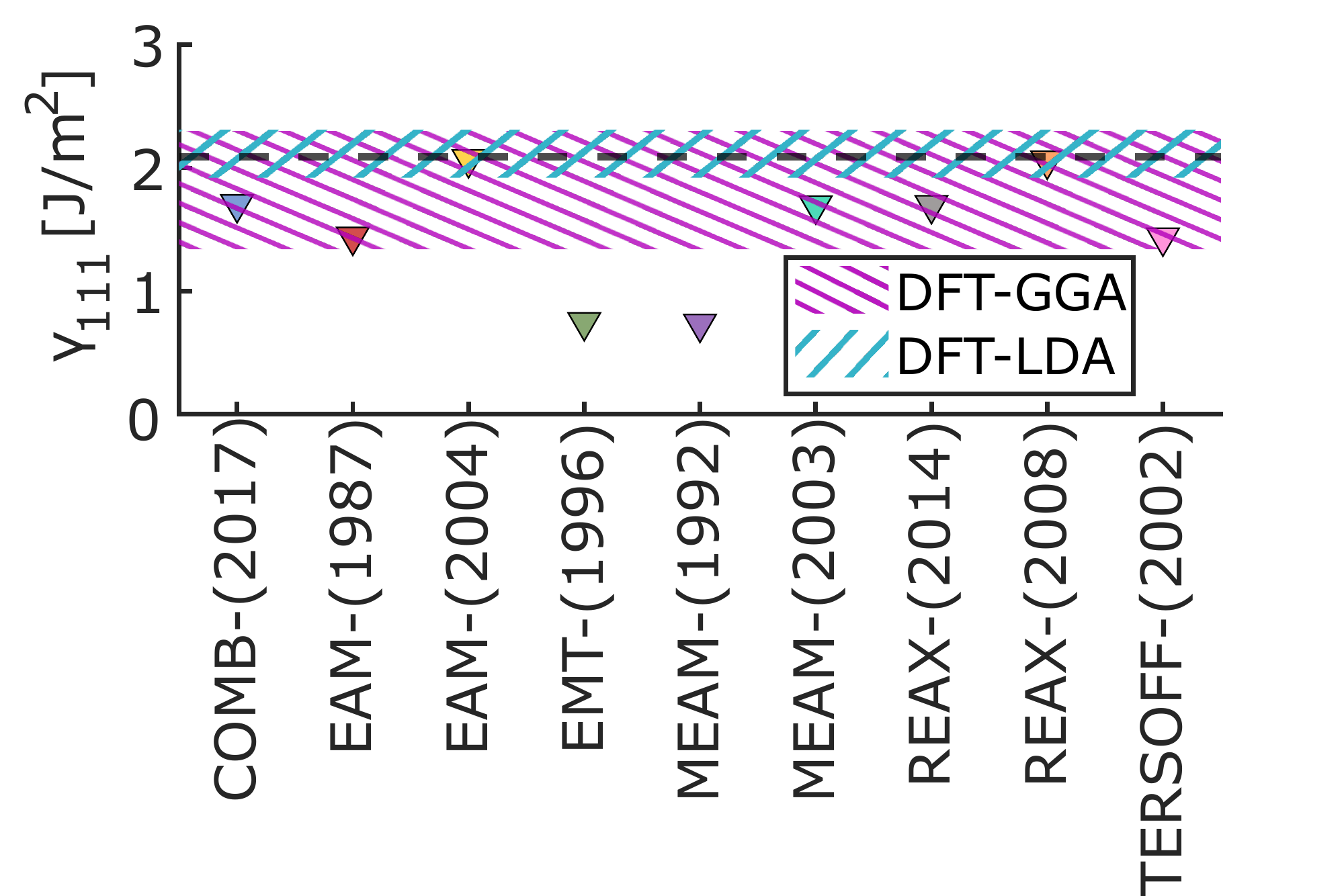}}
&
\subfigure[]{}{\includegraphics[width=0.5\textwidth]{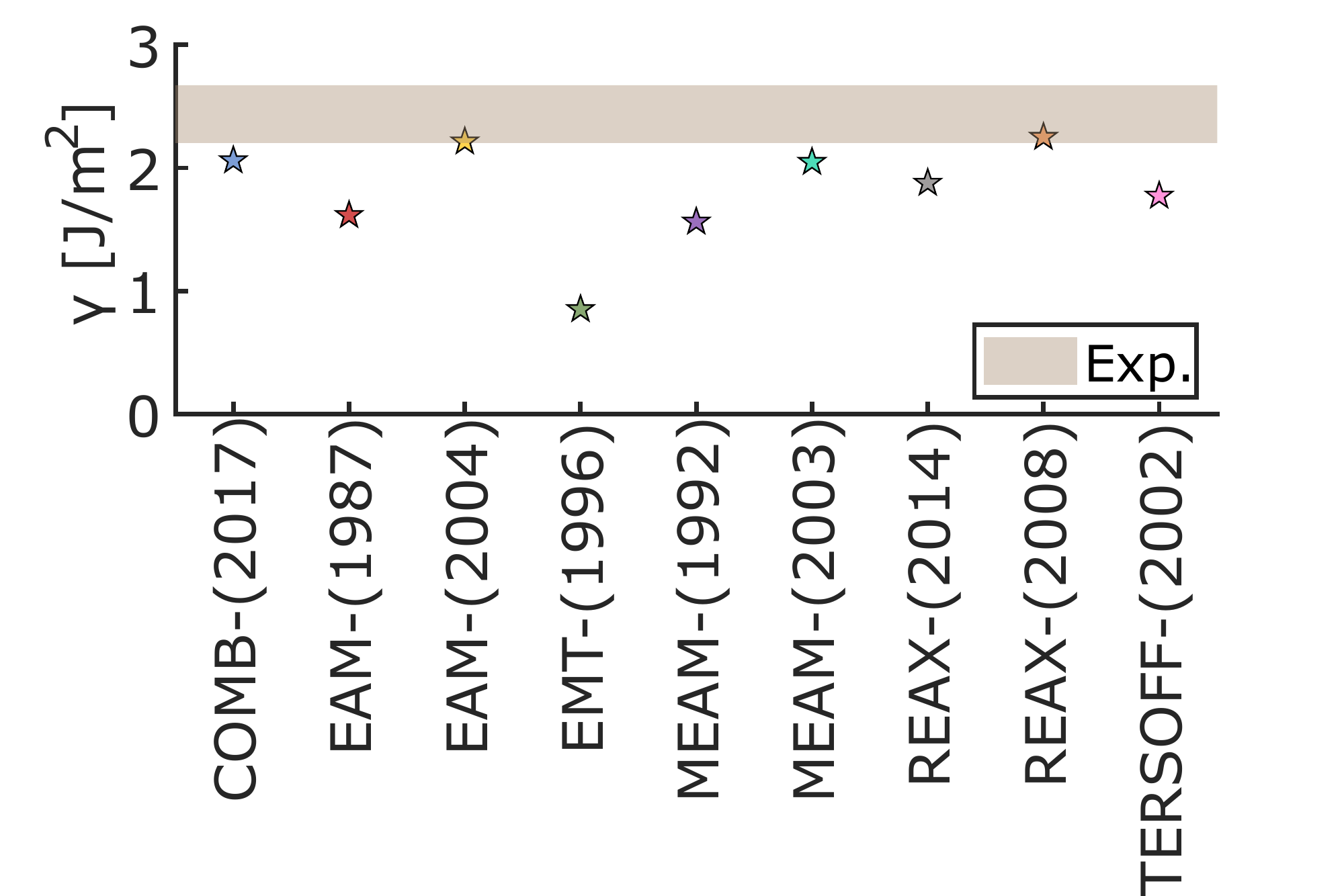}}
\\
\end{tabular}
\caption{(a) $\gamma_{100}$, (b)$\gamma_{110}$, and (c) $\gamma_{111}$ Surface energies from MD simulations (symbols) compared to DFT results (patterned areas). DFT values~\cite{Vega2018} calculated with the VWN functional are shown as dashed black lines. (d) Average surface energies $\gamma$ from MD simulations (symbols) compared to experimental results (shaded area).}  \label{fig:SurfaceEnergy-ff}
\end{figure}

Figure~\ref{fig:SurfaceEnergy-ff} shows the surface energies calculated using the different force fields evaluated in this work. The reference surface energy (DFT with the VWN functional) in each case (a) to (c) is shown as a dashed line. Additionally, energies reported from DFT calculations are shown as patterned areas. An extended literature review on the surface energies of platinum based on DFT and experimental results, and the values calculated from the nine force fields evaluated here are presented in the \href{https://pubs.acs.org/doi/10.1021/acs.jctc.1c00434}{Supporting Information}.

The EMT-(1996) potential severely underestimates the surface energy of all the faces, while the MEAM-(1992) potential calculates low surface energy for the (111) face. Surface energies calculated by the other potentials are within the range of values calculated using DFT with any exchange correlation function. Since the semi-empirical interatomic potentials are obtained by fitting to DFT values, the parameters predicted by the force fields likely correspond to the DFT approximation used in their parameterization. 

Experimental surface energies are commonly calculated from measured liquid-metal surface tension and correspond to average surface energies over crystals with different face orientations,~\cite{Patra2017,DeWaele2016} so Figure 2d shows the average of the surface energies in the three directions evaluated in this study compared to the range of experimental results. Although most force fields underestimate the surface energy, EAM-(2004), MEAM-(2003) and REAX-(2008) predict an average surface energy within the range of experimental values.

In general, the force fields are able to capture the differences between the energies of the different surfaces. Most MD-calculated surface energies follow $\gamma_{111}<\gamma_{100}<\gamma_{110}$, consistent with DFT and experimental results. An exception to this is observed for the MEAM-(1992) potential, where $\gamma_{100}>\gamma_{110}$ and, to a lesser extent, for the MEAM-Lee and TERSOFF-(2002) potentials, where $\gamma_{100}$ is slightly larger than $\gamma_{110}$.

\subsection{Stiffness Constants}
\begin{figure} [ht]
\centering
\begin{tabular}{c c}
\subfigure[]{}{\includegraphics[width=0.5\textwidth]{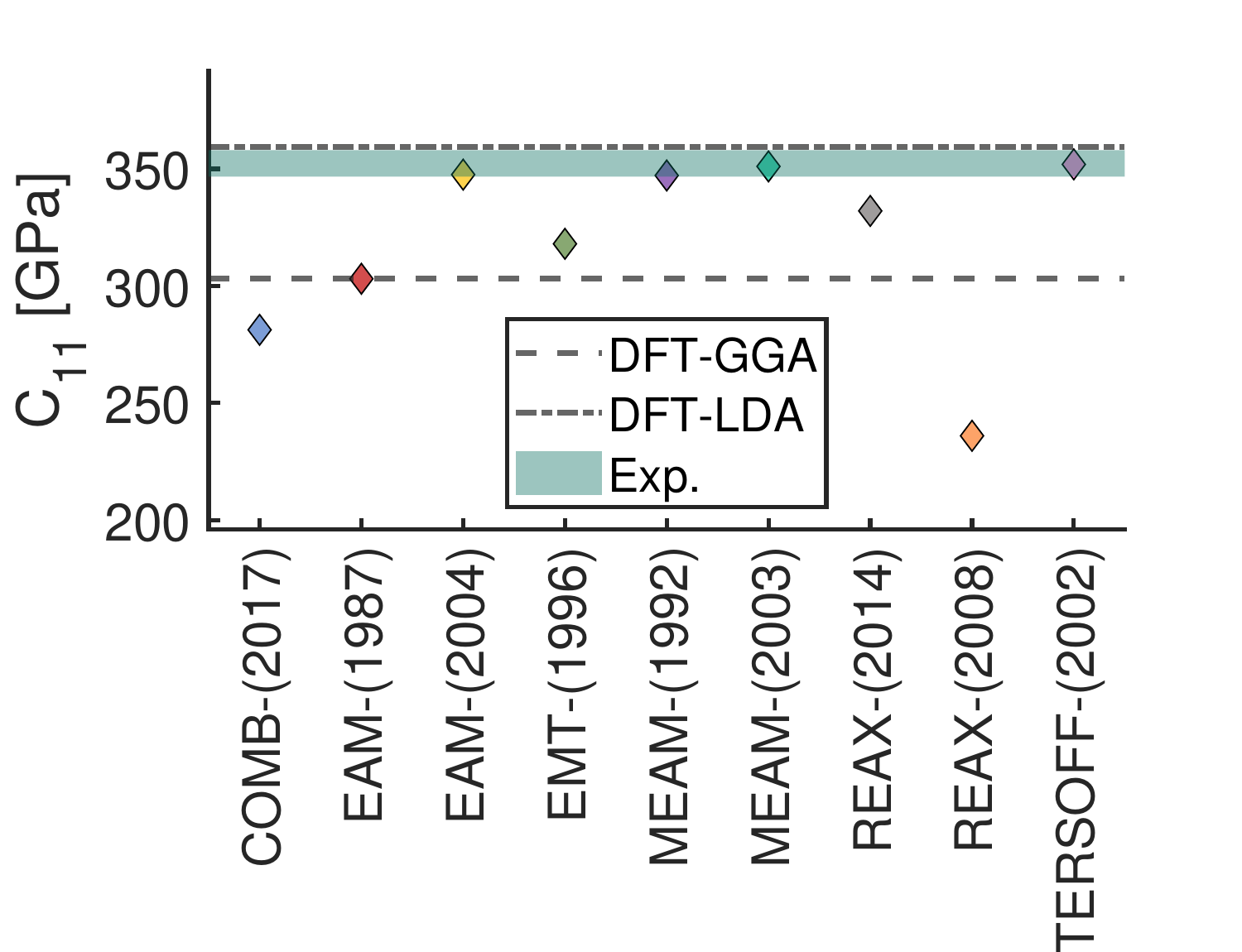}}
&
\subfigure[]{}{\includegraphics[width=0.5\textwidth]{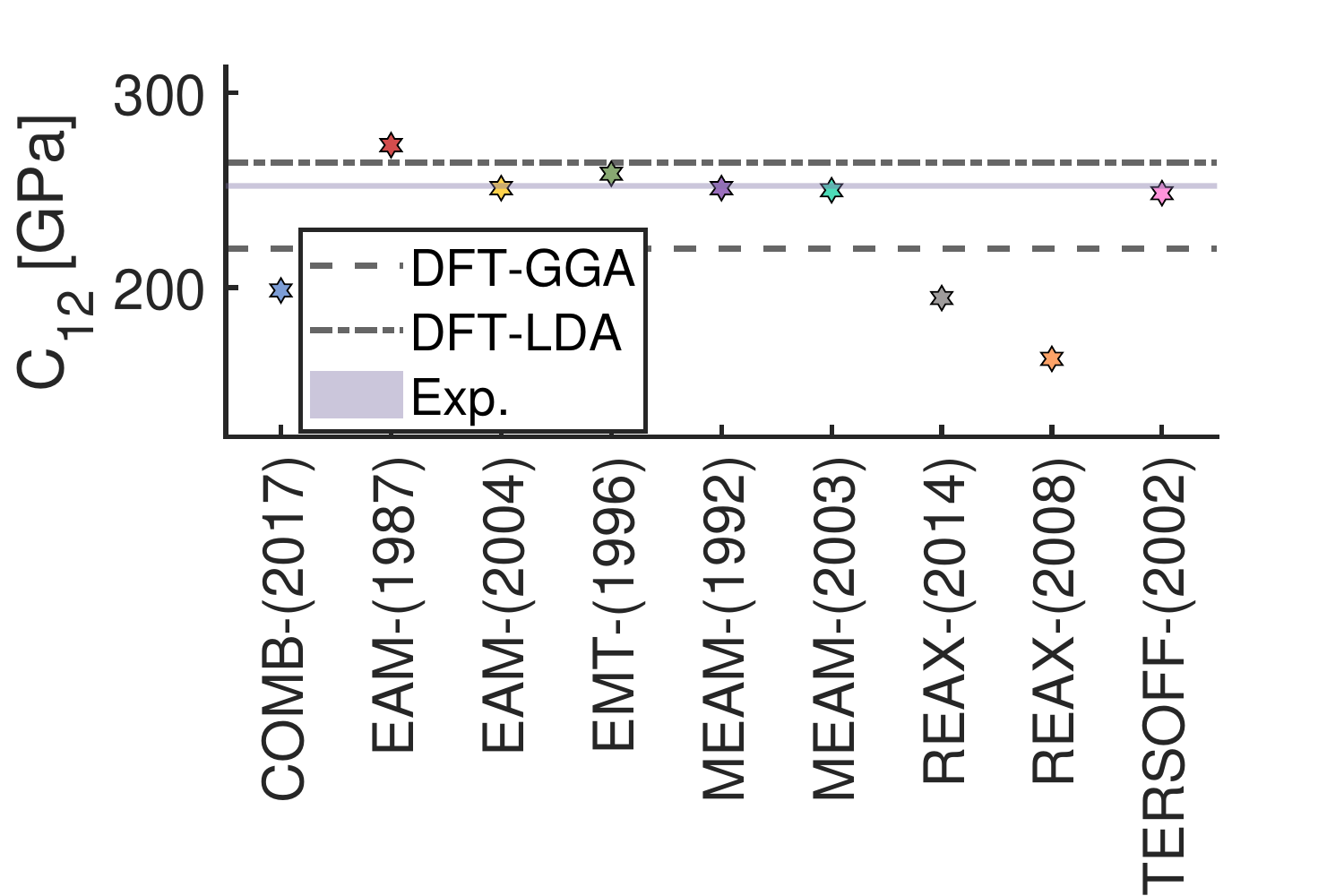}}
\\
\subfigure[]{}{\includegraphics[width=0.5\textwidth]{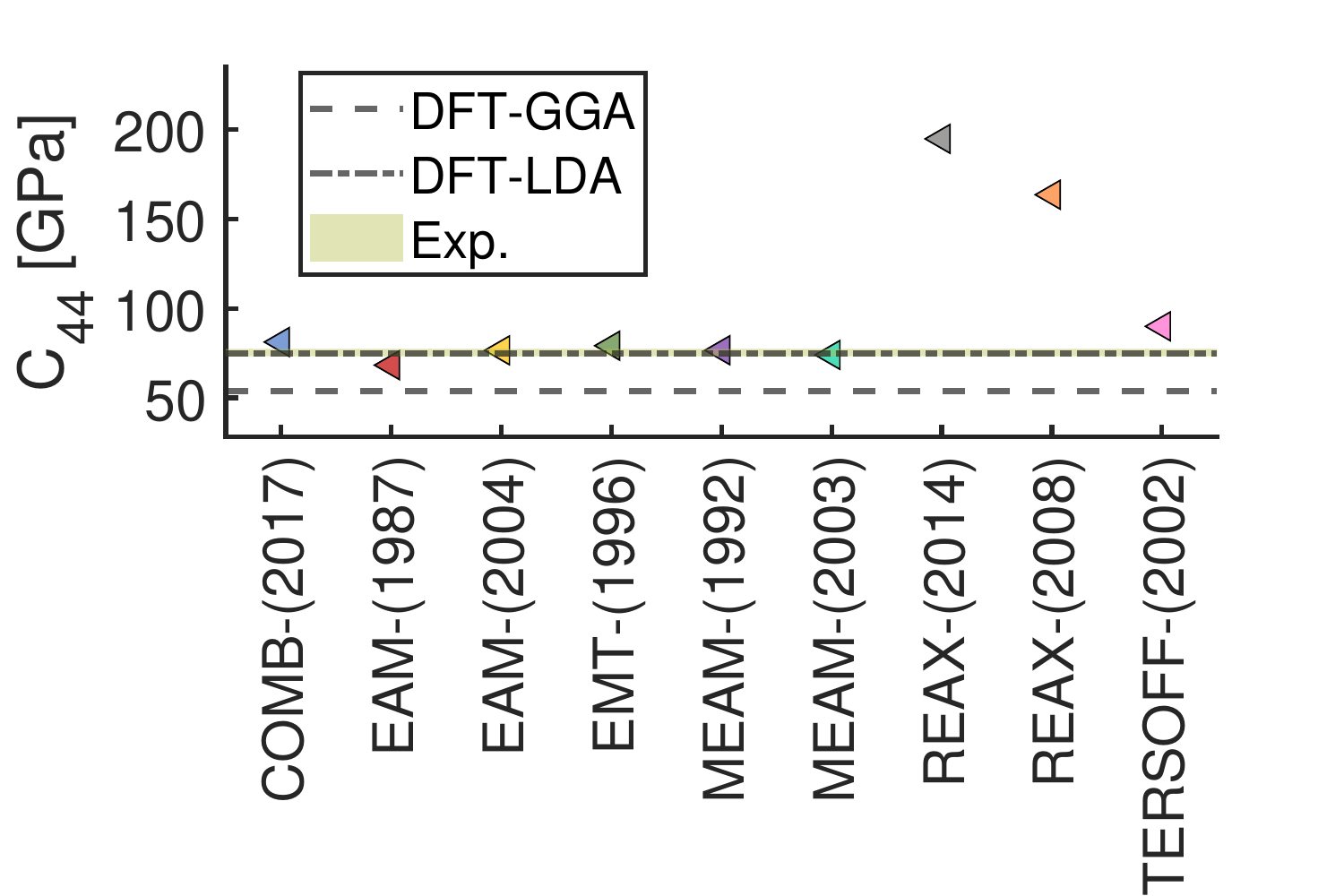}}
&
\subfigure[]{}{\includegraphics[width=0.5\textwidth]{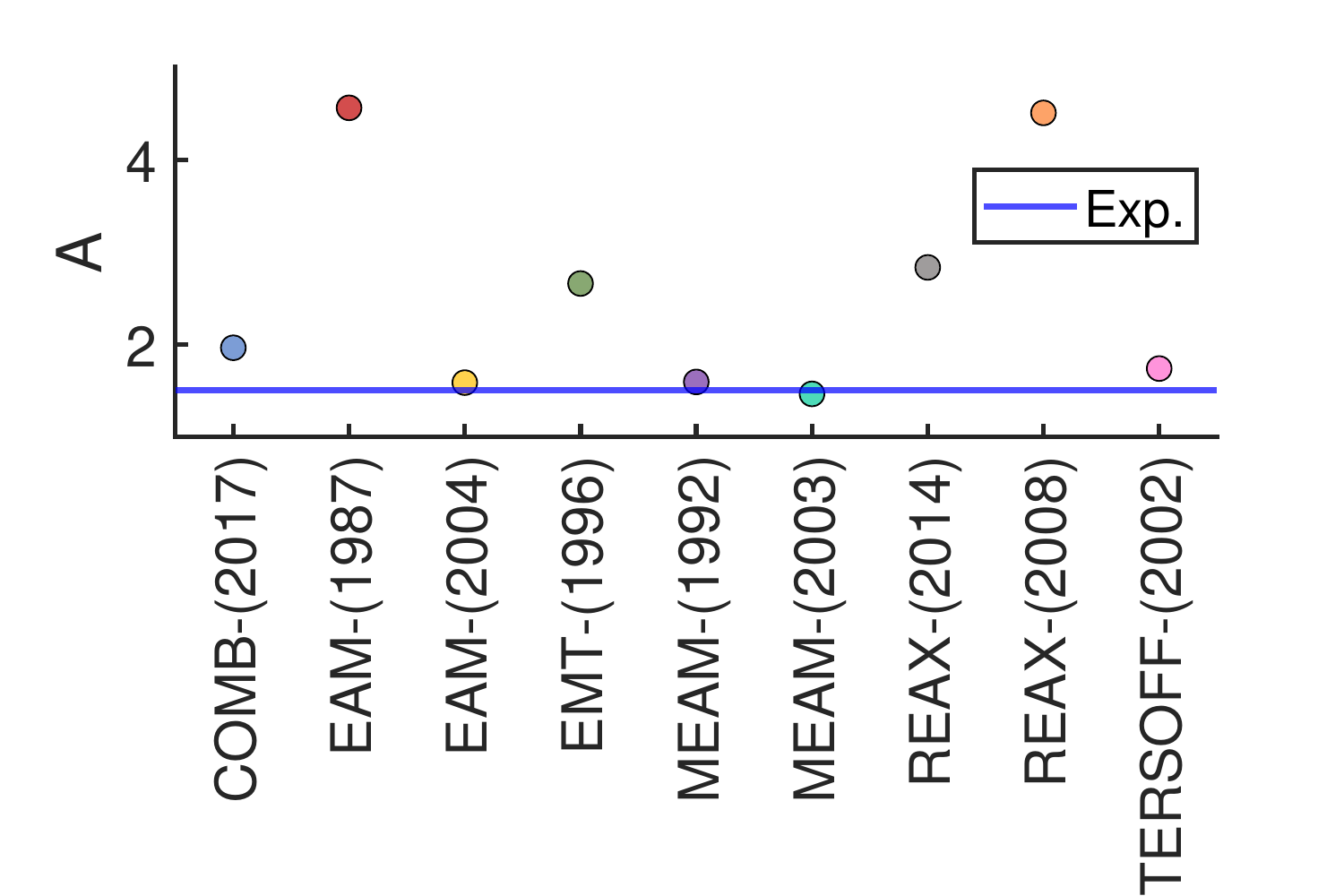}}
\\
\end{tabular}
\caption{(a-c) Stiffness constants from MD simulations (symbols) compared to DFT~\cite{Tran2016,Menendez-Proupin2007} (dashed lines) and experimental results (shaded areas)~\cite{Simmons1965, Landolt1966} (a) $C_{11}$, (b)$C_{12}$, and (c) $C_{44}$. (d) Zener ratio $A$ from MD simulations (symbols) compared to experimental results~\cite{Simmons1965,Landolt1966} (blue line).}
\label{fig:Stiffness-ff}
\end{figure}

The components of the stiffness tensor in Voigt notation~\cite{Hearmon1946} calculated by the different force fields are shown in Figure~\ref{fig:Stiffness-ff}. All the force fields predict the orthotropy of platinum with $C_{11}$ exactly equal to $C_{22}$ and $C_{33}$, $C_{13}$ equal to $C_{23}$ and $C_{12}$, and $C_{44}$ equal to $C_{55}$ and $C_{66}$ for the crystalline FCC Pt. 
The Zener anisotropy ratio $A = 2C_{44}/(C_{11}-C_{12})$ is also shown in Figure~\ref{fig:Stiffness-ff}. The stiffness constant C\textsubscript{44} represents resistance to shear on $\{100\}$ in
$\langle0kl\rangle$, while $ C_{11}-C_{12}/2$ represents resistance to deformation by shear on $\{110\}$ in
$\langle-110\rangle$. Then, \textit{A} represents the ratio of the extreme elastic-shear coefficients. For a perfectly isotropic medium, \textit{A} has a value of 1.~\cite{Ledbetter2006}

The two parametrizations of the REAX force field (2008) and (2014) underestimate the \textit{C\textsubscript{12}} constant, and overestimate \textit{C\textsubscript{44}} compared to the experimental values. The overestimation of \textit{C\textsubscript{44}} is more than two times the experimental constant and predicts that \textit{C\textsubscript{44}} is equal to \textit{C\textsubscript{12}} from both ReaxFF potentials. The relationship of shear response compared to dilation on compression is unrealistic for cubic crystals if \textit{C\textsubscript{44}} equals \textit{C\textsubscript{12}}, as it is shown also from the high value of the Zener ratio.

The original EAM-(1987) potential, the EMT potential, and the COMB potential also predict Zener anisotropic parameters that do not correspond to the experimental relation.
Specifically, they predict a higher ratio between the elastic shear coefficients than that observed experimentally, corresponding to a larger degree of elastic anisotropy. By contrast, the EAM-(2004) potential, the Tersoff-like, and both MEAM potentials predict elastic properties very similar to experimental values. The calculated values of stiffness constants and elastic properties can be found in the \href{https://pubs.acs.org/doi/10.1021/acs.jctc.1c00434}{Supporting Information}. 

\subsection{Equation of State}
Because of the small surface area of platinum nanoparticles, even low loads can exert large pressure. Therefore, a force field that accurately predicts volumetric changes in response to pressure is essential. Furthermore, erroneous phase changes estimated by the computational models reflect inaccurate energetic predictions of the potential and should be avoided.

\begin{figure}[ht]
\centering
\begin{tabular}{c c}
\subfigure[]{}{\includegraphics[width=0.5\textwidth]{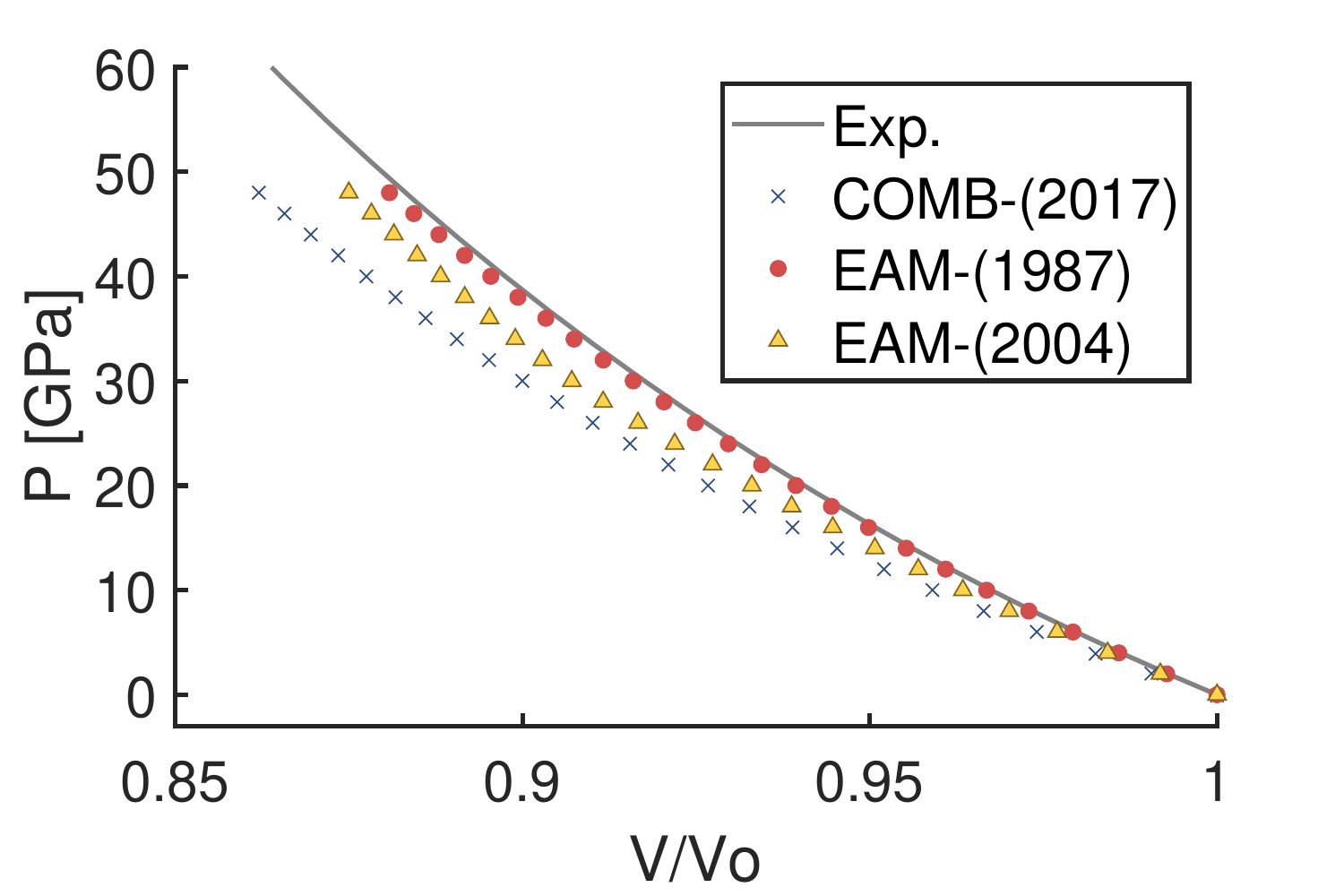}}
&
\subfigure[]{}{\includegraphics[width=0.5\textwidth]{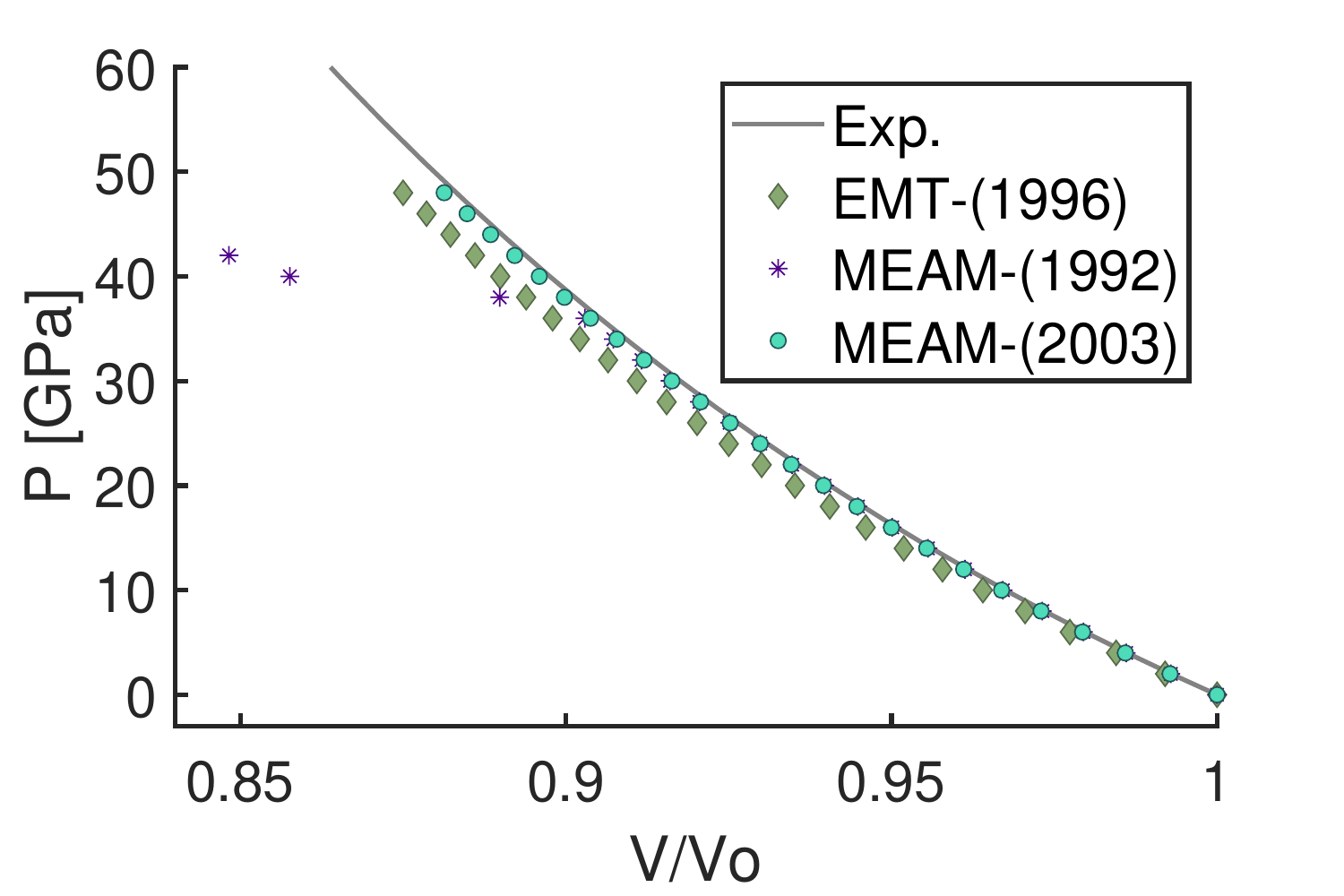}}
\\
\subfigure[]{}{\includegraphics[width=0.5\textwidth]{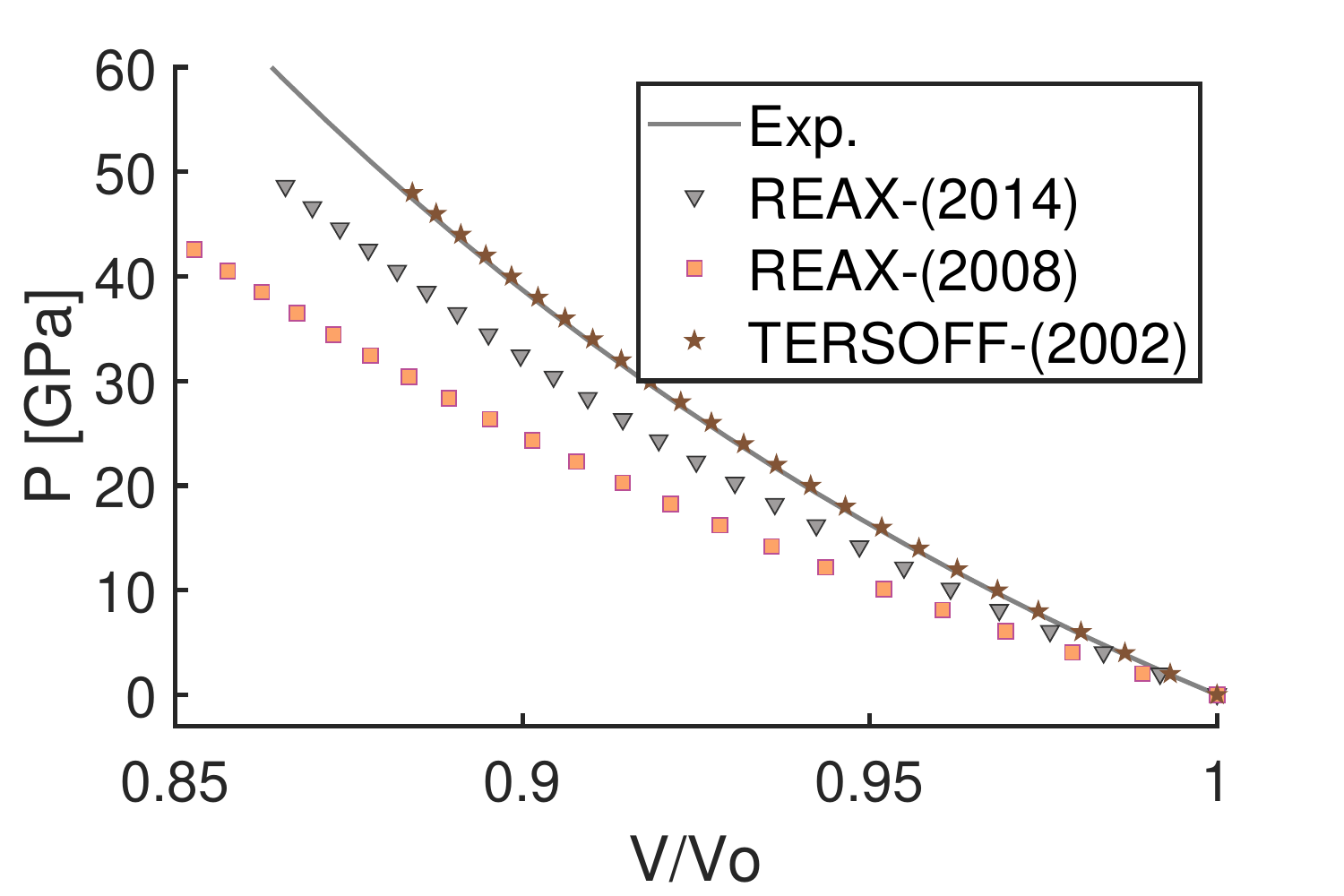}}
\\
\end{tabular}
\caption{Pressure-volume (P-V) relationships for the different force fields (symbols) compared to experimental results~\cite{Holmes1989, Ye2017} (lines).}\label{fig:P-v_curves}
\end{figure}

The  pressure-specific volume isotherms for platinum simulated with the nine different force fields selected for this study are shown in Figures~\ref{fig:P-v_curves}a-c. Each figure shows isotherms predicted by three of the nine force fields compared to the curves from experimental studies.~\cite{Holmes1989, Ye2017} None of the predicted isotherms exhibits discontinuities except for the curve obtained using the MEAM-(1992) potential, shown in Figure~\ref{fig:P-v_curves}b. 
Previous experimental~\cite{Matsui2009} and DFT~\cite{Sun2008} studies have demonstrated that the pressure - specific volume isotherms for platinum do not diverge from the Birch-Murnaghan equation of state, even for pressures as high as 500~GPa, so the discontinuity observed for the MEAM-(1992) potential is unphysical.

\begin{figure}[ht]
\centering
\begin{tabular}{c c}
\subfigure[]{}{\includegraphics[width=0.5\textwidth]{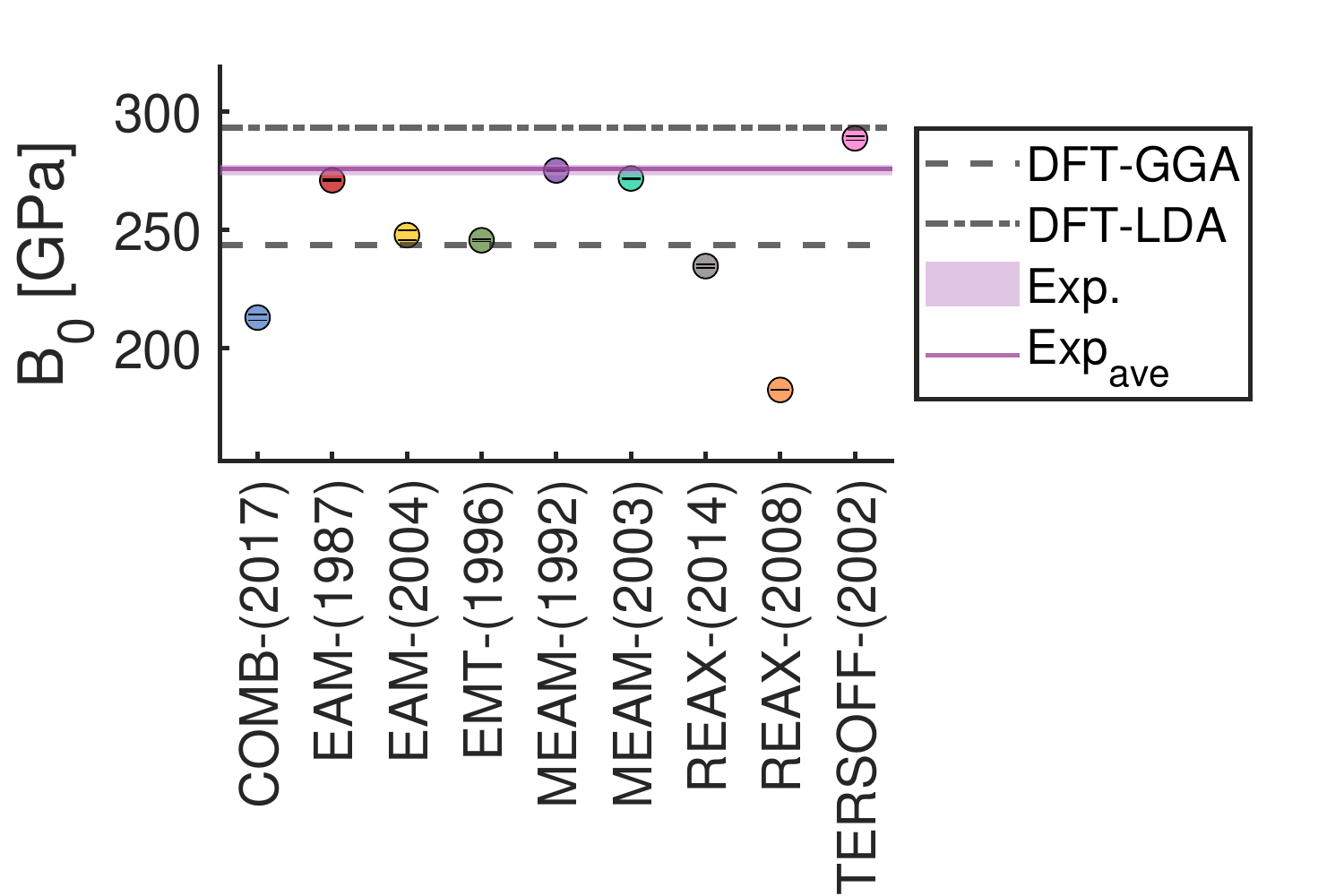}}
&
\subfigure[]{}{\includegraphics[width=0.5\textwidth]{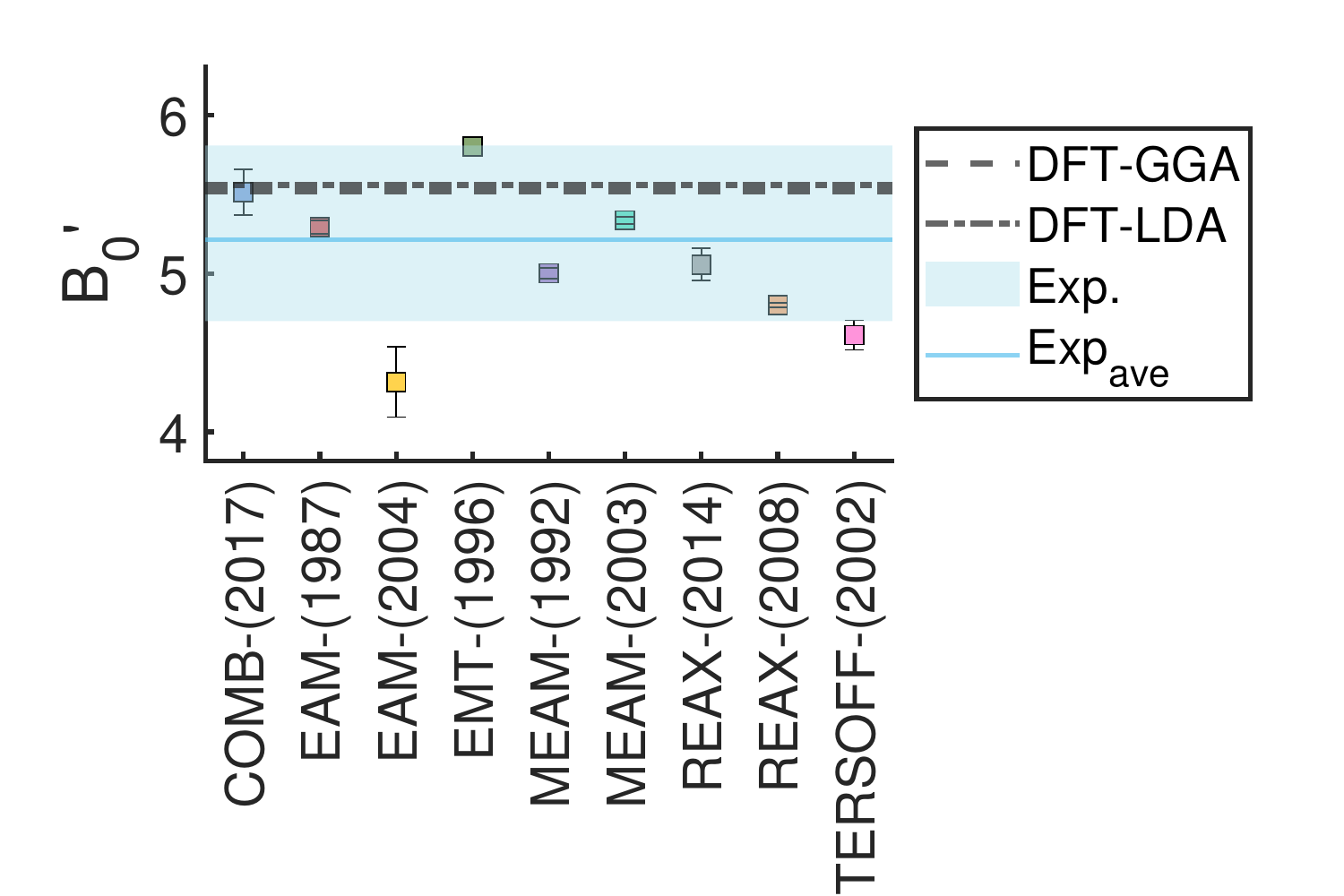}}
\\
\end{tabular}
\caption{Bulk modulus at atmospheric pressure $B_{0}$ and its derivative $B_{0}'$ for platinum obtained by fitting pressure-volume data from different force fields in Figure~\ref{fig:P-v_curves} to the Birch-Murnaghan equation of state (symbols). Error bars reflect the 95\%-confidence intervals. Results are compared to DFT values,~\cite{Sun2008, Tran2016} shown as dash lines, and experimental results,~\cite{Dewaele2004,Fei2004,Zha2008,Matsui2009,Yokoo2009,Dorfman2012,Ye2017,Elkin2020} shown as shaded areas. The average of the experimental results is shown as a solid line.}\label{fig:B0B'}
\end{figure}

The curves shown in Figure~\ref{fig:P-v_curves} were fitted to the Birch-Murnaghan equation of state from 0~GPa to 28~GPa, a range in which the isotherms are continuous for all the force fields. The isothermal bulk modulus $B_{0}$ and the first derivative of the bulk modulus $B_{0}'$ were obtained from the fit and reported in Figure~\ref{fig:B0B'}.
Calculated values and confidence bounds can be seen in the \href{https://pubs.acs.org/doi/10.1021/acs.jctc.1c00434}{Supporting Information}. 
 
Results show that the bulk modulus derived from the equation of state is severely underestimated by the REAX-(2008) and the COMB-(2017) potentials. 
Previous studies of the equation of state of platinum based on DFT with the general gradient approximation, reported in the literature,~\cite{Singh-Miller2009, DaSilva2006, Khein1995} underestimate $B_{0}$ compared to experimental results.~\cite{Matsui2009, Holmes1989, Ye2017} The bulk modulus calculated with the EAM-(2004), EMT-(1996), and REAX-(2014) are similar to those properties calculated with DFT-GGA and the Perdew-Burke-Ernzerhof functional,~\cite{Sun2008} while the TERSOFF-(2002) potential overestimates the bulk modulus, similar to DFT-LDA calculations.~\cite{Menendez-Proupin2007} The bulk modulus calculated by the EAM-(1987), MEAM-(1992), and MEAM-(2003) are comparable to experimental values.

The pressure derivative of the bulk modulus quantifies the increased resistance to compression with increasing pressure. Figure 5b compares our simulation results with experimental values reported in the literature; the exact values are given in the \href{https://pubs.acs.org/doi/10.1021/acs.jctc.1c00434}{Supporting Information} (Table S5). The mean value of $B_{0}'$ for platinum is 5.2$\pm$0.4 although there is a wide range of reported experimental values, likely because of differences in the chosen experimental method, the selection of fitted equation of state,~\cite{Hofmeister1991} and the range of pressures evaluated. Most of the force fields evaluated are within the range of experimental measurements, except for EAM-(2004) which underestimates $B_{0}'$, indicating that this force field less accurately predicts changes in the resistance to compression compared to the other potentials.

\subsection{Force Field Selection}
\begin{figure}[hbt!]
\centering
\begin{tabular}{c c c}
\subfigure[]{}{\includegraphics[width=0.3\textwidth]{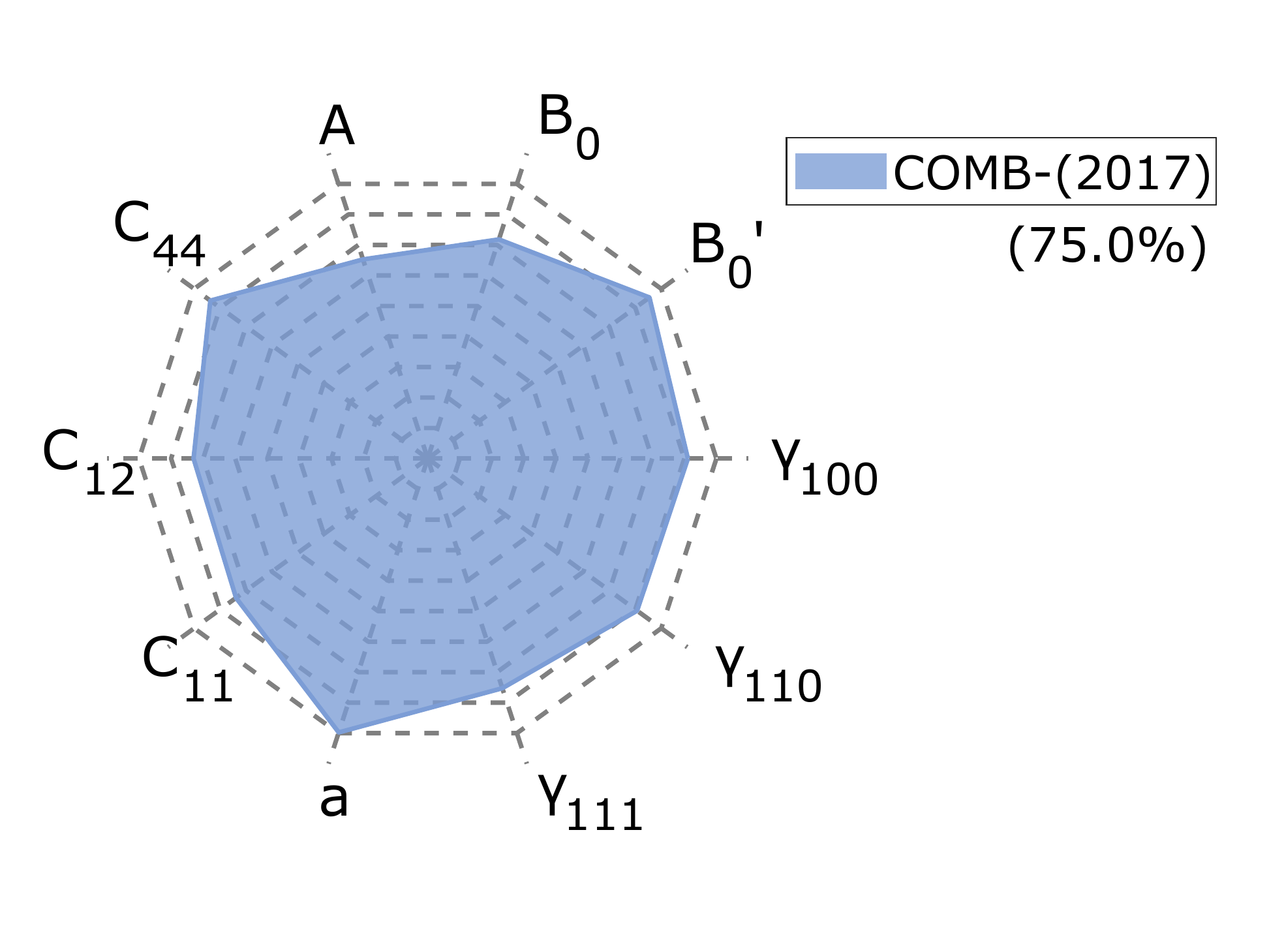}} &
\subfigure[]{}{\includegraphics[width=0.3\textwidth]{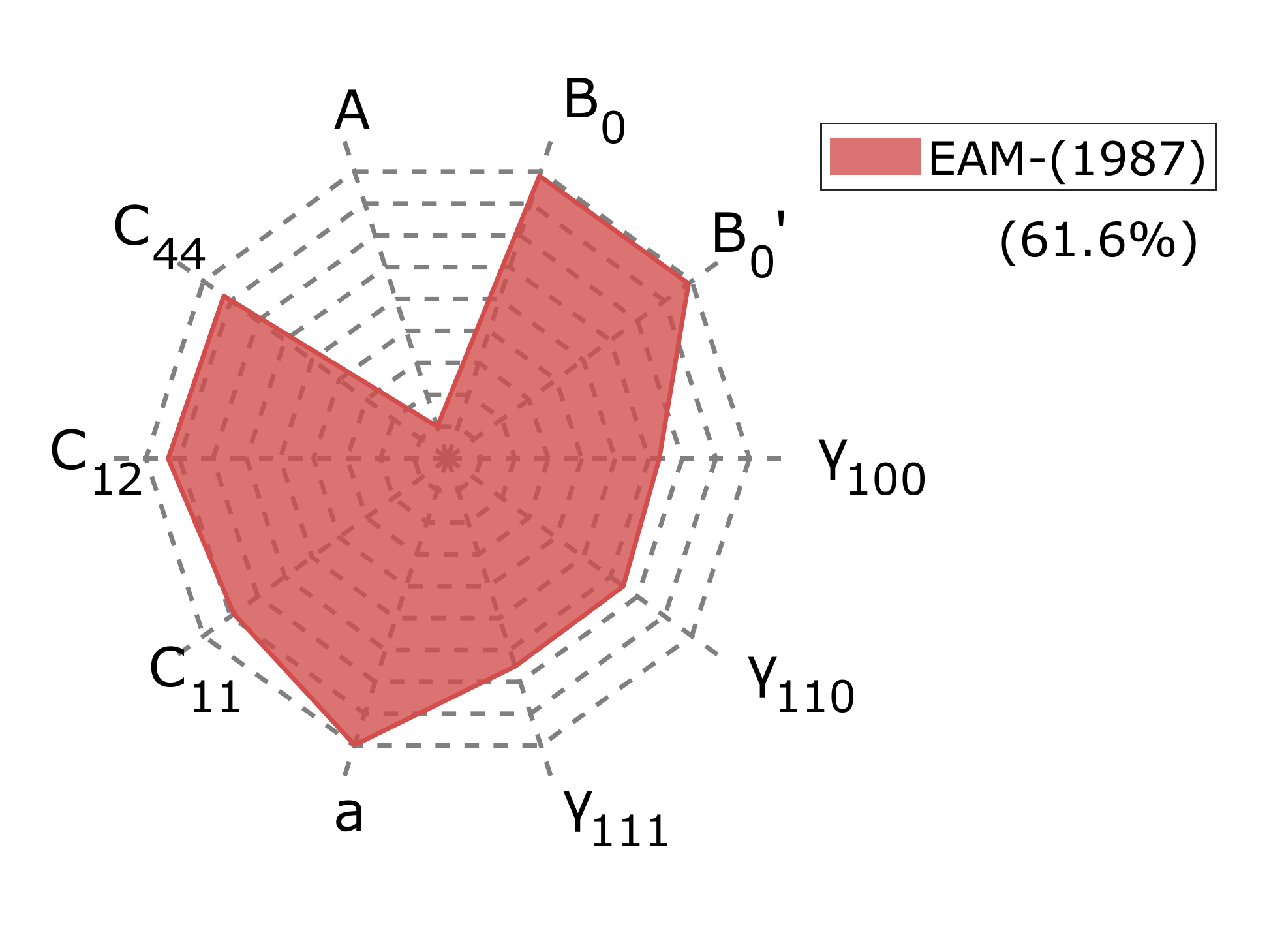}} &
\subfigure[]{}{\includegraphics[width=0.3\textwidth]{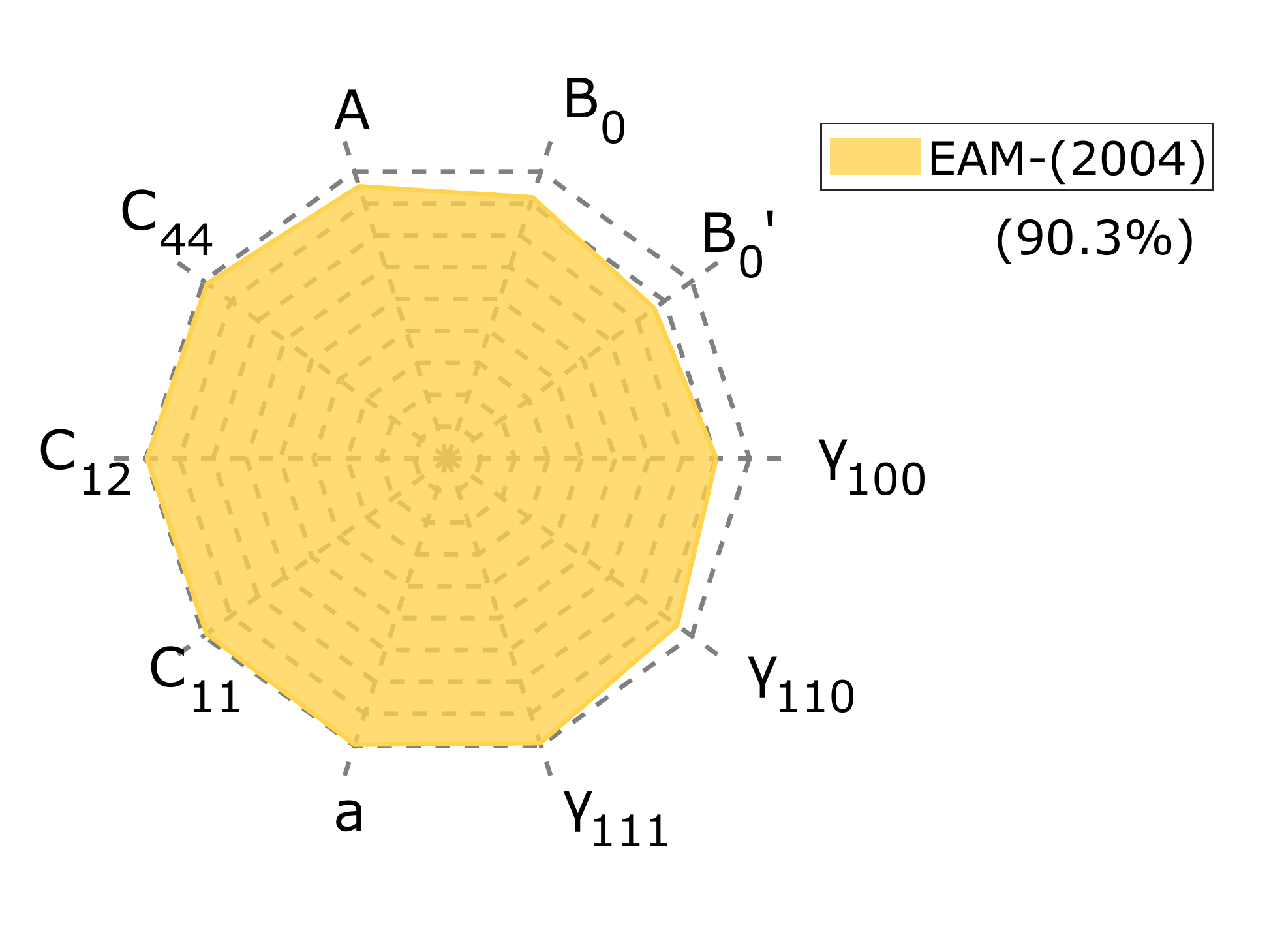}} \\
\subfigure[]{}{\includegraphics[width=0.3\textwidth]{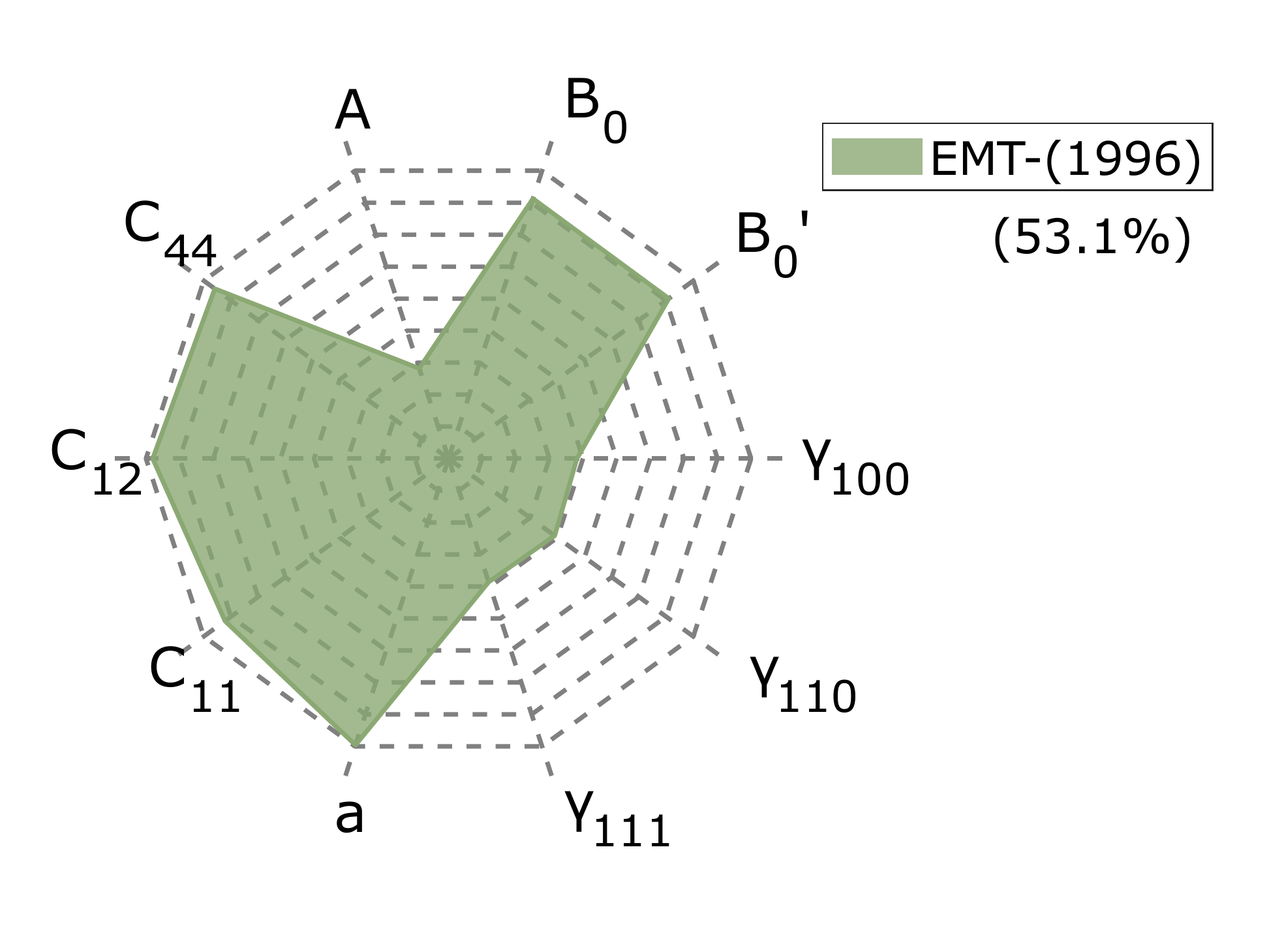}} &
\subfigure[]{}{\includegraphics[width=0.3\textwidth]{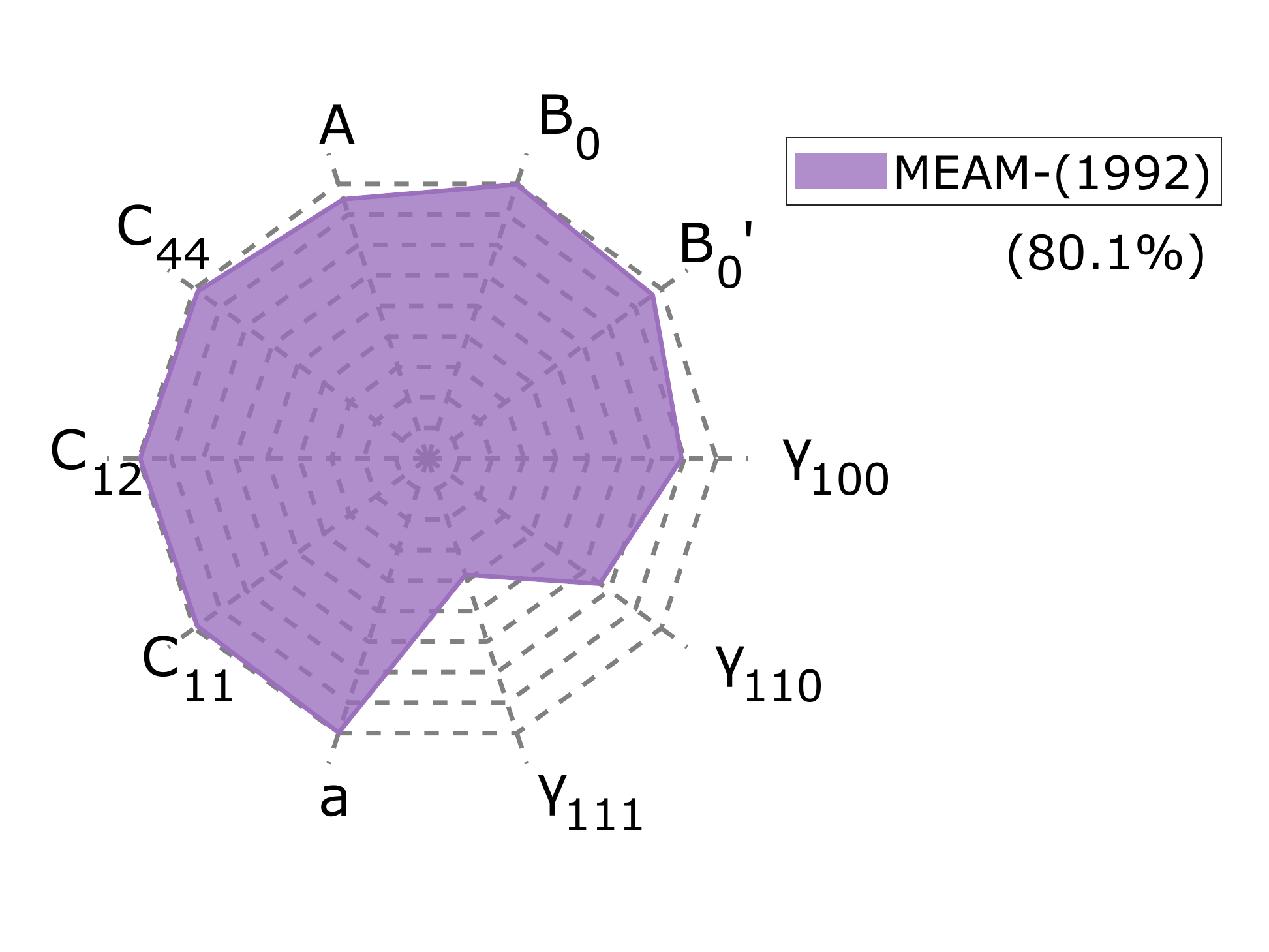}} &
\subfigure[]{}{\includegraphics[width=0.3\textwidth]{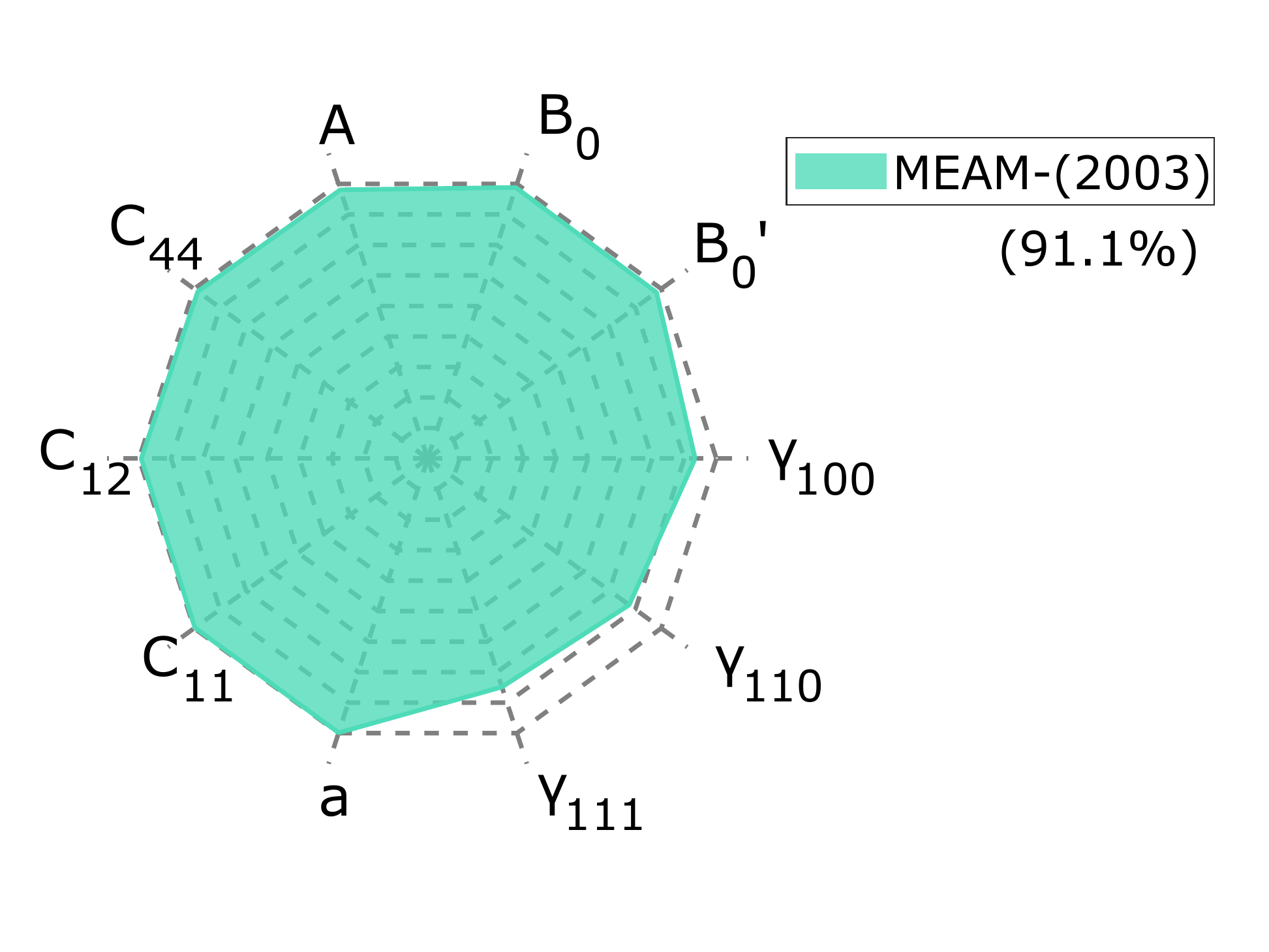}} \\
\subfigure[]{}{\includegraphics[width=0.3\textwidth]{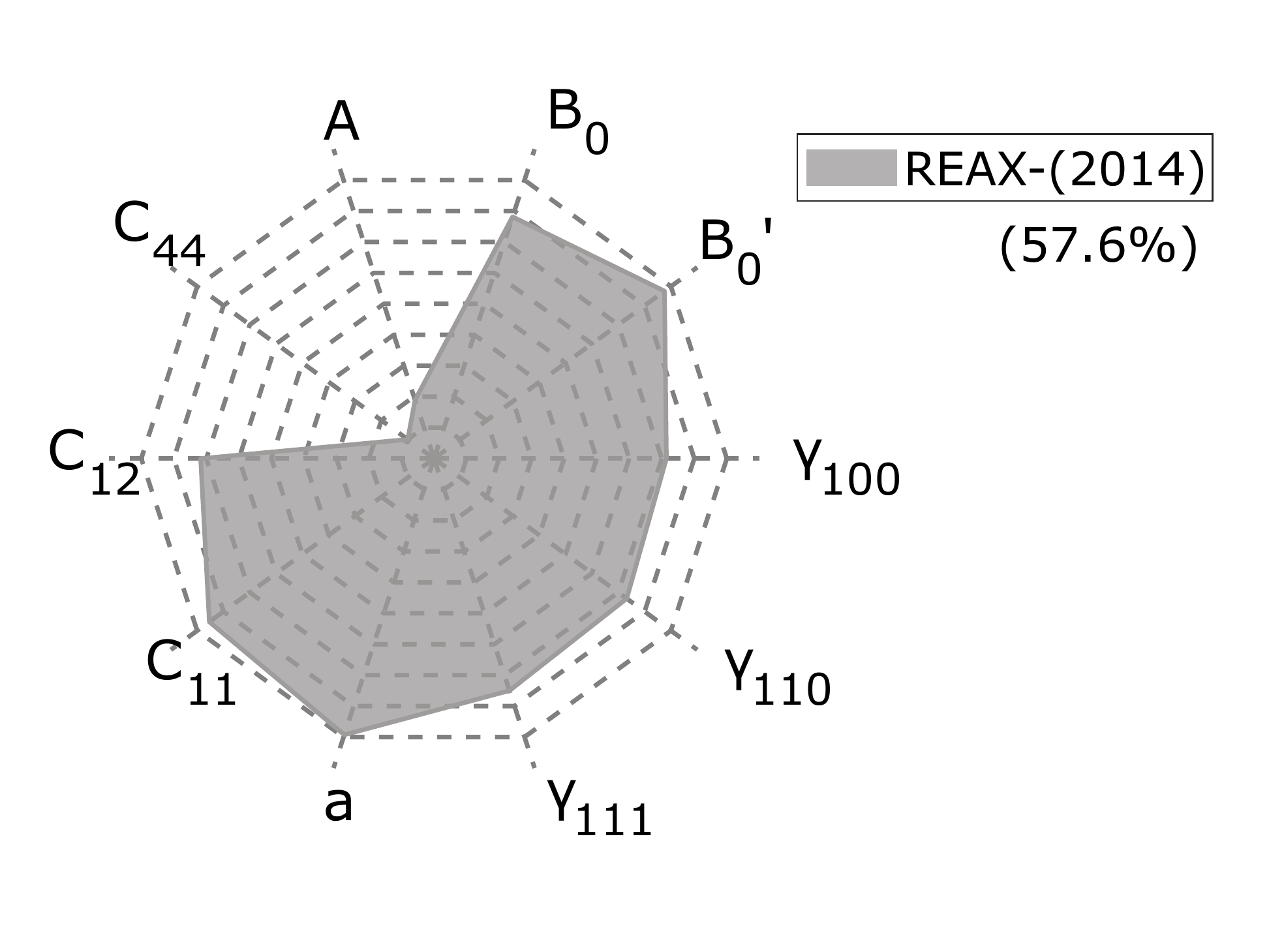}} &
\subfigure[]{}{\includegraphics[width=0.3\textwidth]{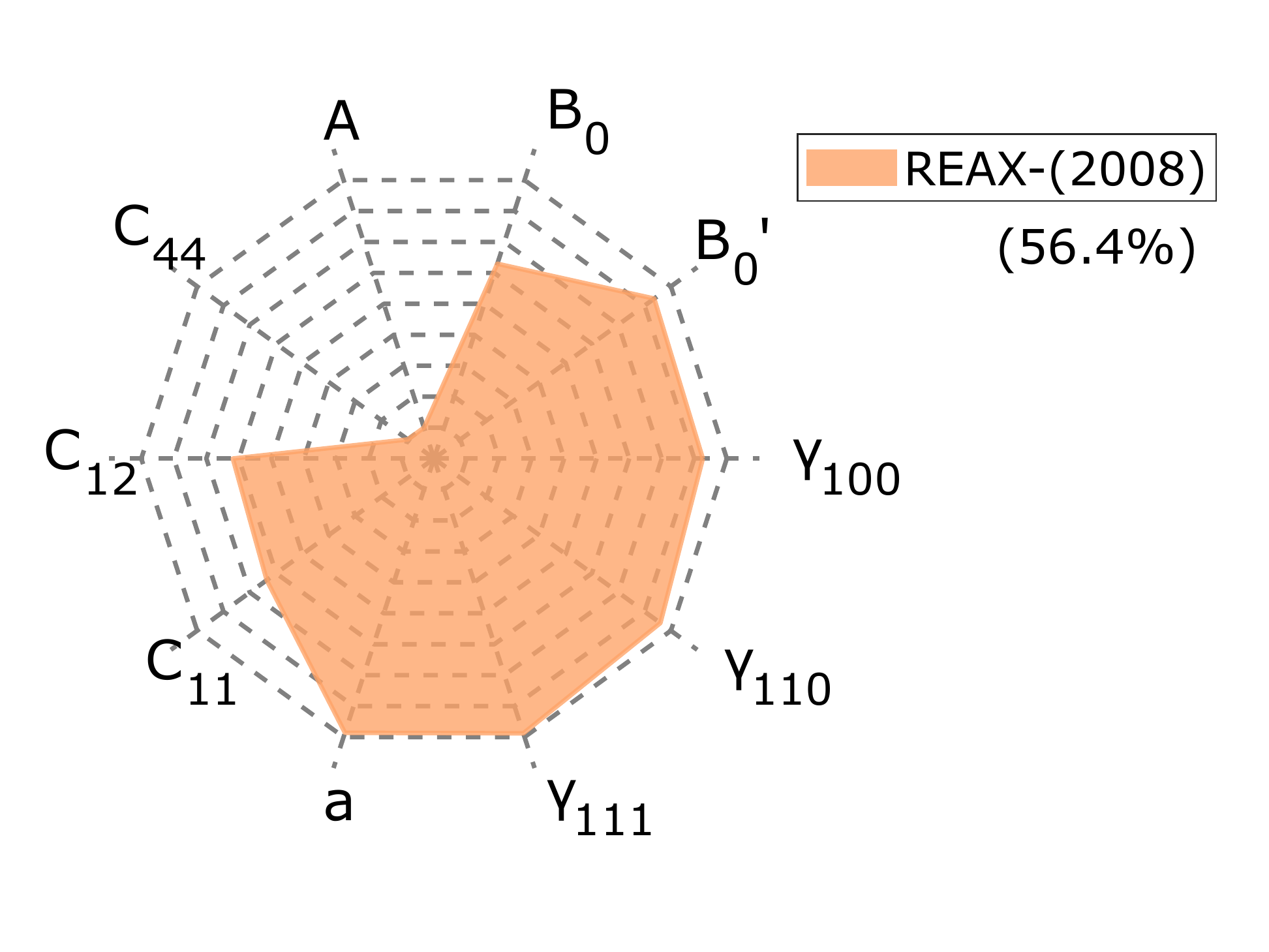}} &
\subfigure[]{}{\includegraphics[width=0.3\textwidth]{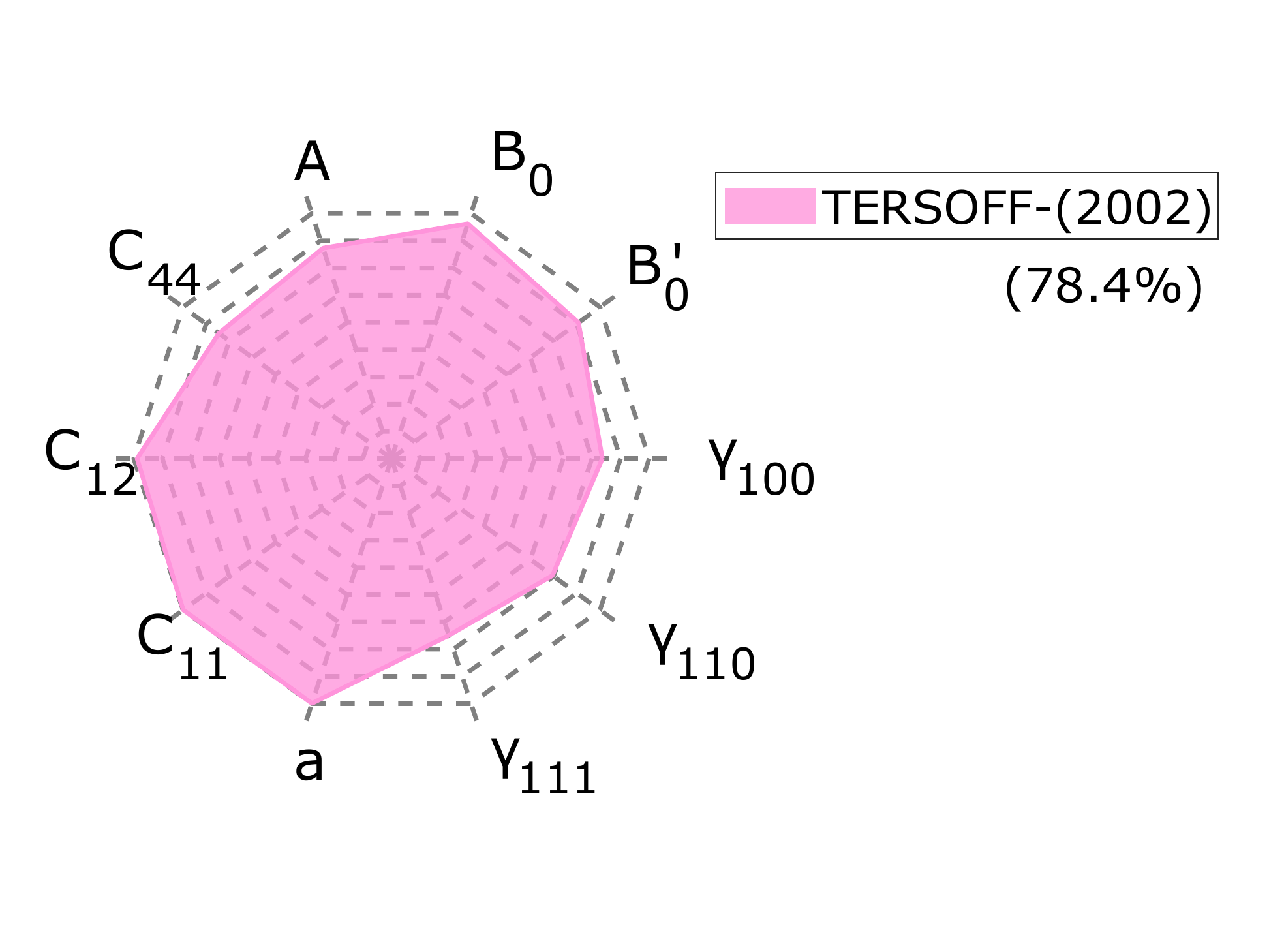}} \\
\end{tabular}
\caption{Radar plots summarizing the accuracy of MD calculations based on comparisons to experimental values and DFT calculations: (a) COMB-(2017), (b) EAM-(1987), (c) EAM-(2004), (d) EMT-(1996), (e) MEAM-(1992), (f) MEAM-(2003), (g) REAX-(2014), (h) REAX-(2008), and (i) TERSOFF-(2002). The percentage of the radar plot area filled, given in parenthesis, quantifies overall accuracy, where a higher percentage reflects greater accuracy.}\label{fig:radarPlot}
\end{figure}

Bulk and surface properties of platinum calculated from the MD models with nine different force fields were compared to experimental results as detailed above: lattice parameter,~\cite{Kittel2005} stiffness constants,~\cite{Simmons1965,Landolt1966} and Birch-Murnaghan equation of state fit parameters (average of values reported in Refs.~\cite{Dewaele2004,Fei2004,Zha2008,Matsui2009,Yokoo2009,Dorfman2012,Ye2017,Elkin2020}). The surface energies calculated from the MD models were compared to DFT calculations with the Vosko-Wilk-Nusair work function.~\cite{Vega2018}
To visualize the differences between the force fields and determine which is most accurate overall, radar plots were created, as shown in Figure~\ref{fig:radarPlot}.
In these figures, a larger fill area means a smaller error, i.e. the perimeter of the radar plot corresponds to zero error, while the center point corresponds to 100\% error. If the error was larger than 100\%, the value was adjusted and is shown as a maximum error at the center point. The percentage of area filled was also calculated and is presented in these figures.

As seen in Figure~\ref{fig:radarPlot}, none of the nine evaluated potentials accurately predicts all the properties of platinum. There are uncertainties inherent to empirical force fields that are responsible for the inaccuracy. First, truncation in the calculation of short- and long-range interactions have to be assumed and are characteristics of the selected functional form of the force field.~\cite{Angelikopoulos2012} Second, force field parameters are calibrated to reproduce specific experimental measurements and quantum mechanics calculations. However, when the potentials are used for simulations outside the scope of the conditions for which they were parameterized, their predictions may become inaccurate.~\cite{Angelikopoulos2012} Additionally, because of the fit to experimental measurements, the parameterization of a force field is susceptible to experimental or observational errors. There are also differences in the material models employed in MD compared to real materials in experiments. In our MD models, the simulated material is based on a perfect crystal, while in reality materials are affected by defects, grain distribution and size, or grain boundary effects. Finally, force field parameters fitted to DFT results depend heavily on the approximation to the exchange correlation energy functional chosen for the DFT calculations and fitting.  

Regardless, visual analysis of Fig.~\ref{fig:radarPlot} indicates that the COMB-(2017), EAM-(2004), MEAM-(2003), MEAM (1992), and TERSOFF-(2002) force fields are the most accurate in terms of their ability to model physical and mechanical properties of platinum in its bulk form using periodic boundary conditions, for pressures up to 28~GPa. This observation is supported quantitatively by the percentage of the total radar plot area filled, which is above 75\% for the above-mentioned five force fields. However, molecular models of nanomaterials have features not present in bulk materials due to the lack of neighboring atoms at the surface of the systems and differences in the boundary conditions of the simulation cell. 
Therefore, the potentials were further evaluated based on the stability of nanoparticles with facets in different orientations, as presented next.

\subsection{Stability of Small Platinum Nanoparticles}
\begin{figure}[ht]
\centering
\begin{tabular}{c c}
\subfigure[]{}{\includegraphics[width=0.45\textwidth]{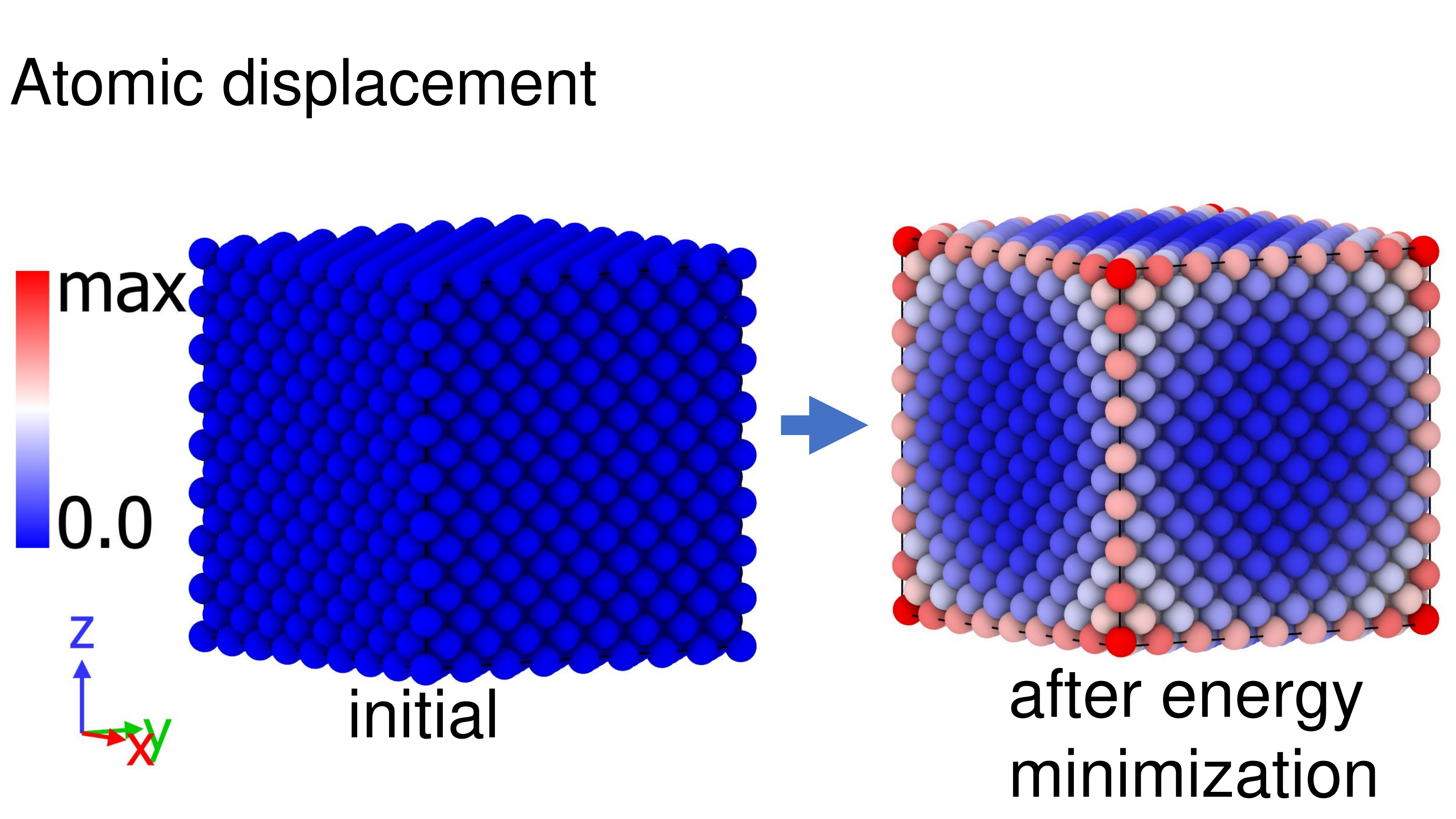}}
&
\subfigure[]{}{\includegraphics[width=0.45\textwidth]{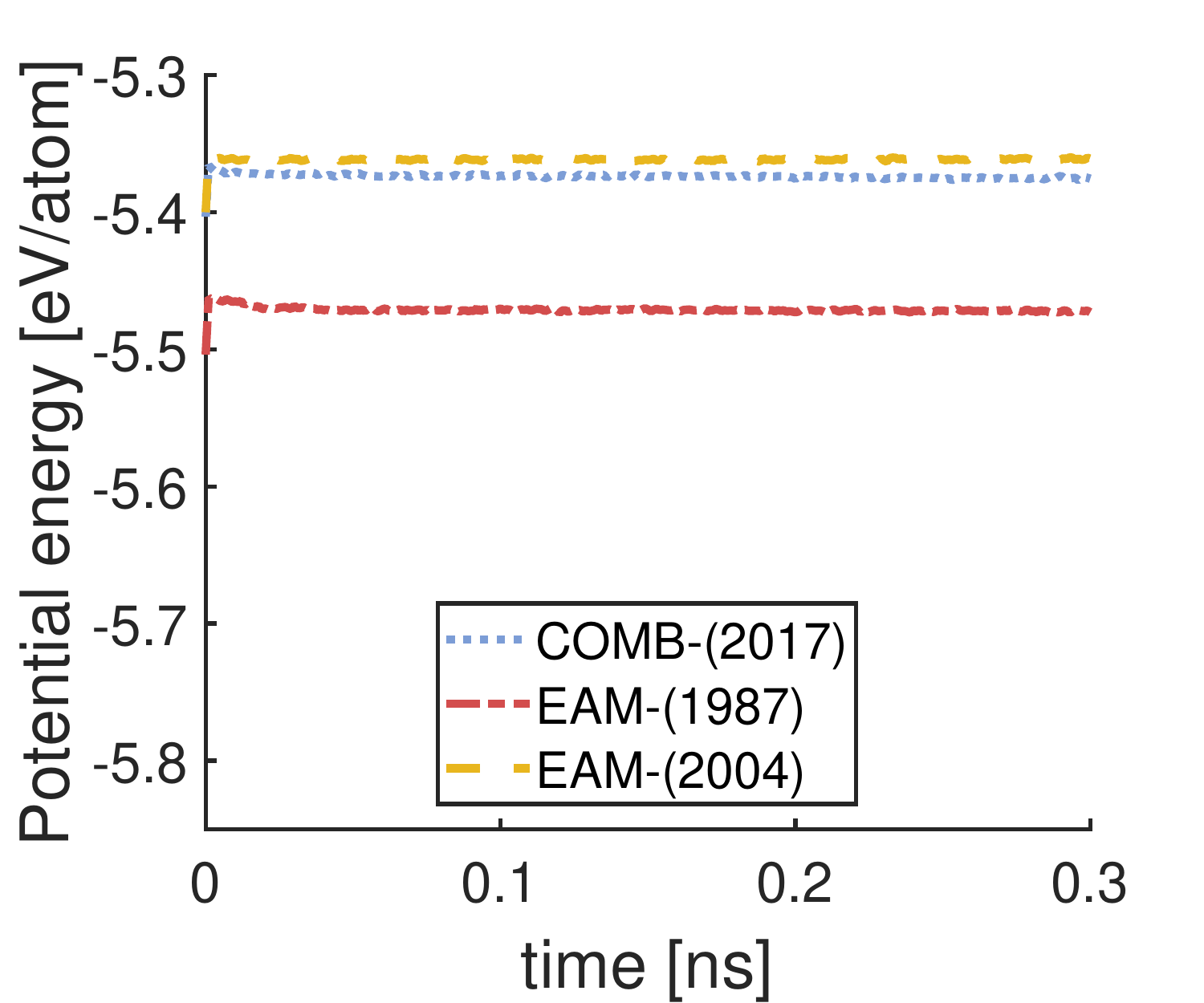}}
\\
\subfigure[]{}{\includegraphics[width=0.45\textwidth]{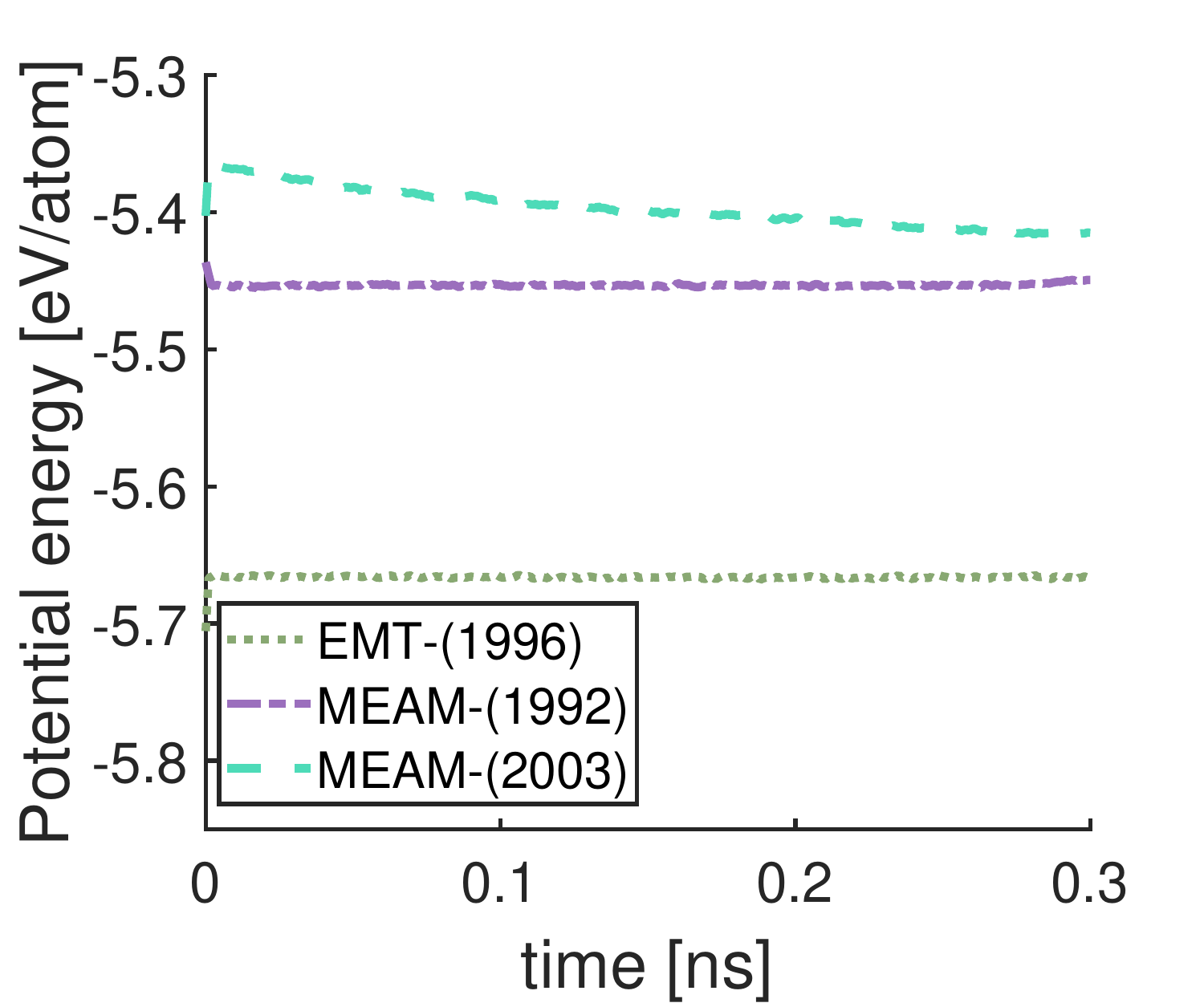}}
&
\subfigure[]{}{\includegraphics[width=0.45\textwidth]{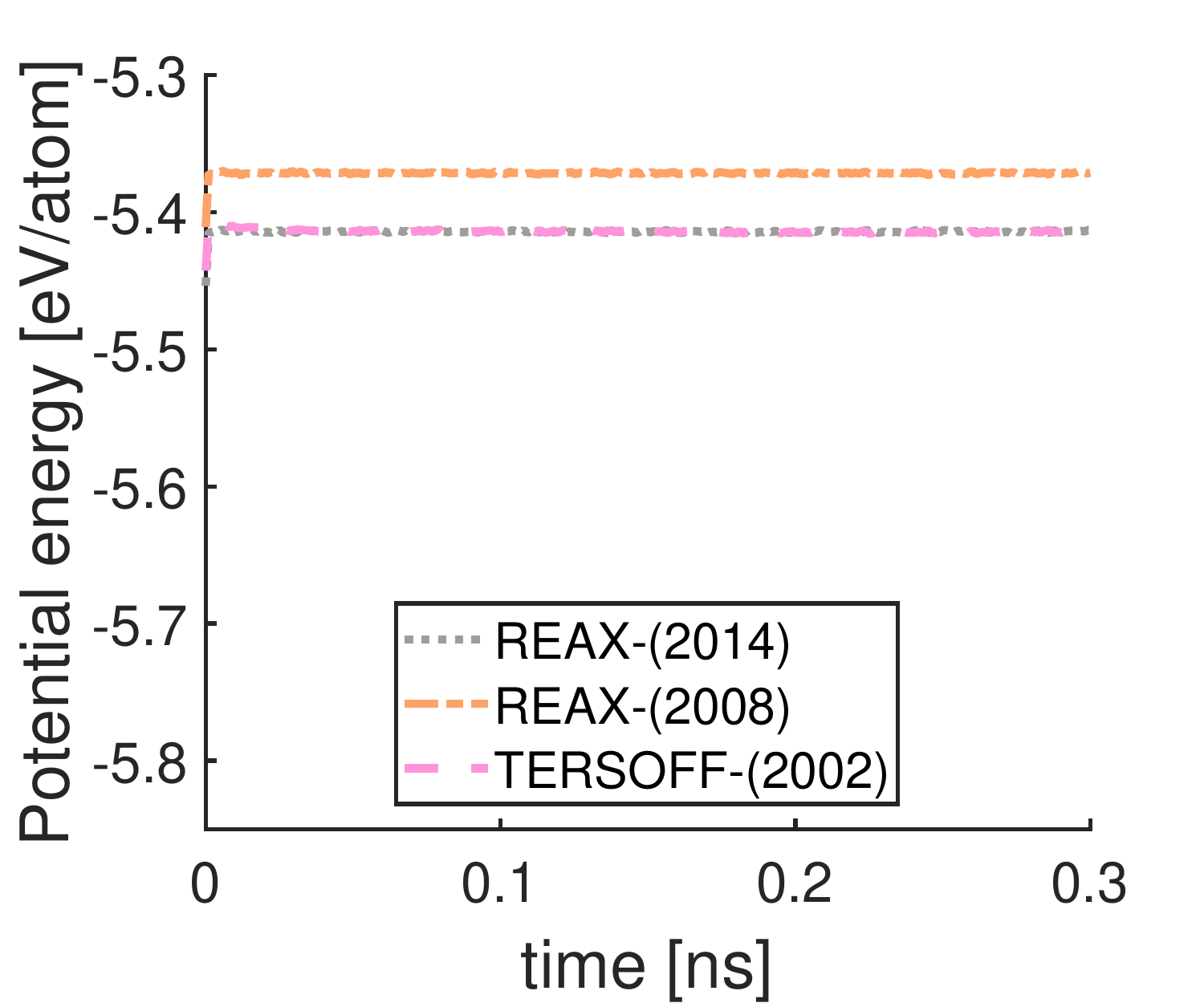}}
\\
\end{tabular}
\caption{(a) Representative snapshot of a 3.2~nm cubic nanoparticle before and after energy minimization with atom colors corresponding to displacement in $10^{-1}$~nm. (b)-(d) Potential energy per atom during thermal equilibration with the different force fields.}\label{fig:PE-time_cubic}
\end{figure}

\begin{figure}[ht]
\centering
\begin{tabular}{c c}
\subfigure[]{}{\includegraphics[width=0.45\textwidth]{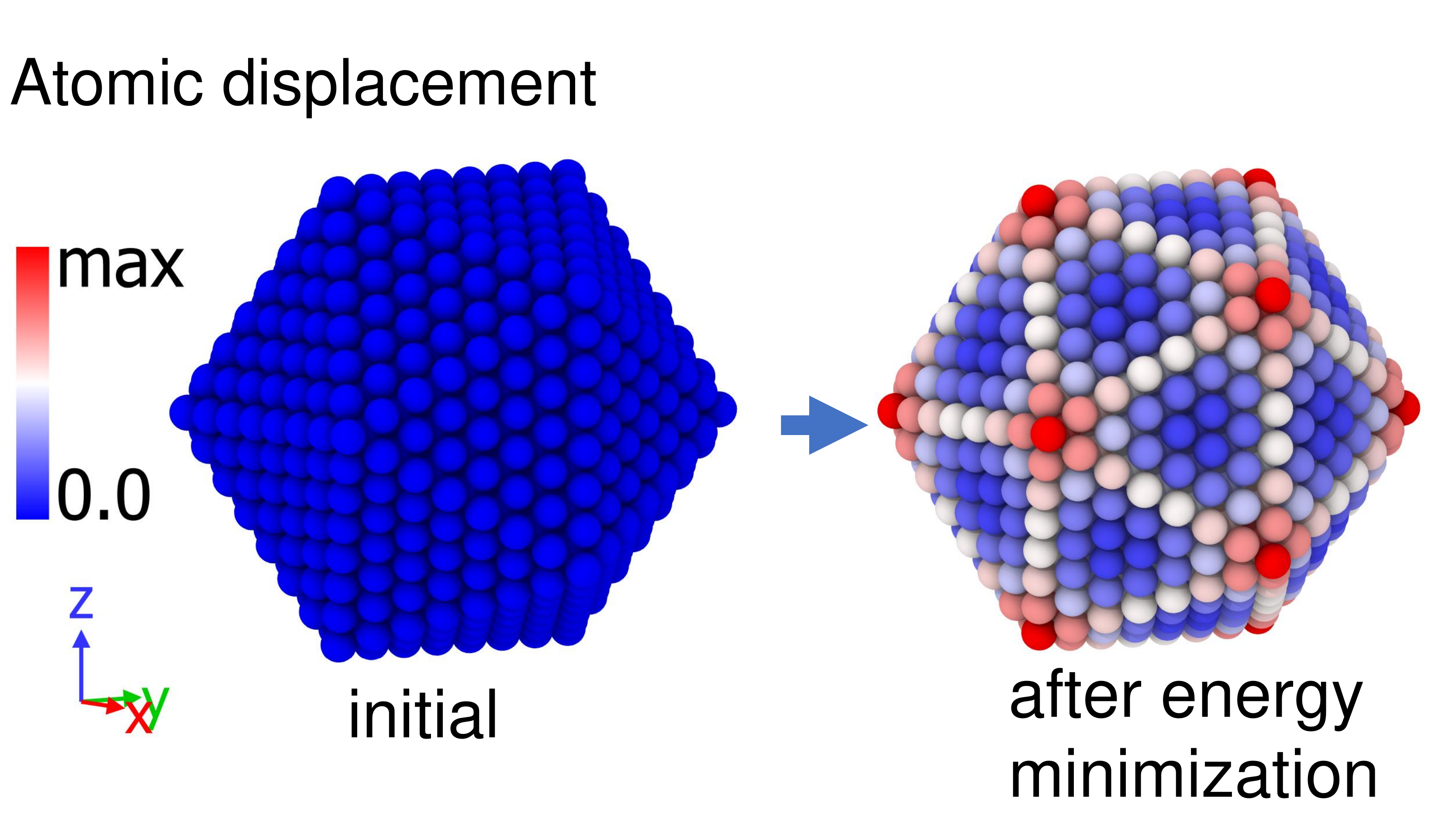}}
&
\subfigure[]{}{\includegraphics[width=0.45\textwidth]{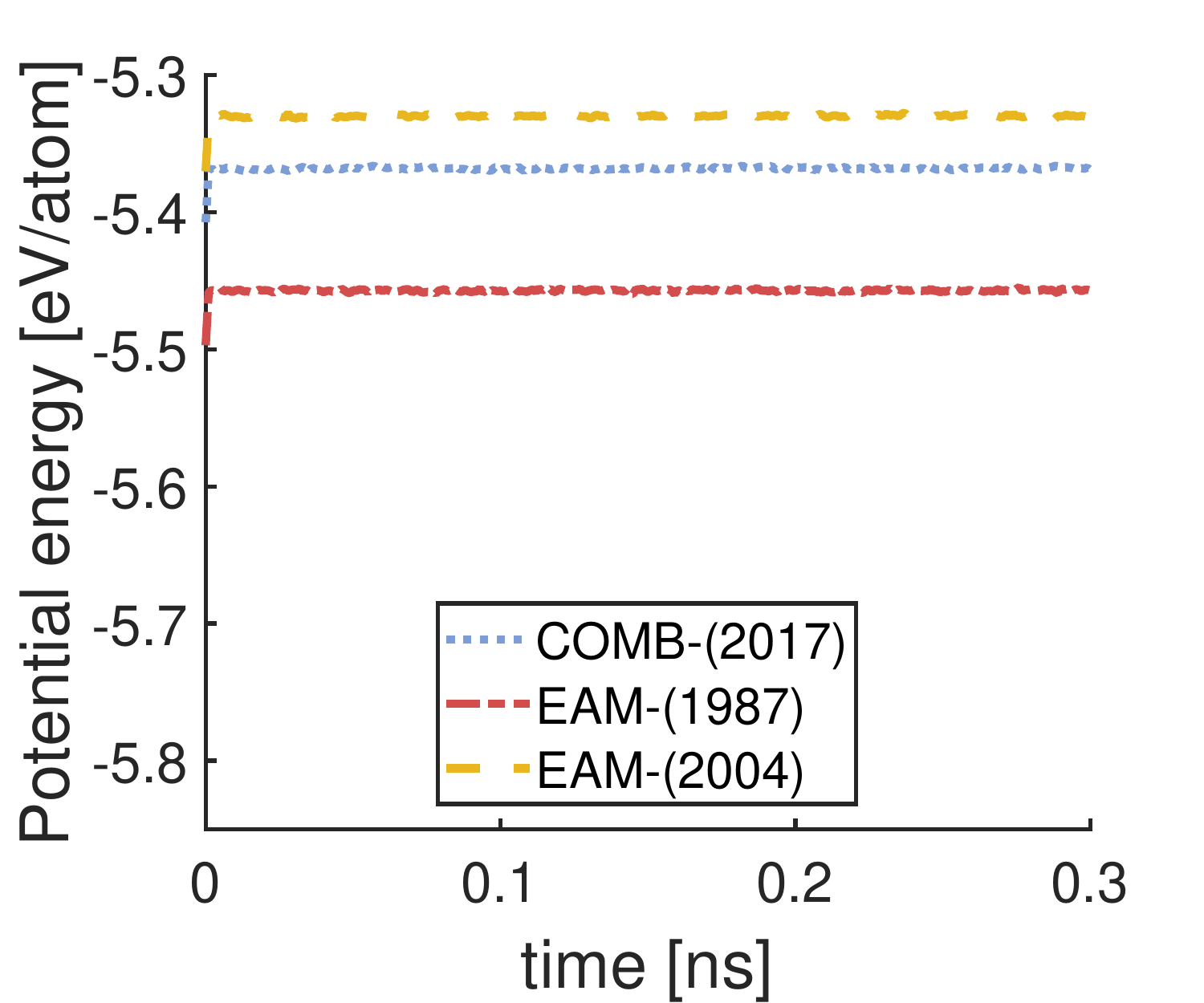}}
\\
\subfigure[]{}{\includegraphics[width=0.45\textwidth]{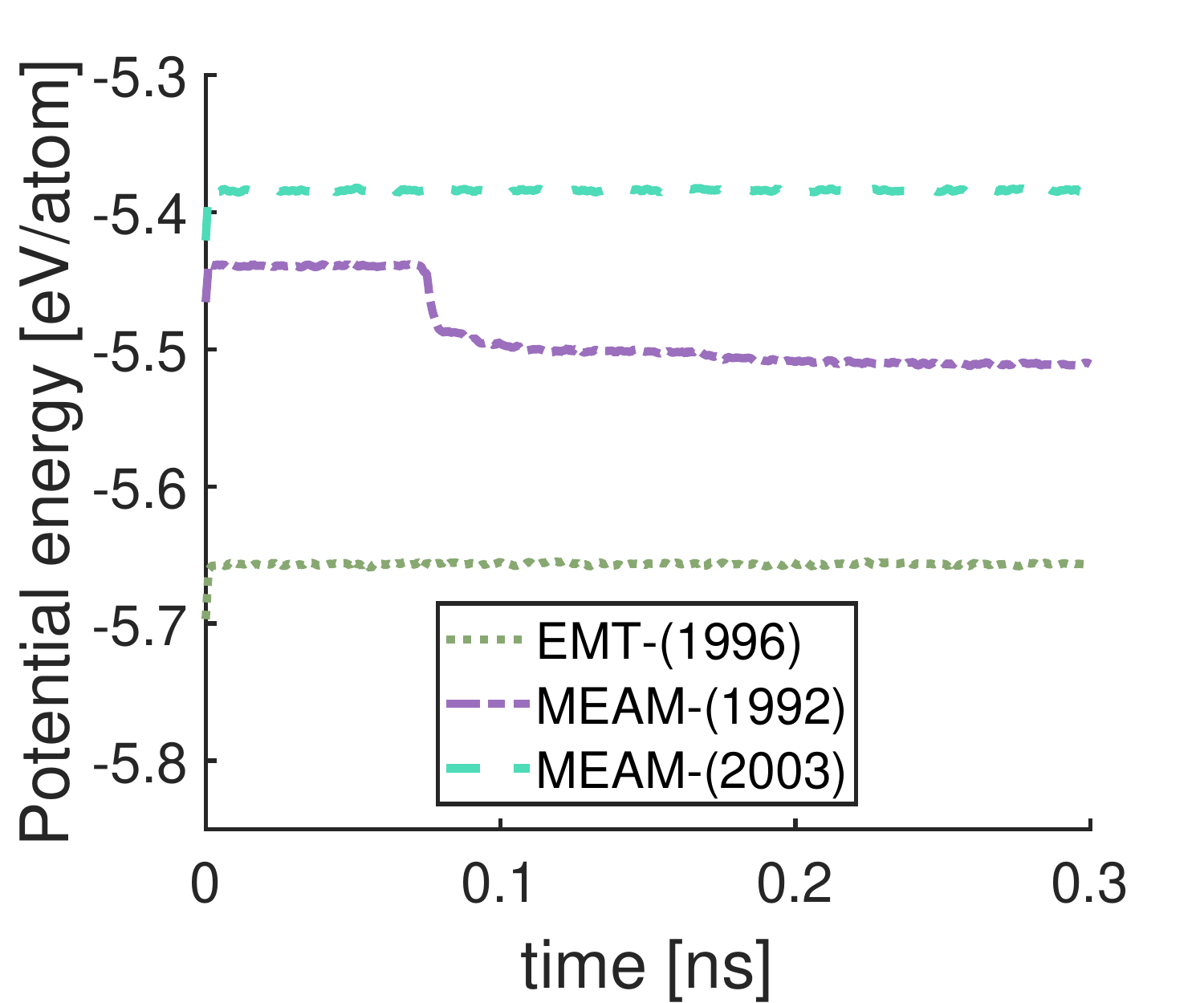}}
&
\subfigure[]{}{\includegraphics[width=0.45\textwidth]{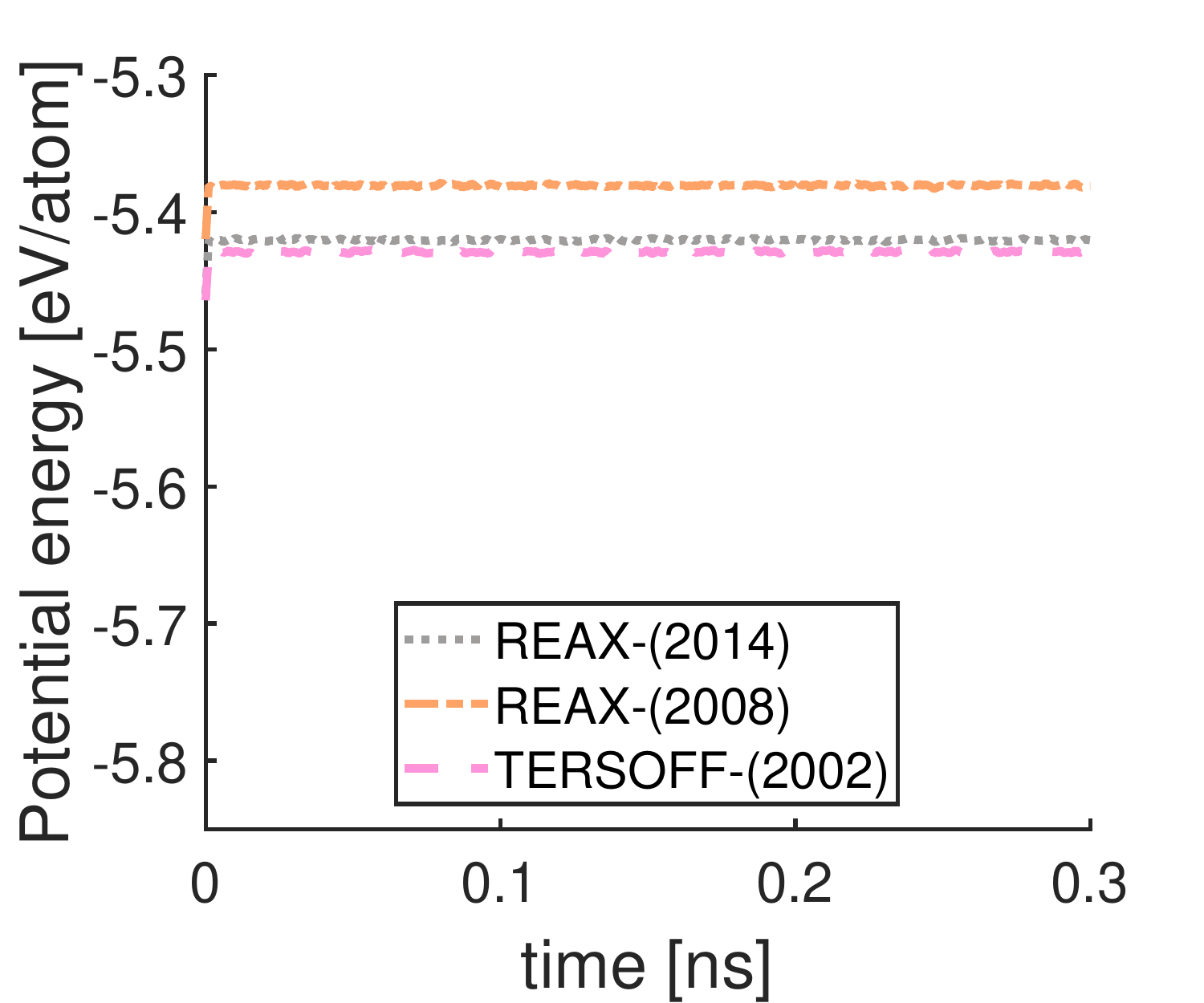}}
\\
\end{tabular}
\caption{a) Representative snapshot of a 3.2~nm icosahedral nanoparticle before and after energy minimization with atom colors corresponding to displacement in $10^{-1}$~nm. (b)-(d) Potential energy per atom during thermal equilibration with the different force fields.}\label{fig:PE-time_ico}
\end{figure}

A cubic platinum nanoparticle with \{100\} facets and an icosahedral nanoparticle with \{111\} facets were evaluated. An initial geometric optimization showed that the displacement of the surface atoms differed from the displacement of the subsurface atoms relative to their initial positions.
Representative snapshots from the optimization with atoms colored according to their atomic displacement are shown in Figure~\ref{fig:PE-time_cubic}a for the cubic nanoparticle and Figure~\ref{fig:PE-time_ico}a for the icosahedral nanoparticle, with visualization performed using OVITO software.~\cite{ovito}  The direction of this atomic displacement followed the same pattern for all the potentials, except for the cubic nanoparticle with both ReaxFF (figures for all the force fields shown in \href{https://pubs.acs.org/doi/10.1021/acs.jctc.1c00434}{Supporting Information}). For most of the force fields, the optimized positions of the subsurface atoms corresponded to the crystalline FCC phase, while the atoms at the surface moved slightly inward as a result of the energy imbalance from the lack of  neighboring atoms. Different behavior was observed for the cubic nanoparticle with both ReaxFF potentials, where the surface atoms remained near their initial positions, and the subsurface atoms moved towards the surface. However, the final configurations for all the force fields had internal atoms corresponding to the crystalline FCC phase and surface atoms displaced from the FCC lattice. 
This effect was more pronounced at the corners and edges of the structures. 

Following geometry optimization, the nanoparticles were equilibrated at room temperature for 0.3~ns.
The stability of the nanoparticle was then evaluated based on the change in potential energy over time, i.e. the potential energy will have small fluctuations around a constant mean value if the structure is stable. 

Figure~\ref{fig:PE-time_cubic} shows the potential energy per atom of the cubic nanoparticle with faces oriented in \{100\} planes during the thermal equilibration process. 
For all force fields except MEAM-(2003), the cubic nanoparticle is stable at room temperature.
With the MEAM-(2003) potential, the potential energy decreases with increasing equilibration time, indicating continuous structural reordering over time. In contrast, all the other potentials show an initial relaxation of the potential energy followed by stability of the nanoparticle in a cubic form. Studies on platinum nanoparticles have shown that nanocubes can be synthesized and are stable at room temperature.~\cite{Wu2017, Fu2013}

Figure~\ref{fig:PE-time_ico} shows the potential energy per atom vs equilibration time for the icosahedral nanoparticle with facets oriented in \{111\} planes. All the force fields predict stable icosahedral nanoparticles, except the MEAM-(1992) force field which exhibits a sudden change in the potential energy without external disturbance. Synthesis of stable icosahedral nanoparticles has been proven to be feasible experimentally.~\cite{Wu2017}

The prediction of nanoparticle stability was evaluated using three different thermostats. The results with the Nosé-Hoover thermostat are shown in figures ~\ref{fig:PE-time_cubic} and ~\ref{fig:PE-time_ico}, but the same trends are exhibited by the other thermostats, as shown in the \href{https://pubs.acs.org/doi/10.1021/acs.jctc.1c00434}{Supporting Information}, Figures S4-S7. With all three thermostats, the MEAM-(2003) force field predicts an unstable cube with facets in \{100\} orientations and the MEAM-(1992) predicts unstable icosahedron with facets in \{111\} orientations, while the other force fields predict stable nanoparticles. This analysis of the stability of nanoparticles demonstrates that the MEAM-(1992) and the MEAM-(2003) force fields are not suitable for modeling faceted platinum nanoparticles, although the MEAM-(2003) accurately predicted the properties of bulk platinum.
\clearpage

\section{Conclusions}
In this study, we evaluated the ability of nine potentials to model small, faceted platinum nanoparticles using molecular dynamics simulations. First, the force fields were evaluated based on their prediction of bulk and surface properties, specifically: stiffness constants, the equation of state, and surface energies.
The simulation-predicted values were compared to results from experiments or quantum mechanics calculations. Five force fields, the EAM-(2004),~\cite{Zhou2004} MEAM-(2003),~\cite{Lee2003}, MEAM-(1992),~\citenum{Baskes1992}, TERSOFF-(2002),~\cite{Albe2002}, and COMB-(2017)~\cite{Antony2017} potentials were found to be the most accurate in terms of their ability to model physical, mechanical and surface properties. 
However, the MEAM-(1992) force field is not recommended to model systems with pressures above 28~GPa.
Further, an analysis of the stability of nanoparticles with surfaces in the \{100\} and \{111\} planes revealed that the MEAM-(2003) and MEAM-(1992) potentials failed to reproduce the structural integrity of nanoparticles that can be synthesized and remain stable at room conditions in experimental conditions.
The EAM-(2004) potential, the TERSOFF-(2002) potential, and the COMB-(2017) potential predicted the expected nanoparticle stability.
Therefore, since the bulk and surface properties were most accurately predicted by the EAM-(2004) potential,~\cite{Zhou2004} this study demonstrates that this EAM potential is the most suitable for molecular dynamics simulations of small, faceted platinum nanoparticles.

\subsection{Acknowledgement}
This material is based upon work supported by the U.S. Department of Energy, Office of Science, Office of Basic Energy Sciences, under Award Number DE-FOA-0002181
Some of the simulations were run using the Extreme Science and Engineering Discovery Environment (XSEDE), which is supported by the National Science Foundation Grant ACI-1548562

\subsection{Supporting Information}
The representation of a slab system for calculation of surface energies, a comprehensive review of the surface energy of platinum and bulk modulus derived from the equation of state, calculated bulk and surface properties, and the atomic displacement for the two nanoparticles shapes after energy minimization process are available in the \href{https://pubs.acs.org/doi/10.1021/acs.jctc.1c00434}{Supporting Information}.

\bibliographystyle{abbrvnat}

\begin{thebibliography}{108}
\providecommand{\natexlab}[1]{#1}
\providecommand{\url}[1]{\texttt{#1}}
\expandafter\ifx\csname urlstyle\endcsname\relax
  \providecommand{\doi}[1]{doi: #1}\else
  \providecommand{\doi}{doi: \begingroup \urlstyle{rm}\Url}\fi

\bibitem[Ahmadi et~al.(2019)Ahmadi, Timoshenko, Behafarid, and {Roldan
  Cuenya}]{Ahmadi2019}
M.~Ahmadi, J.~Timoshenko, F.~Behafarid, and B.~{Roldan Cuenya}.
\newblock {Tuning the Structure of Pt Nanoparticles through Support
  Interactions: An in Situ Polarized X-ray Absorption Study Coupled with
  Atomistic Simulations}.
\newblock \emph{J. Phys. Chem. C}, 123\penalty0 (16):\penalty0 10666--10676,
  mar 2019.
\newblock \doi{10.1021/acs.jpcc.9b00945}.

\bibitem[Albe et~al.(2002)Albe, Nordlund, and Averback]{Albe2002}
K.~Albe, K.~Nordlund, and R.~S. Averback.
\newblock {Modeling the metal-semiconductor interaction: Analytical bond-order
  potential for platinum-carbon}.
\newblock \emph{Phys. Rev. B}, 65\penalty0 (19):\penalty0 195124, may 2002.
\newblock \doi{10.1103/PhysRevB.65.195124}.
\newblock URL \url{https://link.aps.org/doi/10.1103/PhysRevB.65.195124}.

\bibitem[Angelikopoulos et~al.(2012)Angelikopoulos, Papadimitriou, and
  Koumoutsakos]{Angelikopoulos2012}
P.~Angelikopoulos, C.~Papadimitriou, and P.~Koumoutsakos.
\newblock {Bayesian uncertainty quantification and propagation in molecular
  dynamics simulations: A high performance computing framework}.
\newblock \emph{J. Chem. Phys.}, 137\penalty0 (14):\penalty0 144103, oct 2012.
\newblock ISSN 0021-9606.
\newblock \doi{10.1063/1.4757266}.
\newblock URL \url{https://doi.org/10.1063/1.4757266}.

\bibitem[Antony et~al.(2017)Antony, Akhade, Lu, Liang, Janik, Phillpot, and
  Sinnott]{Antony2017}
A.~C. Antony, S.~A. Akhade, Z.~Lu, T.~Liang, M.~J. Janik, S.~R. Phillpot, and
  S.~B. Sinnott.
\newblock {Charge optimized many body (COMB) potentials for Pt and Au}.
\newblock \emph{J. Phys.: Condens. Matter}, 29\penalty0 (22):\penalty0 225901,
  jun 2017.
\newblock \doi{10.1088/1361-648X/aa6d43}.
\newblock URL
  \url{https://iopscience.iop.org/article/10.1088/1361-648X/aa6d43}.

\bibitem[Armor(2011)]{Armor2011}
J.~N. Armor.
\newblock {A history of industrial catalysis}.
\newblock \emph{Catal. Today}, 163\penalty0 (1):\penalty0 3--9, 2011.
\newblock ISSN 0920-5861.
\newblock \doi{https://doi.org/10.1016/j.cattod.2009.11.019}.
\newblock URL
  \url{http://www.sciencedirect.com/science/article/pii/S0920586109006944}.

\bibitem[Baskes(1992)]{Baskes1992}
M.~I. Baskes.
\newblock {Modified embedded-atom potentials for cubic materials and
  impurities}.
\newblock \emph{Phys. Rev. B}, 46\penalty0 (5):\penalty0 2727--2742, aug 1992.
\newblock \doi{10.1103/PhysRevB.46.2727}.
\newblock URL \url{https://link.aps.org/doi/10.1103/PhysRevB.46.2727}.

\bibitem[Baskes and Melius(1979)]{Baskes1979}
M.~I. Baskes and C.~F. Melius.
\newblock {Pair potentials for fcc metals}.
\newblock \emph{Phys. Rev. B}, 20\penalty0 (8):\penalty0 3197--3204, oct 1979.
\newblock \doi{10.1103/PhysRevB.20.3197}.
\newblock URL \url{https://link.aps.org/doi/10.1103/PhysRevB.20.3197}.

\bibitem[Birch(1947)]{Birch1947}
F.~Birch.
\newblock {Finite Elastic Strain of Cubic Crystals}.
\newblock \emph{Phys. Rev.}, 71\penalty0 (11):\penalty0 809--824, jun 1947.
\newblock \doi{10.1103/PhysRev.71.809}.
\newblock URL \url{https://link.aps.org/doi/10.1103/PhysRev.71.809}.

\bibitem[Brenner(1990)]{Brenner1990}
D.~W. Brenner.
\newblock {Empirical potential for hydrocarbons for use in simulating the
  chemical vapor deposition of diamond films}.
\newblock \emph{Phys. Rev. B}, 42\penalty0 (15):\penalty0 9458--9471, nov 1990.
\newblock \doi{10.1103/PhysRevB.42.9458}.
\newblock URL \url{https://link.aps.org/doi/10.1103/PhysRevB.42.9458}.

\bibitem[Brondani et~al.(2009)Brondani, Scheeren, Dupont, and
  Vieira]{Brondani2009}
D.~Brondani, C.~W. Scheeren, J.~Dupont, and I.~C. Vieira.
\newblock {Biosensor based on platinum nanoparticles dispersed in ionic liquid
  and laccase for determination of adrenaline}.
\newblock \emph{Sens. Actuators. B}, 140\penalty0 (1):\penalty0 252--259, 2009.
\newblock ISSN 0925-4005.
\newblock \doi{https://doi.org/10.1016/j.snb.2009.04.037}.
\newblock URL
  \url{http://www.sciencedirect.com/science/article/pii/S0925400509003475}.

\bibitem[Bruix et~al.(2012)Bruix, Rodriguez, Ramírez, Senanayake, Evans, Park,
  Stacchiola, Liu, Hrbek, and Illas]{Bruix2012}
A.~Bruix, J.~A. Rodriguez, P.~J. Ramírez, S.~D. Senanayake, J.~Evans, J.~B.
  Park, D.~Stacchiola, P.~Liu, J.~Hrbek, and F.~Illas.
\newblock A new type of strong metal–support interaction and the production
  of h2 through the transformation of water on pt/ceo2(111) and
  pt/ceox/tio2(110) catalysts.
\newblock \emph{J. Am. Chem. Soc.}, 134\penalty0 (21):\penalty0 8968--8974,
  2012.
\newblock \doi{10.1021/ja302070k}.
\newblock URL \url{https://doi.org/10.1021/ja302070k}.
\newblock PMID: 22563752.

\bibitem[{C. T. van Duin} et~al.(2001){C. T. van Duin}, Dasgupta, Lorant, and
  {A. Goddard}]{C.T.vanDuin2001}
A.~{C. T. van Duin}, S.~Dasgupta, F.~Lorant, and W.~{A. Goddard}.
\newblock {ReaxFF: A Reactive Force Field for Hydrocarbons}.
\newblock \emph{J. Phys. Chem. A}, 105\penalty0 (41):\penalty0 9396--9409, sep
  2001.
\newblock \doi{10.1021/jp004368u}.

\bibitem[Cai and Ye(1996)]{Cai1996}
J.~Cai and Y.~Y. Ye.
\newblock {Simple analytical embedded-atom-potential model including a
  long-range force for fcc metals and their alloys}.
\newblock \emph{Phys. Rev. B}, 54\penalty0 (12):\penalty0 8398--8410, sep 1996.
\newblock \doi{10.1103/PhysRevB.54.8398}.
\newblock URL \url{https://link.aps.org/doi/10.1103/PhysRevB.54.8398}.

\bibitem[Calogero et~al.(2011)Calogero, Calandra, Irrera, Sinopoli, Citro, and
  {Di Marco.}]{Calogero2011}
G.~Calogero, P.~Calandra, A.~Irrera, A.~Sinopoli, I.~Citro, and G.~{Di Marco.}
\newblock {A new type of transparent and low cost counter-electrode based on
  platinum nanoparticles for dye-sensitized solar cells}.
\newblock \emph{Energy Environ. Sci.}, 4\penalty0 (5):\penalty0 1838--1844,
  2011.
\newblock ISSN 1754-5692.
\newblock \doi{10.1039/C0EE00463D}.
\newblock URL \url{http://dx.doi.org/10.1039/C0EE00463D}.

\bibitem[Calvimontes(2017)]{Calvimontes2017}
A.~Calvimontes.
\newblock {The measurement of the surface energy of solids using a laboratory
  drop tower}.
\newblock \emph{npj Microgravity}, 3\penalty0 (1):\penalty0 25, 2017.
\newblock ISSN 2373-8065.
\newblock \doi{10.1038/s41526-017-0031-y}.
\newblock URL \url{https://doi.org/10.1038/s41526-017-0031-y}.

\bibitem[Cao et~al.(2016)Cao, Jiang, Zhu, and Yu]{Cao2016}
S.~Cao, J.~Jiang, B.~Zhu, and J.~Yu.
\newblock {Shape-dependent photocatalytic hydrogen evolution activity over a Pt
  nanoparticle coupled g-C3N4 photocatalyst}.
\newblock \emph{Phys. Chem. Chem. Phys.}, 18\penalty0 (28):\penalty0
  19457--19463, 2016.
\newblock ISSN 1463-9076.
\newblock \doi{10.1039/C6CP02832B}.
\newblock URL \url{http://dx.doi.org/10.1039/C6CP02832B}.

\bibitem[Caro et~al.(2005)Caro, Crowson, and Caro]{Caro2005}
A.~Caro, D.~A. Crowson, and M.~Caro.
\newblock {Classical Many-Body Potential for Concentrated Alloys and the
  Inversion of Order in Iron-Chromium Alloys}.
\newblock \emph{Phys. Rev. Lett.}, 95\penalty0 (7):\penalty0 75702, aug 2005.
\newblock \doi{10.1103/PhysRevLett.95.075702}.
\newblock URL \url{https://link.aps.org/doi/10.1103/PhysRevLett.95.075702}.

\bibitem[Casillas et~al.(2012)Casillas, Palomares-B{\'{a}}ez,
  Rodr{\'{i}}guez-L{\'{o}}pez, Luo, Ponce, Esparza, Vel{\'{a}}zquez-Salazar,
  Hurtado-Macias, Gonz{\'{a}}lez-Hern{\'{a}}ndez, and
  Jos{\'{e}}-Yacaman]{Casillas2012}
G.~Casillas, J.~P. Palomares-B{\'{a}}ez, J.~L. Rodr{\'{i}}guez-L{\'{o}}pez,
  J.~Luo, A.~Ponce, R.~Esparza, J.~J. Vel{\'{a}}zquez-Salazar,
  A.~Hurtado-Macias, J.~Gonz{\'{a}}lez-Hern{\'{a}}ndez, and
  M.~Jos{\'{e}}-Yacaman.
\newblock {In situ TEM study of mechanical behaviour of twinned nanoparticles}.
\newblock \emph{Philos. Mag.}, 92\penalty0 (35):\penalty0 4437--4453, dec 2012.
\newblock ISSN 1478-6435.
\newblock \doi{10.1080/14786435.2012.709951}.
\newblock URL \url{https://doi.org/10.1080/14786435.2012.709951}.

\bibitem[Chapman and Ramprasad(2020)]{Chapman2020}
J.~Chapman and R.~Ramprasad.
\newblock {Multiscale Modeling of Defect Phenomena in Platinum Using Machine
  Learning of Force Fields}.
\newblock \emph{JOM}, 72\penalty0 (12):\penalty0 4346--4358, 2020.
\newblock ISSN 1543-1851.
\newblock \doi{10.1007/s11837-020-04385-0}.
\newblock URL \url{https://doi.org/10.1007/s11837-020-04385-0}.

\bibitem[Chen et~al.(2020)Chen, Gao, and Castell]{Chen2020}
P.~Chen, Y.~Gao, and M.~R. Castell.
\newblock {Experimental determination of the {\{}111{\}}/{\{}001{\}} surface
  energy ratio for Pd crystals}.
\newblock \emph{Appl. Phys. Lett.}, 117\penalty0 (10):\penalty0 101601, sep
  2020.
\newblock ISSN 0003-6951.
\newblock \doi{10.1063/5.0022879}.
\newblock URL \url{https://doi.org/10.1063/5.0022879}.

\bibitem[Chepkasov et~al.(2018)Chepkasov, Visotin, Kovaleva, Manakhov,
  Baidyshev, and Popov]{Chepkasov2018}
I.~V. Chepkasov, M.~A. Visotin, E.~A. Kovaleva, A.~M. Manakhov, V.~S.
  Baidyshev, and Z.~I. Popov.
\newblock {Stability and Electronic Properties of PtPd Nanoparticles via MD and
  DFT Calculations}.
\newblock \emph{J. Phys. Chem. C}, 122\penalty0 (31):\penalty0 18070--18076,
  aug 2018.
\newblock ISSN 1932-7447.
\newblock \doi{10.1021/acs.jpcc.8b04177}.
\newblock URL \url{https://doi.org/10.1021/acs.jpcc.8b04177}.

\bibitem[Chiu et~al.(2011)Chiu, Li, Ruan, Ye, Murray, and Huang]{Chiu2011}
C.-Y. Chiu, Y.~Li, L.~Ruan, X.~Ye, C.~B. Murray, and Y.~Huang.
\newblock {Platinum nanocrystals selectively shaped using facet-specific
  peptide sequences}.
\newblock \emph{Nat. Chem.}, 3\penalty0 (5):\penalty0 393--399, 2011.
\newblock ISSN 1755-4349.
\newblock \doi{10.1038/nchem.1025}.
\newblock URL \url{https://doi.org/10.1038/nchem.1025}.

\bibitem[Choi et~al.(2020)Choi, Chung, Kim, Kim, Yun, Cha, Harder, Kawaguchi,
  Liu, Ulvestad, You, Song, and Kim]{Choi2020}
S.~Choi, M.~Chung, D.~Kim, S.~Kim, K.~Yun, W.~Cha, R.~Harder, T.~Kawaguchi,
  Y.~Liu, A.~Ulvestad, H.~You, M.~K. Song, and H.~Kim.
\newblock {In Situ Strain Evolution on Pt Nanoparticles during Hydrogen
  Peroxide Decomposition}.
\newblock \emph{Nano Lett.}, 20\penalty0 (12):\penalty0 8541--8548, dec 2020.
\newblock ISSN 1530-6984.
\newblock \doi{10.1021/acs.nanolett.0c03005}.
\newblock URL \url{https://doi.org/10.1021/acs.nanolett.0c03005}.

\bibitem[Cleri and Rosato(1993)]{Cleri1993}
F.~Cleri and V.~Rosato.
\newblock {Tight-binding potentials for transition metals and alloys}.
\newblock \emph{Phys. Rev. B}, 48\penalty0 (1):\penalty0 22--33, jul 1993.
\newblock \doi{10.1103/PhysRevB.48.22}.
\newblock URL \url{https://link.aps.org/doi/10.1103/PhysRevB.48.22}.

\bibitem[{Da Silva} et~al.(2006){Da Silva}, Stampfl, and
  Scheffler]{DaSilva2006}
J.~L.~F. {Da Silva}, C.~Stampfl, and M.~Scheffler.
\newblock {Converged properties of clean metal surfaces by all-electron
  first-principles calculations}.
\newblock \emph{Surf. Sci.}, 600\penalty0 (3):\penalty0 703--715, 2006.
\newblock ISSN 0039-6028.
\newblock \doi{https://doi.org/10.1016/j.susc.2005.12.008}.
\newblock URL
  \url{http://www.sciencedirect.com/science/article/pii/S0039602805013154}.

\bibitem[Dai et~al.(2006)Dai, Kong, Li, and Liu]{Dai2006}
X.~D. Dai, Y.~Kong, J.~H. Li, and B.~X. Liu.
\newblock {Extended Finnis–Sinclair potential for bcc and fcc metals and
  alloys}.
\newblock \emph{J. Phys.: Condens. Matter}, 18\penalty0 (19):\penalty0
  4527--4542, may 2006.
\newblock ISSN 0953-8984.
\newblock \doi{10.1088/0953-8984/18/19/008}.
\newblock URL
  \url{https://iopscience.iop.org/article/10.1088/0953-8984/18/19/008}.

\bibitem[Daio et~al.(2015)Daio, Staykov, Guo, Liu, Tanaka, {Matthew Lyth}, and
  Sasaki]{Daio2015}
T.~Daio, A.~Staykov, L.~Guo, J.~Liu, M.~Tanaka, S.~{Matthew Lyth}, and
  K.~Sasaki.
\newblock {Lattice Strain Mapping of Platinum Nanoparticles on Carbon and SnO2
  Supports}.
\newblock \emph{Sci. Rep.}, 5\penalty0 (1):\penalty0 13126, 2015.
\newblock ISSN 2045-2322.
\newblock \doi{10.1038/srep13126}.
\newblock URL \url{https://doi.org/10.1038/srep13126}.

\bibitem[Daw and Baskes(1983)]{Daw1983}
M.~S. Daw and M.~I. Baskes.
\newblock {Semiempirical, Quantum Mechanical Calculation of Hydrogen
  Embrittlement in Metals}.
\newblock \emph{Phys. Rev. Lett.}, 50\penalty0 (17):\penalty0 1285--1288, apr
  1983.
\newblock \doi{10.1103/PhysRevLett.50.1285}.
\newblock URL \url{https://link.aps.org/doi/10.1103/PhysRevLett.50.1285}.

\bibitem[{De Clercq} et~al.(2016){De Clercq}, Giorgio, and
  Mottet]{DeClercq2016}
A.~{De Clercq}, S.~Giorgio, and C.~Mottet.
\newblock {Pd surface and Pt subsurface segregation in Pt1--c Pd c nanoalloys}.
\newblock \emph{J. Phys.: Condens. Matter}, 28\penalty0 (6):\penalty0 064006,
  feb 2016.
\newblock ISSN 0953-8984.
\newblock \doi{10.1088/0953-8984/28/6/064006}.
\newblock URL
  \url{https://iopscience.iop.org/article/10.1088/0953-8984/28/6/064006}.

\bibitem[De~Waele et~al.(2016)De~Waele, Lejaeghere, Sluydts, and
  Cottenier]{DeWaele2016}
S.~De~Waele, K.~Lejaeghere, M.~Sluydts, and S.~Cottenier.
\newblock Error estimates for density-functional theory predictions of surface
  energy and work function.
\newblock \emph{Phys. Rev. B}, 94:\penalty0 235418, Dec 2016.
\newblock \doi{10.1103/PhysRevB.94.235418}.
\newblock URL \url{https://link.aps.org/doi/10.1103/PhysRevB.94.235418}.

\bibitem[{Deneen Nowak} et~al.(2007){Deneen Nowak}, Mook, Minor, Gerberich, and
  Carter]{DeneenNowak2007}
J.~{Deneen Nowak}, W.~M. Mook, A.~M. Minor, W.~W. Gerberich, and C.~B. Carter.
\newblock Fracturing a nanoparticle.
\newblock \emph{Philos. Mag.}, 87\penalty0 (1):\penalty0 29--37, jan 2007.
\newblock ISSN 1478-6435.
\newblock \doi{10.1080/14786430600876585}.
\newblock URL \url{https://doi.org/10.1080/14786430600876585}.

\bibitem[Dewaele et~al.(2004)Dewaele, Loubeyre, and Mezouar]{Dewaele2004}
A.~Dewaele, P.~Loubeyre, and M.~Mezouar.
\newblock Equations of state of six metals above
  $94\phantom{\rule{0.3em}{0ex}}\mathrm{GPa}$.
\newblock \emph{Phys. Rev. B}, 70:\penalty0 094112, Sep 2004.
\newblock \doi{10.1103/PhysRevB.70.094112}.
\newblock URL \url{https://link.aps.org/doi/10.1103/PhysRevB.70.094112}.

\bibitem[Dorfman et~al.({2012})Dorfman, Prakapenka, Meng, and
  Duffy]{Dorfman2012}
S.~M. Dorfman, V.~B. Prakapenka, Y.~Meng, and T.~S. Duffy.
\newblock {Intercomparison of pressure standards (Au, Pt, Mo, MgO, NaCl and Ne)
  to 2.5 Mbar}.
\newblock \emph{{J Geophys Res-Sol EA }}, {117}, {AUG 30} {2012}.
\newblock ISSN {2169-9313}.
\newblock \doi{{10.1029/2012JB009292}}.

\bibitem[Elkin et~al.(2020)Elkin, Mikhaylov, Ovechkin, and Smirnov]{Elkin2020}
V.~M. Elkin, V.~N. Mikhaylov, A.~A. Ovechkin, and N.~A. Smirnov.
\newblock {A wide-range multiphase equation of state for platinum}.
\newblock \emph{J Phys-Condens Mat}, 32\penalty0 (43):\penalty0 435403, oct
  2020.
\newblock ISSN 0953-8984.
\newblock \doi{10.1088/1361-648X/aba428}.
\newblock URL
  \url{https://iopscience.iop.org/article/10.1088/1361-648X/aba428}.

\bibitem[Fantauzzi et~al.(2014)Fantauzzi, Bandlow, Sabo, Mueller, van Duin, and
  Jacob]{Fantauzzi2014}
D.~Fantauzzi, J.~Bandlow, L.~Sabo, J.~E. Mueller, A.~C.~T. van Duin, and
  T.~Jacob.
\newblock Development of a reaxff potential for pt–o systems describing the
  energetics and dynamics of pt-oxide formation.
\newblock \emph{Phys. Chem. Chem. Phys.}, 16\penalty0 (42):\penalty0
  23118--23133, 2014.
\newblock \doi{10.1039/C4CP03111C}.
\newblock URL \url{http://dx.doi.org/10.1039/C4CP03111C}.

\bibitem[Fei et~al.(2004)Fei, Li, Hirose, Minarik, {Van Orman}, Sanloup, {van
  Westrenen}, Komabayashi, and ichi Funakoshi]{Fei2004}
Y.~Fei, J.~Li, K.~Hirose, W.~Minarik, J.~{Van Orman}, C.~Sanloup, W.~{van
  Westrenen}, T.~Komabayashi, and K.~ichi Funakoshi.
\newblock A critical evaluation of pressure scales at high temperatures by in
  situ x-ray diffraction measurements.
\newblock \emph{Phys Earth Planet In}, 143-144:\penalty0 515--526, 2004.
\newblock ISSN 0031-9201.
\newblock \doi{https://doi.org/10.1016/j.pepi.2003.09.018}.
\newblock URL
  \url{https://www.sciencedirect.com/science/article/pii/S0031920104000858}.
\newblock New Developments in High-Pressure Mineral Physics and Applications to
  the Earth's Interior.

\bibitem[Finnis and Sinclair(1984)]{Finnis1984}
M.~W. Finnis and J.~E. Sinclair.
\newblock {A simple empirical N-body potential for transition metals}.
\newblock \emph{Philos. Mag. A}, 50\penalty0 (1):\penalty0 45--55, jul 1984.
\newblock ISSN 0141-8610.
\newblock \doi{10.1080/01418618408244210}.
\newblock URL \url{https://doi.org/10.1080/01418618408244210}.

\bibitem[Firouzjaie and Mustain(2020)]{Firouzjaie2020}
H.~A. Firouzjaie and W.~E. Mustain.
\newblock {Catalytic Advantages, Challenges, and Priorities in Alkaline
  Membrane Fuel Cells}.
\newblock \emph{ACS Catal.}, 10\penalty0 (1):\penalty0 225--234, jan 2020.
\newblock \doi{10.1021/acscatal.9b03892}.
\newblock URL \url{https://doi.org/10.1021/acscatal.9b03892}.

\bibitem[Foiles(1987)]{Foiles1987}
S.~M. Foiles.
\newblock {Reconstruction of fcc (110) surfaces}.
\newblock \emph{Surf. Sci.}, 191\penalty0 (1):\penalty0 L779--L786, 1987.
\newblock ISSN 0167-2584.
\newblock \doi{https://doi.org/10.1016/0167-2584(87)90889-9}.
\newblock URL
  \url{http://www.sciencedirect.com/science/article/pii/0167258487908899}.

\bibitem[Foiles et~al.(1986)Foiles, Baskes, and Daw]{Foiles1986}
S.~M. Foiles, M.~I. Baskes, and M.~S. Daw.
\newblock {Embedded-atom-method functions for the fcc metals Cu, Ag, Au, Ni,
  Pd, Pt, and their alloys}.
\newblock \emph{Phys. Rev. B}, 33\penalty0 (12):\penalty0 7983--7991, jun 1986.
\newblock \doi{10.1103/PhysRevB.33.7983}.
\newblock URL \url{https://link.aps.org/doi/10.1103/PhysRevB.33.7983}.

\bibitem[Fu et~al.(2013)Fu, Wu, Jiang, Tao, Chen, Lin, Zhou, Wei, Tang, Lu, and
  Xia]{Fu2013}
G.~Fu, K.~Wu, X.~Jiang, L.~Tao, Y.~Chen, J.~Lin, Y.~Zhou, S.~Wei, Y.~Tang,
  T.~Lu, and X.~Xia.
\newblock Polyallylamine-directed green synthesis of platinum nanocubes. shape
  and electronic effect codependent enhanced electrocatalytic activity.
\newblock \emph{Phys. Chem. Chem. Phys.}, 15:\penalty0 3793--3802, 2013.
\newblock \doi{10.1039/C3CP44191A}.
\newblock URL \url{http://dx.doi.org/10.1039/C3CP44191A}.

\bibitem[Garlyyev et~al.(2019)Garlyyev, Kratzl, Rück, Michalička, Fichtner,
  Macak, Kratky, Günther, Cokoja, Bandarenka, Gagliardi, and
  Fischer]{Garlyyev2019}
B.~Garlyyev, K.~Kratzl, M.~Rück, J.~Michalička, J.~Fichtner, J.~M. Macak,
  T.~Kratky, S.~Günther, M.~Cokoja, A.~S. Bandarenka, A.~Gagliardi, and R.~A.
  Fischer.
\newblock Optimizing the size of platinum nanoparticles for enhanced mass
  activity in the electrochemical oxygen reduction reaction.
\newblock \emph{Angew. Chem., Int. Ed.}, 58\penalty0 (28):\penalty0 9596--9600,
  2019.
\newblock \doi{https://doi.org/10.1002/anie.201904492}.
\newblock URL
  \url{https://onlinelibrary.wiley.com/doi/abs/10.1002/anie.201904492}.

\bibitem[Gezelter et~al.(2010)Gezelter, Kuang, Marr, Stocker, Li, Vardeman,
  Lin, Fennell, Sun, Daily, et~al.]{gezelter2010openmd}
J.~Gezelter, S.~Kuang, J.~Marr, K.~Stocker, C.~Li, C.~Vardeman, T.~Lin,
  C.~Fennell, X.~Sun, K.~Daily, et~al.
\newblock Openmd, an open source engine for molecular dynamics.
\newblock \emph{University of Notre Dame, Notre Dame, IN}, 2010.

\bibitem[H{\"{a}}berlen et~al.(1997)H{\"{a}}berlen, Chung, Stener, and
  R{\"{o}}sch]{Haberlen1997}
O.~D. H{\"{a}}berlen, S.-C. Chung, M.~Stener, and N.~R{\"{o}}sch.
\newblock {From clusters to bulk: A relativistic density functional
  investigation on a series of gold clusters Aun, n=6,{\ldots},147}.
\newblock \emph{J. Chem. Phys.}, 106\penalty0 (12):\penalty0 5189--5201, mar
  1997.
\newblock ISSN 0021-9606.
\newblock \doi{10.1063/1.473518}.
\newblock URL \url{https://doi.org/10.1063/1.473518}.

\bibitem[Hale et~al.(2018)Hale, Trautt, and Becker]{Hale2018}
L.~M. Hale, Z.~T. Trautt, and C.~A. Becker.
\newblock {Evaluating variability with atomistic simulations: The effect of
  potential and calculation methodology on the modeling of lattice and elastic
  constants}.
\newblock \emph{Modell. Simul. Mater. Sci. Eng.}, 26\penalty0 (5), may 2018.
\newblock ISSN 1361651X.
\newblock \doi{10.1088/1361-651X/aabc05}.

\bibitem[Harrison et~al.(2018)Harrison, Schall, Maskey, Mikulski, Knippenberg,
  and Morrow]{Harrison2018}
J.~A. Harrison, J.~D. Schall, S.~Maskey, P.~T. Mikulski, M.~T. Knippenberg, and
  B.~H. Morrow.
\newblock {Review of force fields and intermolecular potentials used in
  atomistic computational materials research}.
\newblock \emph{Appl. Phys. Rev.}, 5\penalty0 (3):\penalty0 31104, aug 2018.
\newblock \doi{10.1063/1.5020808}.
\newblock URL \url{https://doi.org/10.1063/1.5020808}.

\bibitem[Hearmon(1946)]{Hearmon1946}
R.~F.~S. Hearmon.
\newblock The elastic constants of anisotropic materials.
\newblock \emph{Rev. Mod. Phys.}, 18:\penalty0 409--440, Jul 1946.
\newblock \doi{10.1103/RevModPhys.18.409}.
\newblock URL \url{https://link.aps.org/doi/10.1103/RevModPhys.18.409}.

\bibitem[Hernandez et~al.(2019)Hernandez, Balasubramanian, Yuan, Mason, and
  Mueller]{Hernandez2019}
A.~Hernandez, A.~Balasubramanian, F.~Yuan, S.~A.~M. Mason, and T.~Mueller.
\newblock {Fast, accurate, and transferable many-body interatomic potentials by
  symbolic regression}.
\newblock \emph{npj Comput. Mater.}, 5\penalty0 (1):\penalty0 112, 2019.
\newblock ISSN 2057-3960.
\newblock \doi{10.1038/s41524-019-0249-1}.
\newblock URL \url{https://doi.org/10.1038/s41524-019-0249-1}.

\bibitem[Hofmeister(1991)]{Hofmeister1991}
A.~M. Hofmeister.
\newblock Pressure derivatives of the bulk modulus.
\newblock \emph{J Geophys Res-Sol EA}, 96\penalty0 (B13):\penalty0
  21893--21907, 1991.
\newblock \doi{https://doi.org/10.1029/91JB02157}.
\newblock URL
  \url{https://agupubs.onlinelibrary.wiley.com/doi/abs/10.1029/91JB02157}.

\bibitem[Holmes et~al.(1989)Holmes, Moriarty, Gathers, and Nellis]{Holmes1989}
N.~C. Holmes, J.~A. Moriarty, G.~R. Gathers, and W.~J. Nellis.
\newblock {The equation of state of platinum to 660 GPa (6.6 Mbar)}.
\newblock \emph{J. Appl. Phys.}, 66\penalty0 (7):\penalty0 2962--2967, oct
  1989.
\newblock ISSN 0021-8979.
\newblock \doi{10.1063/1.344177}.
\newblock URL \url{https://doi.org/10.1063/1.344177}.

\bibitem[Jacobsen et~al.(1987)Jacobsen, Norskov, and Puska]{Jacobsen1987}
K.~W. Jacobsen, J.~K. Norskov, and M.~J. Puska.
\newblock {Interatomic interactions in the effective-medium theory}.
\newblock \emph{Phys. Rev. B}, 35\penalty0 (14):\penalty0 7423--7442, may 1987.
\newblock \doi{10.1103/PhysRevB.35.7423}.
\newblock URL \url{https://link.aps.org/doi/10.1103/PhysRevB.35.7423}.

\bibitem[Jacobsen et~al.(1996)Jacobsen, Stoltze, and N{\o}rskov]{Jacobsen1996}
K.~W. Jacobsen, P.~Stoltze, and J.~K. N{\o}rskov.
\newblock {A semi-empirical effective medium theory for metals and alloys}.
\newblock \emph{Surf. Sci.}, 366\penalty0 (2):\penalty0 394--402, 1996.
\newblock ISSN 0039-6028.
\newblock \doi{https://doi.org/10.1016/0039-6028(96)00816-3}.
\newblock URL
  \url{http://www.sciencedirect.com/science/article/pii/0039602896008163}.

\bibitem[Januszko and Bose(2015)]{Januszko2015}
A.~Januszko and S.~K. Bose.
\newblock {Phonon spectra and temperature variation of bulk properties of Cu,
  Ag, Au and Pt using Sutton–Chen and modified Sutton–Chen potentials}.
\newblock \emph{J. Phys. Chem. Solids}, 82:\penalty0 67--75, 2015.
\newblock ISSN 0022-3697.
\newblock \doi{https://doi.org/10.1016/j.jpcs.2015.03.008}.
\newblock URL
  \url{http://www.sciencedirect.com/science/article/pii/S0022369715000633}.

\bibitem[Jung et~al.(2019)Jung, Braunwarth, and Jacob]{Jung2019}
C.~K. Jung, L.~Braunwarth, and T.~Jacob.
\newblock {Grand Canonical ReaxFF Molecular Dynamics Simulations for Catalytic
  Reactions}.
\newblock \emph{J. Chem. Theory Comput.}, 15\penalty0 (11):\penalty0
  5810--5816, nov 2019.
\newblock ISSN 1549-9618.
\newblock \doi{10.1021/acs.jctc.9b00687}.
\newblock URL \url{https://doi.org/10.1021/acs.jctc.9b00687}.

\bibitem[Khein et~al.(1995)Khein, Singh, and Umrigar]{Khein1995}
A.~Khein, D.~J. Singh, and C.~J. Umrigar.
\newblock {All-electron study of gradient corrections to the local-density
  functional in metallic systems}.
\newblock \emph{Phys. Rev. B}, 51\penalty0 (7):\penalty0 4105--4109, feb 1995.
\newblock \doi{10.1103/PhysRevB.51.4105}.
\newblock URL \url{https://link.aps.org/doi/10.1103/PhysRevB.51.4105}.

\bibitem[Kittel(2005)]{Kittel2005}
C.~Kittel.
\newblock \emph{{Introduction to solid state physics}}.
\newblock Wiley, Hoboken, NJ, 2005.
\newblock ISBN 047141526X 9780471415268 0471680575 9780471680574.

\bibitem[Landolt et~al.(1966)Landolt, Hellwege, Bornstein, and
  Madelung]{Landolt1966}
H.~Landolt, K.~H. Hellwege, R.~Bornstein, and O.~Madelung.
\newblock \emph{{Landolt-Bornstein numerical data and functional relationships
  in science and technology. Group III., Group III.,}}.
\newblock Springer-Verlag, Berlin, 1966.

\bibitem[Ledbetter and Migliori(2006)]{Ledbetter2006}
H.~Ledbetter and A.~Migliori.
\newblock {A general elastic-anisotropy measure}.
\newblock \emph{J. Appl. Phys.}, 100\penalty0 (6):\penalty0 63516, sep 2006.
\newblock ISSN 0021-8979.
\newblock \doi{10.1063/1.2338835}.
\newblock URL \url{https://doi.org/10.1063/1.2338835}.

\bibitem[Lee et~al.(2003)Lee, Shim, and Baskes]{Lee2003}
B.-J. Lee, J.-H. Shim, and M.~I. Baskes.
\newblock {Semiempirical atomic potentials for the fcc metals Cu, Ag, Au, Ni,
  Pd, Pt, Al, and Pb based on first and second nearest-neighbor modified
  embedded atom method}.
\newblock \emph{Phys. Rev. B}, 68\penalty0 (14):\penalty0 144112, oct 2003.
\newblock \doi{10.1103/PhysRevB.68.144112}.
\newblock URL \url{https://link.aps.org/doi/10.1103/PhysRevB.68.144112}.

\bibitem[Li et~al.(2000)Li, Petroski, and El-Sayed]{Li2000}
Y.~Li, J.~Petroski, and M.~A. El-Sayed.
\newblock Activation energy of the reaction between hexacyanoferrate(iii) and
  thiosulfate ions catalyzed by platinum nanoparticles.
\newblock \emph{J. Phys. Chem. B}, 104\penalty0 (47):\penalty0 10956--10959,
  2000.
\newblock \doi{10.1021/jp002569s}.
\newblock URL \url{https://doi.org/10.1021/jp002569s}.

\bibitem[Ludwig et~al.(2006)Ludwig, {G. Vlachos}, {C. T. van Duin}, and {A.
  Goddard}]{Ludwig2006}
J.~Ludwig, D.~{G. Vlachos}, A.~{C. T. van Duin}, and W.~{A. Goddard}.
\newblock {Dynamics of the Dissociation of Hydrogen on Stepped Platinum
  Surfaces Using the ReaxFF Reactive Force Field}.
\newblock \emph{J. Phys. Chem. B}, 110\penalty0 (9):\penalty0 4274--4282, feb
  2006.
\newblock \doi{10.1021/jp0561064}.

\bibitem[Matsui et~al.(2009)Matsui, Ito, Katsura, Yamazaki, Yoshino, Yokoyama,
  and Funakoshi]{Matsui2009}
M.~Matsui, E.~Ito, T.~Katsura, D.~Yamazaki, T.~Yoshino, A.~Yokoyama, and K.-i.
  Funakoshi.
\newblock {The temperature-pressure-volume equation of state of platinum}.
\newblock \emph{J. Appl. Phys.}, 105\penalty0 (1):\penalty0 13505, jan 2009.
\newblock ISSN 0021--8979.
\newblock \doi{10.1063/1.3054331}.
\newblock URL \url{https://doi.org/10.1063/1.3054331}.

\bibitem[Men{\'{e}}ndez-Proupin and Singh(2007)]{Menendez-Proupin2007}
E.~Men{\'{e}}ndez-Proupin and A.~K. Singh.
\newblock {Ab initio calculations of elastic properties of compressed Pt}.
\newblock \emph{Phys. Rev. B}, 76\penalty0 (5):\penalty0 54117, aug 2007.
\newblock \doi{10.1103/PhysRevB.76.054117}.
\newblock URL \url{https://link.aps.org/doi/10.1103/PhysRevB.76.054117}.

\bibitem[Murnaghan(1944)]{Murnaghan1944}
F.~D. Murnaghan.
\newblock {The Compressibility of Media under Extreme Pressures}.
\newblock \emph{Proc. Natl. Acad. Sci.}, 30\penalty0 (9):\penalty0 244--247,
  sep 1944.
\newblock ISSN 0027-8424.
\newblock \doi{10.1073/PNAS.30.9.244}.
\newblock URL \url{https://www.pnas.org/content/30/9/244}.

\bibitem[Nanba et~al.(2017)Nanba, Ishimoto, and Koyama]{Nanba2017}
Y.~Nanba, T.~Ishimoto, and M.~Koyama.
\newblock {Structural Stability of Ruthenium Nanoparticles: A Density
  Functional Theory Study}.
\newblock \emph{J. Phys. Chem. C}, 121\penalty0 (49):\penalty0 27445--27452,
  dec 2017.
\newblock ISSN 1932-7447.
\newblock \doi{10.1021/acs.jpcc.7b08672}.
\newblock URL \url{https://doi.org/10.1021/acs.jpcc.7b08672}.

\bibitem[Narayanan et~al.(2016)Narayanan, Kinaci, {G. Sen}, {J. Davis}, {K.
  Gray}, {K. Y. Chan}, and {K. R. S. Sankaranarayanan}]{Narayanan2016}
B.~Narayanan, A.~Kinaci, F.~{G. Sen}, M.~{J. Davis}, S.~{K. Gray}, M.~{K. Y.
  Chan}, and S.~{K. R. S. Sankaranarayanan}.
\newblock {Describing the Diverse Geometries of Gold from Nanoclusters to
  Bulk—A First-Principles-Based Hybrid Bond-Order Potential}.
\newblock \emph{J. Phys. Chem. C}, 120\penalty0 (25):\penalty0 13787--13800,
  jun 2016.
\newblock \doi{10.1021/acs.jpcc.6b02934}.

\bibitem[{Nilsson Pingel} et~al.(2018){Nilsson Pingel}, J{\o}rgensen,
  Yankovich, Gr{\"{o}}nbeck, and Olsson]{NilssonPingel2018}
T.~{Nilsson Pingel}, M.~J{\o}rgensen, A.~B. Yankovich, H.~Gr{\"{o}}nbeck, and
  E.~Olsson.
\newblock {Influence of atomic site-specific strain on catalytic activity of
  supported nanoparticles}.
\newblock \emph{Nat. Commun.}, 9\penalty0 (1):\penalty0 2722, 2018.
\newblock ISSN 2041-1723.
\newblock \doi{10.1038/s41467-018-05055-1}.
\newblock URL \url{https://doi.org/10.1038/s41467-018-05055-1}.

\bibitem[{P. Yl{\"{a}}-M{\"{a}}ih{\"{a}}niemi} et~al.(2008){P.
  Yl{\"{a}}-M{\"{a}}ih{\"{a}}niemi}, {Y. Y. Heng}, Thielmann, and {R.
  Williams}]{P.Yla-Maihaniemi2008}
P.~{P. Yl{\"{a}}-M{\"{a}}ih{\"{a}}niemi}, J.~{Y. Y. Heng}, F.~Thielmann, and
  D.~{R. Williams}.
\newblock {Inverse Gas Chromatographic Method for Measuring the Dispersive
  Surface Energy Distribution for Particulates}.
\newblock \emph{Langmuir}, 24\penalty0 (17):\penalty0 9551--9557, aug 2008.
\newblock \doi{10.1021/la801676n}.

\bibitem[Pal and Pal(2015)]{Pal2015}
J.~Pal and T.~Pal.
\newblock {Faceted metal and metal oxide nanoparticles: design, fabrication and
  catalysis}.
\newblock \emph{Nanoscale}, 7\penalty0 (34):\penalty0 14159--14190, 2015.
\newblock ISSN 2040-3364.
\newblock \doi{10.1039/C5NR03395K}.
\newblock URL \url{http://dx.doi.org/10.1039/C5NR03395K}.

\bibitem[Panagiotides and Papanicolaou(2010)]{Panagiotides2010}
N.~Panagiotides and N.~I. Papanicolaou.
\newblock {Diffusion of platinum adatoms and dimers on Pt(111) surface by
  molecular-dynamics simulation}.
\newblock \emph{Int. J. Quantum Chem.}, 110\penalty0 (1):\penalty0 202--209,
  jan 2010.
\newblock ISSN 0020-7608.
\newblock \doi{10.1002/qua.22045}.
\newblock URL \url{https://doi.org/10.1002/qua.22045}.

\bibitem[Panizon and Ferrando(2015)]{Panizon2015}
E.~Panizon and R.~Ferrando.
\newblock {Solid--solid transitions in Pd-Pt nanoalloys}.
\newblock \emph{Phys. Rev. B}, 92\penalty0 (20):\penalty0 205417, nov 2015.
\newblock \doi{10.1103/PhysRevB.92.205417}.
\newblock URL \url{https://link.aps.org/doi/10.1103/PhysRevB.92.205417}.

\bibitem[Papanicolaou and Panagiotides(2009)]{Papanicolaou2009}
N.~I. Papanicolaou and N.~Panagiotides.
\newblock Interatomic potential for platinum and self-diffusion on pt(111)
  surface by molecular-dynamics simulation.
\newblock In N.~Russo, V.~Y. Antonchenko, and E.~S. Kryachko, editors,
  \emph{SelfOrganization of Molecular Systems}, pages 335--344, Dordrecht,
  2009. Springer Netherlands.
\newblock ISBN 978-90-481-2590-6.

\bibitem[Parakh et~al.(2020)Parakh, Lee, Harkins, Kiani, Doan, Kunz, Doran,
  Hanson, Ryu, and Gu]{Parakh2020}
A.~Parakh, S.~Lee, K.~A. Harkins, M.~T. Kiani, D.~Doan, M.~Kunz, A.~Doran,
  L.~Hanson, S.~Ryu, and X.~W. Gu.
\newblock {Nucleation of Dislocations in 3.9 nm Nanocrystals at High Pressure}.
\newblock \emph{Phys. Rev. Lett.}, 124\penalty0 (10):\penalty0 106104, mar
  2020.
\newblock \doi{10.1103/PhysRevLett.124.106104}.
\newblock URL \url{https://link.aps.org/doi/10.1103/PhysRevLett.124.106104}.

\bibitem[Patra et~al.(2017)Patra, Bates, Sun, and Perdew]{Patra2017}
A.~Patra, J.~E. Bates, J.~Sun, and J.~P. Perdew.
\newblock Properties of real metallic surfaces: Effects of density functional
  semilocality and van der waals nonlocality.
\newblock \emph{Proceedings of the National Academy of Sciences}, 114\penalty0
  (44):\penalty0 E9188--E9196, 2017.
\newblock ISSN 0027-8424.
\newblock \doi{10.1073/pnas.1713320114}.
\newblock URL \url{https://www.pnas.org/content/114/44/E9188}.

\bibitem[Plimpton(1995)]{Plimpton1995}
S.~Plimpton.
\newblock {Fast Parallel Algorithms for Short-Range Molecular Dynamics}.
\newblock \emph{J. Comput. Phys.}, 117\penalty0 (1):\penalty0 1--19, 1995.
\newblock ISSN 0021-9991.
\newblock \doi{https://doi.org/10.1006/jcph.1995.1039}.
\newblock URL
  \url{http://www.sciencedirect.com/science/article/pii/S002199918571039X}.

\bibitem[Rioux et~al.(2008)Rioux, Song, Yang, and Somorjai]{Rioux2008}
R.~M. Rioux, H.~Song, P.~Yang, and G.~A. Somorjai.
\newblock Chapter 7 - platinum nanoclusters’ size and surface structure
  sensitivity of catalytic reactions.
\newblock In B.~Corain, G.~Schmid, and N.~Toshima, editors, \emph{Met.
  Nanoclusters Catal. Mater. Sci.: Issue Size Control}, pages 149--166.
  Elsevier, Amsterdam, 2008.
\newblock ISBN 978-0-444-53057-8.
\newblock \doi{https://doi.org/10.1016/B978-044453057-8.50009-X}.
\newblock URL
  \url{https://www.sciencedirect.com/science/article/pii/B978044453057850009X}.

\bibitem[Rodriguez et~al.(1996)Rodriguez, Amiens, Chaudret, Casanove, Lecante,
  and {S. Bradley}]{Rodriguez1996}
A.~Rodriguez, C.~Amiens, B.~Chaudret, M.-J. Casanove, P.~Lecante, and J.~{S.
  Bradley}.
\newblock {Synthesis and Isolation of Cuboctahedral and Icosahedral Platinum
  Nanoparticles. Ligand-Dependent Structures}.
\newblock \emph{Chem. Mater.}, 8\penalty0 (8):\penalty0 1978--1986, aug 1996.
\newblock \doi{10.1021/cm960338l}.
\newblock URL \url{https://pubs.acs.org/sharingguidelines}.

\bibitem[{Roldan Cuenya}(2013)]{RoldanCuenya2013}
B.~{Roldan Cuenya}.
\newblock {Metal Nanoparticle Catalysts Beginning to Shape-up}.
\newblock \emph{Acc. Chem. Res.}, 46\penalty0 (8):\penalty0 1682--1691, aug
  2013.
\newblock ISSN 0001-4842.
\newblock \doi{10.1021/ar300226p}.
\newblock URL \url{https://doi.org/10.1021/ar300226p}.

\bibitem[Sanz-Navarro et~al.(2008)Sanz-Navarro, {\AA}strand, Chen, R{\o}nning,
  van Duin, Jacob, and Goddard]{Sanz-Navarro2008}
C.~F. Sanz-Navarro, P.-O. {\AA}strand, D.~Chen, M.~R{\o}nning, A.~C.~T. van
  Duin, T.~Jacob, and W.~A. Goddard.
\newblock {Molecular Dynamics Simulations of the Interactions between Platinum
  Clusters and Carbon Platelets}.
\newblock \emph{J. Phys. Chem. A}, 112\penalty0 (7):\penalty0 1392--1402, feb
  2008.
\newblock ISSN 1089-5639.
\newblock \doi{10.1021/jp074806y}.
\newblock URL \url{https://pubs.acs.org/doi/10.1021/jp074806y}.

\bibitem[Scheerschmidt(2007)]{Scheerschmidt2007}
K.~Scheerschmidt.
\newblock \emph{Empirical Molecular Dynamics: Possibilities, Requirements, and
  Limitations}, pages 213--244.
\newblock Springer Berlin Heidelberg, Berlin, Heidelberg, 2007.
\newblock ISBN 978-3-540-33401-9.
\newblock \doi{10.1007/11690320_10}.
\newblock URL \url{https://doi.org/10.1007/11690320_10}.

\bibitem[Shan et~al.(2010)Shan, Devine, Kemper, Sinnott, and
  Phillpot]{Shan2010}
T.-R. Shan, B.~D. Devine, T.~W. Kemper, S.~B. Sinnott, and S.~R. Phillpot.
\newblock {Charge-optimized many-body potential for the hafnium/hafnium oxide
  system}.
\newblock \emph{Phys. Rev. B}, 81\penalty0 (12):\penalty0 125328, mar 2010.
\newblock \doi{10.1103/PhysRevB.81.125328}.
\newblock URL \url{https://link.aps.org/doi/10.1103/PhysRevB.81.125328}.

\bibitem[Simmons(1965)]{Simmons1965}
G.~Simmons.
\newblock \emph{{Single Crystal Elastic Constants and Calculated Aggregate
  Properties}}.
\newblock J. Grad. Res. Cent. Southern Methodist University Press, 1965.
\newblock URL \url{https://books.google.com/books?id=4q5esM1iatMC}.

\bibitem[Singh and Vannice(2001)]{Singh2001}
U.~K. Singh and M.~Vannice.
\newblock {Kinetics of liquid-phase hydrogenation reactions over supported
  metal catalysts — a review}.
\newblock \emph{Appl. Catal., A}, 213\penalty0 (1):\penalty0 1--24, 2001.
\newblock ISSN 0926-860X.
\newblock \doi{https://doi.org/10.1016/S0926-860X(00)00885-1}.
\newblock URL
  \url{http://www.sciencedirect.com/science/article/pii/S0926860X00008851}.

\bibitem[Singh-Miller and Marzari(2009)]{Singh-Miller2009}
N.~E. Singh-Miller and N.~Marzari.
\newblock {Surface energies, work functions, and surface relaxations of
  low-index metallic surfaces from first principles}.
\newblock \emph{Phys. Rev. B}, 80\penalty0 (23):\penalty0 235407, dec 2009.
\newblock \doi{10.1103/PhysRevB.80.235407}.
\newblock URL \url{https://link.aps.org/doi/10.1103/PhysRevB.80.235407}.

\bibitem[Song et~al.(2004)Song, Kim, Connor, {A. Somorjai}, and Yang]{Song2004}
H.~Song, F.~Kim, S.~Connor, G.~{A. Somorjai}, and P.~Yang.
\newblock {Pt Nanocrystals: Shape Control and Langmuir-Blodgett Monolayer
  Formation}.
\newblock \emph{J. Phys. Chem. B}, 109\penalty0 (1):\penalty0 188--193, dec
  2004.
\newblock \doi{10.1021/jp0464775}.

\bibitem[Steinhauser and Hiermaier(2009)]{Steinhauser2009}
M.~O. Steinhauser and S.~Hiermaier.
\newblock A review of computational methods in materials science: Examples from
  shock-wave and polymer physics.
\newblock \emph{Int J Mol Sci}, 10\penalty0 (12):\penalty0 5135--5216, 2009.
\newblock ISSN 1422-0067.
\newblock \doi{10.3390/ijms10125135}.
\newblock URL \url{https://www.mdpi.com/1422-0067/10/12/5135}.

\bibitem[Stephen et~al.(2019)Stephen, Rees, Mikheenko, and
  Macaskie]{Stephen2019}
A.~J. Stephen, N.~V. Rees, I.~Mikheenko, and L.~E. Macaskie.
\newblock Platinum and palladium bio-synthesized nanoparticles as sustainable
  fuel cell catalysts.
\newblock \emph{Front. Energy Res.}, 7:\penalty0 66, 2019.
\newblock ISSN 2296-598X.
\newblock \doi{10.3389/fenrg.2019.00066}.
\newblock URL
  \url{https://www.frontiersin.org/article/10.3389/fenrg.2019.00066}.

\bibitem[Stukowski({2010})]{ovito}
A.~Stukowski.
\newblock {Visualization and analysis of atomistic simulation data with
  OVITO-the Open Visualization Tool}.
\newblock \emph{{Model. Simul. Mater. SC}}, {18}\penalty0 ({1}), {JAN} {2010}.
\newblock ISSN {0965-0393}.
\newblock \doi{{10.1088/0965-0393/18/1/015012}}.

\bibitem[Sun et~al.(2008)Sun, Umemoto, Wu, Zheng, and Wentzcovitch]{Sun2008}
T.~Sun, K.~Umemoto, Z.~Wu, J.-C. Zheng, and R.~M. Wentzcovitch.
\newblock {Lattice dynamics and thermal equation of state of platinum}.
\newblock \emph{Phys. Rev. B}, 78\penalty0 (2):\penalty0 24304, jul 2008.
\newblock \doi{10.1103/PhysRevB.78.024304}.
\newblock URL \url{https://link.aps.org/doi/10.1103/PhysRevB.78.024304}.

\bibitem[Tadmor et~al.(2011)Tadmor, Elliott, Sethna, Miller, and
  Becker]{Tadmor2011}
E.~B. Tadmor, R.~S. Elliott, J.~P. Sethna, R.~E. Miller, and C.~A. Becker.
\newblock {The potential of atomistic simulations and the knowledgebase of
  interatomic models}.
\newblock \emph{JOM}, 63\penalty0 (7):\penalty0 17, 2011.
\newblock ISSN 1543-1851.
\newblock \doi{10.1007/s11837-011-0102-6}.
\newblock URL \url{https://doi.org/10.1007/s11837-011-0102-6}.

\bibitem[Tersoff(1988)]{Tersoff1988}
J.~Tersoff.
\newblock {New empirical approach for the structure and energy of covalent
  systems}.
\newblock \emph{Phys. Rev. B}, 37\penalty0 (12):\penalty0 6991--7000, apr 1988.
\newblock \doi{10.1103/PhysRevB.37.6991}.
\newblock URL \url{https://link.aps.org/doi/10.1103/PhysRevB.37.6991}.

\bibitem[Tran et~al.(2016)Tran, Xu, Radhakrishnan, Winston, Sun, Persson, and
  Ong]{Tran2016}
R.~Tran, Z.~Xu, B.~Radhakrishnan, D.~Winston, W.~Sun, K.~A. Persson, and S.~P.
  Ong.
\newblock {Surface energies of elemental crystals}.
\newblock \emph{Sci. Data}, 3\penalty0 (1):\penalty0 160080, 2016.
\newblock ISSN 2052-4463.
\newblock \doi{10.1038/sdata.2016.80}.
\newblock URL \url{https://doi.org/10.1038/sdata.2016.80}.

\bibitem[Tyson and Miller(1977)]{Tyson1977}
W.~R. Tyson and W.~A. Miller.
\newblock {Surface free energies of solid metals: Estimation from liquid
  surface tension measurements}.
\newblock \emph{Surf. Sci.}, 62\penalty0 (1):\penalty0 267--276, 1977.
\newblock ISSN 0039-6028.
\newblock \doi{https://doi.org/10.1016/0039-6028(77)90442-3}.
\newblock URL
  \url{http://www.sciencedirect.com/science/article/pii/0039602877904423}.

\bibitem[Vega et~al.(2018)Vega, Ruvireta, Viñes, and Illas]{Vega2018}
L.~Vega, J.~Ruvireta, F.~Viñes, and F.~Illas.
\newblock Jacob’s ladder as sketched by escher: Assessing the performance of
  broadly used density functionals on transition metal surface properties.
\newblock \emph{J. Chem. Theory Comput.}, 14\penalty0 (1):\penalty0 395--403,
  2018.
\newblock \doi{10.1021/acs.jctc.7b01047}.
\newblock URL \url{https://doi.org/10.1021/acs.jctc.7b01047}.
\newblock PMID: 29182868.

\bibitem[Verga and Skylaris(2018)]{Verga2018}
L.~G. Verga and C.-K. Skylaris.
\newblock Chapter 8 - dft modeling of metallic nanoparticles.
\newblock In S.~T. Bromley and S.~M. Woodley, editors, \emph{Computational
  Modelling of Nanoparticles}, volume~12 of \emph{Frontiers of Nanoscience},
  pages 239--293. Elsevier, 2018.
\newblock \doi{https://doi.org/10.1016/B978-0-08-102232-0.00008-7}.
\newblock URL
  \url{https://www.sciencedirect.com/science/article/pii/B9780081022320000087}.

\bibitem[Vosko et~al.(1980)Vosko, Wilk, and Nusair]{Vosko1980}
S.~H. Vosko, L.~Wilk, and M.~Nusair.
\newblock {Accurate spin-dependent electron liquid correlation energies for
  local spin density calculations: a critical analysis}.
\newblock \emph{Can. J. Phys.}, 59:\penalty0 1200, aug 1980.
\newblock ISSN 0008-4204.
\newblock \doi{10.1139/p80-159}.
\newblock URL \url{https://ui.adsabs.harvard.edu/abs/1980CaJPh..58.1200V}.

\bibitem[Wang et~al.(2009)Wang, Inada, Wu, Zhu, Choi, Liu, Zhou, and
  Adzic]{Wang2009}
J.~X. Wang, H.~Inada, L.~Wu, Y.~Zhu, Y.~Choi, P.~Liu, W.-P. Zhou, and R.~R.
  Adzic.
\newblock {Oxygen Reduction on Well--Defined Core--Shell Nanocatalysts:
  Particle Size, Facet, and Pt Shell Thickness Effects}.
\newblock \emph{J. Am. Chem. Soc.}, 131\penalty0 (47):\penalty0 17298--17302,
  dec 2009.
\newblock ISSN 0002-7863.
\newblock \doi{10.1021/ja9067645}.
\newblock URL \url{https://doi.org/10.1021/ja9067645}.

\bibitem[Wang et~al.(2020)Wang, Fugetsu, Sakata, Fujisue, Kabayama, Tahara, and
  Morisawa]{Wang2020}
Y.~Wang, B.~Fugetsu, I.~Sakata, C.~Fujisue, S.~Kabayama, N.~Tahara, and
  S.~Morisawa.
\newblock {Monolayered Platinum Nanoparticles as Efficient Electrocatalysts for
  the Mass Production of Electrolyzed Hydrogen Water}.
\newblock \emph{Sci. Rep.}, 10\penalty0 (1):\penalty0 10126, 2020.
\newblock ISSN 2045-2322.
\newblock \doi{10.1038/s41598-020-67107-1}.
\newblock URL \url{https://doi.org/10.1038/s41598-020-67107-1}.

\bibitem[Wei et~al.(2016)Wei, Zhao, Fisher, Zhu, and Cheng]{Wei2016}
C.~Wei, Z.~Zhao, A.~Fisher, J.~Zhu, and D.~Cheng.
\newblock {Theoretical Study on the Structures and Thermal Properties of
  Ag–Pt–Ni Trimetallic Clusters}.
\newblock \emph{J. Cluster Sci.}, 27\penalty0 (6):\penalty0 1849--1861, 2016.
\newblock ISSN 1572-8862.
\newblock \doi{10.1007/s10876-016-1068-x}.
\newblock URL \url{https://doi.org/10.1007/s10876-016-1068-x}.

\bibitem[Wu et~al.(2017)Wu, Gao, Yang, and Zuo]{Wu2017}
J.~Wu, W.~Gao, H.~Yang, and J.-M. Zuo.
\newblock {Dissolution Kinetics of Oxidative Etching of Cubic and Icosahedral
  Platinum Nanoparticles Revealed by in Situ Liquid Transmission Electron
  Microscopy}.
\newblock \emph{ACS Nano}, 11\penalty0 (2):\penalty0 1696--1703, feb 2017.
\newblock \doi{10.1021/acsnano.6b07541}.

\bibitem[Xiao and Wang(2004)]{Xiao2004}
L.~Xiao and L.~Wang.
\newblock {Structures of Platinum Clusters: Planar or Spherical?}
\newblock \emph{J. Phys. Chem. A}, 108\penalty0 (41):\penalty0 8605--8614, oct
  2004.
\newblock ISSN 1089-5639.
\newblock \doi{10.1021/jp0485035}.
\newblock URL \url{https://doi.org/10.1021/jp0485035}.

\bibitem[Ye et~al.(2017)Ye, Prakapenka, Meng, and Shim]{Ye2017}
Y.~Ye, V.~Prakapenka, Y.~Meng, and S.-H. Shim.
\newblock {Intercomparison of the gold, platinum, and MgO pressure scales up to
  140 GPa and 2500 K}.
\newblock \emph{J. Geophys. Res.: Solid Earth}, 122\penalty0 (5):\penalty0
  3450--3464, may 2017.
\newblock ISSN 2169-9313.
\newblock \doi{https://doi.org/10.1002/2016JB013811}.
\newblock URL \url{https://doi.org/10.1002/2016JB013811}.

\bibitem[Yokoo et~al.({2009})Yokoo, Kawai, Nakamura, Kondo, Tange, and
  Tsuchiya]{Yokoo2009}
M.~Yokoo, N.~Kawai, K.~G. Nakamura, K.-i. Kondo, Y.~Tange, and T.~Tsuchiya.
\newblock {Ultrahigh-pressure scales for gold and platinum at pressures up to
  550 GPa}.
\newblock \emph{{Phys. Rev. B}}, {80}\penalty0 ({10}), {SEP} {2009}.
\newblock ISSN {2469-9950}.
\newblock \doi{{10.1103/PhysRevB.80.104114}}.

\bibitem[Yun et~al.(2012)Yun, Cho, Cha, Lee, Nam, Oh, Choi, and Lee]{Yun2012}
K.~Yun, Y.-H. Cho, P.-R. Cha, J.~Lee, H.-S. Nam, J.~S. Oh, J.-H. Choi, and
  S.-C. Lee.
\newblock {Monte Carlo simulations of the structure of Pt-based bimetallic
  nanoparticles}.
\newblock \emph{Acta Mater.}, 60\penalty0 (12):\penalty0 4908--4916, 2012.
\newblock ISSN 1359-6454.
\newblock \doi{https://doi.org/10.1016/j.actamat.2012.05.032}.
\newblock URL
  \url{http://www.sciencedirect.com/science/article/pii/S1359645412003527}.

\bibitem[Zang et~al.(2015)Zang, Li, Wang, and Zhang]{Zang2015}
W.~Zang, G.~Li, L.~Wang, and X.~Zhang.
\newblock {Catalytic hydrogenation by noble-metal nanocrystals with
  well-defined facets: a review}.
\newblock \emph{Catal. Sci. Technol.}, 5\penalty0 (5):\penalty0 2532--2553,
  2015.
\newblock ISSN 2044-4753.
\newblock \doi{10.1039/C4CY01619J}.
\newblock URL \url{http://dx.doi.org/10.1039/C4CY01619J}.

\bibitem[Zha et~al.(2008)Zha, Mibe, Bassett, Tschauner, Mao, and
  Hemley]{Zha2008}
C.-S. Zha, K.~Mibe, W.~A. Bassett, O.~Tschauner, H.-K. Mao, and R.~J. Hemley.
\newblock P-v-t equation of state of platinum to 80gpa and 1900k from internal
  resistive heating/x-ray diffraction measurements.
\newblock \emph{J. Appl. Phys.}, 103\penalty0 (5):\penalty0 054908, 2008.
\newblock \doi{10.1063/1.2844358}.
\newblock URL \url{https://doi.org/10.1063/1.2844358}.

\bibitem[Zhou(2003)]{Zhou-2003}
M.~Zhou.
\newblock A new look at the atomic level virial stress: on continuum-molecular
  system equivalence.
\newblock \emph{P Roy Soc Lond A Mat}, 459\penalty0 (2037):\penalty0
  2347--2392, 2003.
\newblock \doi{10.1098/rspa.2003.1127}.
\newblock URL
  \url{https://royalsocietypublishing.org/doi/abs/10.1098/rspa.2003.1127}.

\bibitem[Zhou et~al.(2004)Zhou, Johnson, and Wadley]{Zhou2004}
X.~W. Zhou, R.~A. Johnson, and H.~N.~G. Wadley.
\newblock {Misfit-energy-increasing dislocations in vapor-deposited CoFe/NiFe
  multilayers}.
\newblock \emph{Phys. Rev. B}, 69\penalty0 (14):\penalty0 144113, apr 2004.
\newblock \doi{10.1103/PhysRevB.69.144113}.
\newblock URL \url{https://link.aps.org/doi/10.1103/PhysRevB.69.144113}.

\end{thebibliography}

\end{document}